\documentclass{ws-ijmpd}

\usepackage{mathrsfs}

\newcommand{\diff}{\mathrm{d}}
\newcommand{\Diff}{\mathrm{D}}
\renewcommand{\vec}[1]{\mbox{\protect\boldmath$#1$}}

\begin{document}
\bibliographystyle{plain}

\markboth{Jan Schee and Zden\v{e}k Stuchl\'{i}k}
{Optical phenomena in brany Kerr spacetimes}

\title{Optical phenomena in the field of braneworld Kerr black holes}

\author{Jan Schee}

\address{
        Institute of Physics, Faculty of Philosophy and Science, Silesian University in Opava, Bezru\v{c}ovo n\'{a}m. 13, 
CZ-746 01 Opava, Czech Republic\\
schee@email.cz
}

\author{Zden\v{e}k Stuchl\'{i}k}

\address{
        Institute of Physics, Faculty of Philosophy and Science, Silesian University in Opava, Bezru\v{c}ovo n\'{a}m. 13, 
CZ-746 01 Opava, Czech Republic\\
Zdenek.Stuchlik@fpf.slu.cz
}

\maketitle

\begin{abstract}
We study the influence of the tidal charge parameter of the braneworld models onto some optical phenomena in rotating black hole spacetimes. The escape photon cones are determined for special families of locally non-rotating, circular geodetical and radially free falling observers. The silhuette of a rotating black hole, the shape of an equatorial thin accretion disk and time delay effect for direct and indirect images of a radiation hot spot orbiting the black hole are given and classified in terms of the black hole rotational and tidal parameters. It is shown that rising of negatively-valued tidal parameter, with rotational parameter fixed, generally strenghtens the relativistic effects and suppresses the rotation induced asymmetries in the optical phenomena.
\end{abstract}



\section{\label{sec:Intro}Introduction}

One way of realizing theories describing gravity as a truly
higher-dimensional interaction becoming effectively 4D at low-enough 
energies  is represented by the braneworld models, where the
observable universe is a 3-brane (domain wall) to which the
standard model (non-gravitational) matter fields are confined,
while gravity field enters the extra spatial dimensions the size
of which may be much larger than the Planck length scale
$l_\mathrm{P}\sim 10^{-33}\, \mathrm{cm}$ \cite{Ark-Dim-Dva:1998:}.

As shown by Randall and Sundrum \cite{Ran-Sun:1999:}, gravity can be localized near the brane at low energies even with
a non-compact, infinite size extra dimension 
with the warped spacetime satisfying the 5D Einstein equations
with negative cosmological constant. Then an arbitrary energy-momentum tensor
could be allowed on the brane \cite{Shi-Mae-Sas:2000:}.

The Randall-Sundrum model gives 4D Einstein gravity in low energy
limit, and the conventional potential of Newtonian gravity appears
on the 3-brane with high accuracy \cite{Ran-Sun:1999:}. Significant deviations from the
Einstein gravity occur at very high energies, e.g., in the very
early universe, and in vicinity of compact objects
\cite{Maa:2004:,Dad-etal:2000:,Ger-Maa:2001:,Ali-Gum:2005:}.
Gravitational collapse of matter trapped on the brane results in
black holes mainly localized on the brane, but their horizon could
be extended into the extra dimension.  The high-energy effects
produced by the gravitational collapse are disconnected from the
outside space by the horizon, but they could have a signature on
the brane, influencing properties of black holes
\cite{Maa:2004:}. There are high-energy effects of local
character influencing pressure in collapsing matter, and also
non-local corrections of ``backreaction'' character arising from
the influence of the Weyl curvature of the bulk space on the brane
-- the matter on the brane induces Weyl curvature in the bulk
which makes influence on the structures on the brane due to the bulk
graviton stresses \cite{Maa:2004:}. The combination of
high-energy (local) and bulk stress (non-local) effects alters
significantly the matching problem on the brane, compared to the
4D Einstein gravity; for spherical objects, matching no longer
leads to a Schwarzschild exterior in general
\cite{Maa:2004:,Ger-Maa:2001:}. The Weyl stresses
induced by bulk gravitons imply that the matching conditions do
not have unique solution on the brane; in fact, knowledge of the
5D Weyl tensor is needed as a minimum condition for uniqueness
\cite{Ger-Maa:2001:}.\footnote{At present, no exact 5D
solution in the braneworld model is known.} Some solutions for
spherically symmetric black holes \cite{Dad-etal:2000:} and
uniform density stars \cite{Ger-Maa:2001:} have been discussed. It has been shown
that in the black hole case the matching conditions could be satisfied and the bulk effects 
on the black hole spacetimes could be represented by a single ``brany`` parameter. 

Assuming spherically symmetric metric induced on
the 3-brane, the constrained effective gravitational field equations on the
brane could be solved, giving  Reissner-Nordstr\"{o}m static
black hole solutions endowed with a braneworld parameter $b$ having character of a ``tidal'' charge,
 instead of the standard electric charge
parameter $Q^2$ \cite{Dad-etal:2000:}. The tidal charge  can be both positive and negative, however, there are some
indications that negative tidal charge should properly represent the
``backreaction'' effects of the bulk space Weyl tensor on the
brane \cite{Dad-etal:2000:}.

The stationary and axisymmetric solutions describing
rotating black holes localized in the Randall-Sundrum braneworld
were derived in \cite{Ali-Gum:2005:}, having the metric tensor of the
Kerr-Newman form with a tidal charge describing the 5D correction
term generated by the 5D Weyl tensor stresses. The tidal charge
has an ``electric'' character again and arises due to the 5D
gravitational coupling between the brane and the bulk, reflected
on the brane through the ``electric'' part of the bulk Weyl tensor
\cite{Ali-Gum:2005:}, in analogy with the spherically
symmetric case \cite{Dad-etal:2000:}.

When both the tidal and electric charge are present the
black hole spacetime structure is much more complex and additional off-diagonal metric components $g_{r\phi}$,
 $g_{rt}$ are relevant along with the standard $g_{\phi t}$ component, 
due to the combination of the local bulk effects and the
rotational dragging. This distorts the event horizon which
becomes a stack of non-uniformly rotating null circles having
different radii at fixed $\theta$ while going from the equatorial
plane to the poles \cite{Ali-Gum:2005:}. The uniformly
rotating horizon is recovered for the rotation
parameter $a$ small enough  where Kerr-Newman form of the metric tensor is allowed describing charged and slowly rotating black holes \cite{Ali-Gum:2005:}. In the absence of rotation, the metric tensor reduces
to the Reissner-Nordstr\"{o}m form with correction term of non-local origin
 \cite{Cha-etal:2001:}.

Here we restrict our attention to the Kerr-Newman type of
solutions describing the braneworld rotating black holes with no
electric charge, since in astrophysically relevant situations the
electric charge of the black hole must be exactly zero, or very
small \cite{MTW}. Then the results obtained in analysing the
behaviour of test particles and photons or test fields around the
Kerr-Newman black holes could be used assuming both positive and
negative values of the braneworld tidal parameter $b$ (used instead of
charge parameter $Q^2$).

The information on the properties of strong gravitational fields in vicinity of compact objects, namely
of black holes, is encoded into optical phenomena of different kind that enable us to make estimates of the black hole parameters, including its tidal charge, when predictions of the theoretical models are confronted with the observed data. From this point of view, the spectral profiles of accretion discs around the black holes in galactic binaries, e.g., in microquasars, are most promising \cite{Nar-Mcl-Sha:2007:,McCli-Nar-Sha:2007:}, along with profiled spectral lines in the X-ray flux \cite{Laor:1991:,Bao-Stu:1992:,Stu-Bao:1992:,Kar-Vok-Pol:1992:,Mat-Fab-Ros:1993:,Zak:2003:}. Important information could also be obtained from the quasiperiodic oscillations observed in the X-ray flux of some low-mass black hole binaries of Galactic origin \cite{Rem-McCli:2006:ARASTRA:}, some expected intermediate black hole sources \cite{Stroh:2007a:}, or those observed in Galactic nuclei \cite{Asch:2004:ASTRA:,Asch:2007:}. In the case of our Galaxy centre black hole Sgr A$^*$, we could be able to measure the optical phenomena in more detailed form as compared with the other sources, since it is the nearest supermassive black hole with mass estimated to be $\sim 4\times 10^6 M_\odot$ \cite{Ghez:2005:}, enabling to measure the "silhuette`` of the black hole and other subtle GR phenomena \cite{Bardeen:1973:,Cun-Bar:1973:}.
\par
In the present paper, we give an introductory study of the tidal charge influence on the optical phenomena near a rotating black hole. We focus our attention to some characteristic phenomena in close vicinity of the black-hole horizon, where the effects of the tidal charge could be in principle of the same order as those of the black hole mass and spin, contrary to the case of weak lensing effects. The light escape cones are given for families of astrophysically interesting sources, namely in locally non-rotating frames, and frames related to circular geodetical motion and radially free-falling sources in section 4 \cite{SSJ:RAGTime:2005:Proceedings}. The silhuette of the black hole is determined in section 5. Images of the accretion discs are determined in section 6 using the transfer-function method. In Section 7, time delay of hot spot radiation is determined for direct and indirect images assuming circular geodetical motion in close vicinity of the black hole horizon. In Section 8 relevance of some effects is estimated for the Galaxy centre Sgr $A^*$ supermassive black hole. Concluding remarks are presented in Section 9.

\section{\label{sec:GravFielEqOnBrane}Gravitational field equations on the brane}

In the 5D warped space models of Randall and Sundrum, involving a non-compact extra dimension, the gravitational field equations in the bulk can be expressed in the form \cite{Shi-Mae-Sas:2000:,Dad-etal:2000:}

\begin{equation}
 \tilde{G}_{AB}=\tilde{k}^2[-\tilde\Lambda g_{AB}+\delta(\chi)(-\lambda g_{AB}+T_{AB})],\label{beq1}
\end{equation}
where the fundamental 5D Planck mass $\tilde M_P$ enters via $\tilde{k}^2=8\pi/\tilde{M}_p^3 $, $\lambda$ is the brane tension, and $\tilde\Lambda$ is the negative bulk cosmological constant. Denoting $\chi=x^4$ as the fifth dimension coordinate , $\chi=0$ determines location of the brane in the bulk space, at the point of $Z_2$ symmetry; $g_{AB}=\tilde{g}_{AB}-n_A n_B$ is the induced metric on the brane, with $n_A$ being the unit vector normal to the brane.
\par
The effective gravitational field equations induced on the brane are determined by the bulk field equations (\ref{beq1}), the Gauss - Codazzi equations and the generalised matching Israel conditions with $Z_2$-symmetry. They can be expressed as modified standard Einstein's equations containing additional terms reflecting bulk effects onto the brane \cite{Shi-Mae-Sas:2000:}

\begin{equation}
 G_{\mu\nu}=-\Lambda g_{\mu\nu}+k^2 T_{\mu\nu} + \tilde{k}^2 S_{\mu\nu} -\mathcal{E}_{\mu\nu},\label{beq2}
\end{equation}
where $k^2=8\pi/M_P^2$, with $M_P$ being the braneworld Planck mass. The relations of the energy scales and cosmological constants are given in the form

\begin{equation}
 M_P=\sqrt{\frac{3}{4\pi}}\left(\frac{\tilde{M}_P^2}{\sqrt{\lambda}}\right)\tilde{M}_P;\quad \Lambda=\frac{4\pi}{\tilde{M}_P^3}\left[\tilde\Lambda+\left(\frac{4\pi}{3\tilde{M}_P^3}\right)\lambda^2\right].\label{beq3}
\end{equation}
Local bulk effects on the matter are determined by the ``squared energy-momentum'' tensor $S_{\mu\nu}$, 
that reads
\begin{equation}
 S_{\mu\nu}=\frac{1}{12}T T_{\mu\nu}-\frac{1}{4}T_\mu^{\phantom{\mu}\alpha}T_{\nu\alpha}+\frac{1}{24}g_{\mu\nu}\left(3T^{\alpha\beta}T_{\alpha\beta}-T^2\right),
\end{equation}
while the non-local bulk effects are given by the tensor $\mathcal{E}_{\mu\nu}$ representing the bulk Weyl tensor $\tilde{C}_{ABCD}$ projected onto the brane, whereas

\begin{equation}
 \mathcal{E}_{AB}=\tilde{C}_{ABCD}n^C n^D.\label{beq4}
\end{equation}

Symmetries of the Weyl tensor imply that $\mathcal{E}_{[AB]}=\mathcal{E}_A^{\phantom{A}A}=0$ and $\mathcal{E}_{AB}n^B=0$. Therefore, on the brane, $\chi\rightarrow 0$, there is $\mathcal{E}_{AB}\rightarrow \mathcal{E}_{\mu\nu}\delta_A^{\phantom{A}\mu}\delta_B^{\phantom{B}\nu}$. The $\mathcal{E}_{\mu\nu}$ tensor reflects influence of the non-local gravitational effects in the bulk, including the tidal (``Coulomb``) and transverse traceless (gravitational wave) imprints of the free gravitational field of the bulk.
\par
We restrict our attention to the vacuum (at both bulk and brane) solutions of the gravitational field equations on the brane. Assuming zero cosmological constant on the brane ($\Lambda=0$) we arrive to the condition

\begin{equation}
 \tilde\Lambda=-\frac{4\pi\lambda^2}{3\tilde{M}_P^2}.\label{beq5}
\end{equation}
In the absence of matter fields, there is $T_{\mu\nu}=0=S_{\mu\nu}$, i.e., we are not interested in the properties of the squared energy-momentum $S_{\mu\nu}$ representing local effects of the bulk. In the vacuum case, the effective gravitational field equations on the brane reduce to the form \cite{Shi-Mae-Sas:2000:}
\begin{equation}
 R_{\mu\nu}=-\mathcal{E}_{\mu\nu},\quad R_\mu^{\phantom{\mu}\mu}=0=\mathcal{E}_\mu^{\phantom{\mu}\mu}\label{beq6}
\end{equation}
implying divergence constraint \cite{Shi-Mae-Sas:2000:}

\begin{equation}
 \nabla^\mu\mathcal{E}_{\mu\nu}=0\label{beq7}
\end{equation}
where $\nabla_{\mu}$ denotes the covariant derivative on the brane.
\par
The equation (\ref{beq7}) represents Bianchi identities on the brane, i.e., an integrability condition for the field equations $R_{\mu\nu}=-\mathcal{E}_{\mu\nu}$\cite{Ali-Gum:2005:}. For stationary and axisymmetric (or static, spherically symmetric) solutions Eqs. (\ref{beq6}) and (\ref{beq7}) form a closed system of equations on the brane. 
\par
The 4D general relativity energy-momentum tensor $T_{\mu\nu}$ (with $T_\mu^{\phantom{\mu}\mu}=0$) can be formally identified to the bulk Weyl term on the brane due to the correspondence 

\begin{equation}
 k^2 T_{\mu\nu}\quad\leftrightarrow\quad -\mathcal{E}_{\mu\nu}.\label{beq8}
\end{equation}
The general relativity conservation law $\nabla^\mu T_{\mu\nu}=0$ then corresponds to the constraints equation on the brane (\ref{beq7}). This behaviour indicates that Einstein-Maxwell solutions in general relativity should correspond to braneworld vacuum solutions. This was indeed shown in the case of Schwarzchild  (R-N) \cite{Maa:2004:,Dad-etal:2000:} and Kerr (K-N) spacetimes \cite{Ali-Gum:2005:}. In both of these solutions the influence of the non-local gravitational effects of the bulk on the brane are represented by a single "braneworld" parameter $b$. The Coulomb-like behaviour in the Newtonian potential

\begin{equation}
 \Phi=-\frac{M}{M^2_{P}r}+\frac{b}{2r^2}\label{beq9}
\end{equation}
inspired the name tidal charge \cite{Dad-etal:2000:}.
 \par

\section{\label{sec:NullGeo}Null geodesics in Kerr spacetime with a tidal charge}

\subsection{Geometry}

Following the work of \cite{Ali-Gum:2005:}, and using the standard Boyer-Linquist coordinates ($t$, $r$, $\theta$, $\varphi$), we can write the line element of Kerr black-hole (or naked singularity) spacetime on the three-brane  in the form 
\begin{eqnarray}
	\diff s^2 &=& -(1-\frac{2Mr - b}{\Sigma})\diff t^2 + \frac{\Sigma}{\Delta}\diff r^2 + \Sigma \diff \theta^2  + \frac{A}{\Sigma}\diff\varphi^2 \nonumber\\
	&&- 2\frac{2Mr-b}{\Sigma}\sin^2\theta\diff t\diff\phi,\label{eq1}
\end{eqnarray}
where 

\begin{eqnarray}
	\Sigma &=& r^2 + a^2\cos^2\theta\label{eq2}\\
	\Delta &=& r^2 - 2Mr + a^2 +b\label{eq3}\\
	A &=& (r^2 + a^2)^2 - a^2\Delta\sin^2\theta\label{eq4}.
\end{eqnarray}
M is the mass parameter, $a=J/M$ is the specific angular momentum and the braneworld prarameter $b$ is the \emph{tidal charge} representing imprint of non-local gravitational effects from the bulk space. The metric (\ref{eq1}) has the 
same form as the Kerr-Newman metric, where the tidal charge is replaced by the squared electric charge, $Q^2$. 
The stress tensor on the brane $E_{\mu\nu}$ takes the form \cite{Ali-Gum:2005:}

\begin{eqnarray}
 E_t^{\phantom{t}t}&=&-E_\varphi^{\phantom{\varphi}\varphi}=-\frac{b}{\Sigma^3}[\Sigma-2(r^2+a^2)],\\
 E_r^{\phantom{r}r}&=&-E_\theta^{\phantom{\theta}\theta}=-\frac{b}{\Sigma^2},\\
 E_\varphi^{\phantom{\varphi}t}&=&-(r^2+a^2)\sin^2\theta,\\ 
E_t^{\phantom{t}\varphi}&=&-\frac{2ba}{\Sigma^3}(r^2+a^2)\sin^2\theta
\end{eqnarray}
that is fully analogical ($b\rightarrow Q^2$) to the components of energy-momentum tensor for Kerr-Newman spacetimes in Einstein's general relativity \cite{Ali-Gum:2005:}.

The roots of $\Delta = 0$ identify the type of braneworld Kerr spacetime. There are two possibilities, a black hole or a naked singularity. By introducing $a^2/M^2\rightarrow a^2$, $b/M^2\rightarrow b$, $r_+/M\rightarrow r_+$, or putting $M=1$, we write the roots of $\Delta = 0$ in the form

\begin{equation}
 r_+ = 1+\sqrt{1-a^2-b},\quad\textrm{(outer horizon)}\label{horeq2}
\end{equation}
and
\begin{equation}
 r_- = 1-\sqrt{1-a^2-b},\quad\textrm{(inner horizon)}.\label{horeq3}
\end{equation}
The metric given by the line element (\ref{eq1}) determines the geometry of rotating black hole in braneworld universe if

\begin{equation}
 	1\ge a^2+b.\label{horeq1}
\end{equation}
The strong inequality refers to the case of two horizonts $r_+$ and $r_-$. For extreme black holes ($1=a^2+b$) the horizons coincide $r_+ = r_- = 1$.

\begin{figure}[!th]
\includegraphics[width=10cm]{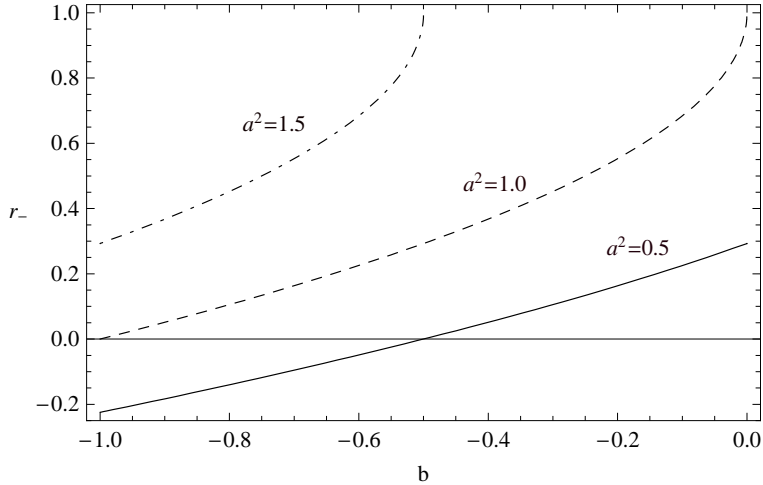}
\caption{\label{fig_1}The plot of the inner horizont radius $r_-$ as a function of tidal charge parameter $b$ for three representative values of rotational parameter $a^2=0.5$, $a^2=1.0$ and $a^2=1.5$.}
\end{figure}
It is clear that for $b\ge 0$ the loci of the inner horizon  $r_-$ are always
positive. But for $b<0$, the loci of the inner horizon can also be at
negative $r$, as illustrated in  Figure \ref{fig_1}. 
\par
Notice that $a^2>1$ is not allowed for standard black holes and for $b>0$ \cite{MTW}, but such a possibility appears for $b<0$. The rotational parameter of extreme black holes is given by $a^2=1-b$.
The case of $1<a^2+b$ refers to the braneworld Kerr naked singularities. 
\par
In this paper we focus on astrophysicaly interesting case of black holes, with emitting sources and observers located above the outer horizont.

\subsection{Carter's equations}

In order to study the optical effects in braneworld Kerr spacetimes, we have to solve  equations of motion of  photons given by the null geodesics of the spacetime under consideration. 
The geodesic equation reads

\begin{equation}
  \frac{\Diff k^\mu}{\diff w}=0,\label{ce1}
\end{equation}
where $k^\mu=\frac{\diff x^\mu}{\diff w}$ is the wave vector tangent to the null geodesic and $w$ is the affine parameter. The normalization condition reads $g_{\mu\nu}k^\mu k^\nu = 0$. 
Since the components of the metric tensor do not depend on $\varphi$ and $t$ coordinates, the conjugate momenta

\begin{eqnarray}
 	k_\varphi=g_{\varphi\nu}k^\nu\equiv \Phi,\label{ce4}\\
 	k_t =g_{t\nu}k^\nu\equiv -E,\label{ce5}
\end{eqnarray}
are the integrals of motion. Carter found another integral of motion $K$ as a separation constant when solving Hamilton-Jacobi equation 

\begin{equation}
 	g^{\mu\nu}\frac{\partial S}{\partial x^\mu}\frac{\partial S}{\partial x^\nu} = 0,\label{ce6}
\end{equation}
where he assumed the action $S$ in separated form

\begin{equation}
	S=-Et+\Phi\varphi + S_r(r)+S_{\theta}(\theta).\label{ce6a}
\end{equation}

The equations of motion can be integrated and written separatelly in the form

\begin{eqnarray}
 	\Sigma\frac{\diff r}{\diff w}&=&\pm\sqrt{R(r)},\label{ce7}\\
 	\Sigma\frac{\diff \theta}{\diff w}&=&\pm\sqrt{W(\theta)},\label{ce8}\\
 	\Sigma\frac{\diff \varphi}{\diff w}&=&-\frac{P_W}{\sin^2\theta}+\frac{a P_R}{\Delta},\label{ce9}\\
 	\Sigma\frac{\diff t}{\diff w}&=&-a P_W + \frac{(r^2+a^2)P_R}{\Delta},\label{ce10}
\end{eqnarray}
where 

\begin{eqnarray}
  R(r)&=&P^2_R-\Delta K,\label{ce11}\\
  W(\theta)&=&K-\left(\frac{P_w}{\sin\theta}\right)^2,\label{ce12}\\
  P_R(r)&=&E(r^2+a^2)-a\Phi,\label{ce13}\\
  P_W(\theta)&=&aE\sin^2\theta - \Phi.\label{ce14}
\end{eqnarray}
It is usefull to introduce integral of motion $Q$ by the formula

\begin{equation}
 Q=K-(E- a\phi)^2.\label{ce15}
\end{equation}
Its relevance comes from the fact that in the case of astrophysically most important motion in the equatorial plane ($\Theta = \pi/2$) there is $Q=0$.

\subsection{Radial and latitudinal motion}
The photon motion (with fixed constants of motion $E$, $\Phi$, $Q$) is allowed in regions where $R(r;E,\Phi,Q)\ge 0$ and $W(\theta; E,\Phi,Q)\ge 0$. The conditions $R(r;E,\Phi,Q)=0$ and $W(\theta;E,\Phi,Q)=0$ determine turning points of the radial and latitudinal motion, respectively, giving boundaries of the region allowed for the motion.
Detailed analysis of the $\theta$-motion can be found in \cite{Bic-Stu:1976:,Fel-Cal:1972:}, while the radial motion was analysed (with restrictions implied by the $\theta$-motion) in \cite{Stu:1981a:} and \cite{Stu:1981b:}. Here we extend this analysis to the case of $b < 0$.
\par

The radial and latitudinal Carter equations read

\begin{eqnarray}
	\Sigma^2\left(\frac{\diff r}{\diff w'} \right)^2 &=& [r^2+a^2-a\lambda]^2-\Delta[\mathcal{L}-2a\lambda + a^2],\label{eq7}\\
	\Sigma^2\left(\frac{\diff \theta}{\diff w'} \right)^2 &=& \mathcal{L} + a^2  \cos^2\theta - \frac{\lambda^2}{\sin^2\theta}\label{eq8}
\end{eqnarray}
where we have introduced impact parameters

\begin{eqnarray}
	\lambda &=& \frac{\Phi}{E},\label{eq9}\\
	\mathcal{L} &=& \frac{L}{E^2} = \frac{Q+\Phi^2}{E^2} = q + \lambda^2,\label{eq10}
\end{eqnarray}
and rescaled the affine parameter by $w^\prime = E w$. We assume $a>0$.

The reality conditions $(\diff r/\diff w')^2 \ge 0$ and $(\diff\theta/\diff w')^2 \ge 0$ lead to the restrictions on the impact parameter $\mathcal{L}$

\begin{equation}
	\mathcal{L}_{min} \leq \mathcal{L} \leq \mathcal{L}_{max},\label{eq12}
\end{equation}
where

\begin{equation}
	\mathcal{L}_{max} \equiv \frac{(a\lambda -2r +b)^2}{\Delta}+ r^2+2r-b,\label{eq13}
\end{equation}
and
\begin{equation}
	\mathcal{L}_{min}\equiv\left\{ \begin{array}{lcr} 
				\lambda^2 & \textrm{for} & |\lambda|\geq a,\\
				2a|\lambda|-a^2 & \textrm{for} & |\lambda|\leq a. 
			   \end{array}\right.\label{eq14}
\end{equation}
The upper(lower) constraint, $\mathcal{L}_{max}$($\mathcal{L}_{min}$), comes from the radial-motion (latitudinal-motion) reality condition. The properties of the photon motion are determined by the behaviour of the surface $\mathcal{L}_{max}(r;\lambda,a,b)$, as given by (\ref{eq13}). The extrema of the surface $\mathcal{L}_{max}$ (giving spherical photon orbits) are determined by

\begin{eqnarray}
	\lambda=\lambda_+ &\equiv& \frac{r^2+a^2}{a},\label{eq15}\\
	\lambda=\lambda_- &\equiv& \frac{r^2-b r - a^2 - r\Delta}{a(r-1)}.\label{eq16}
\end{eqnarray}
The values of $\mathcal{L}_{max}$ at these extreme points are given by

\begin{eqnarray}
	\mathcal{L}_{max}(\lambda_{+})\equiv\mathcal{L}_+ &=& 2r^2+a^2,\label{eq17}\\
	\mathcal{L}_{max}(\lambda_{-})\equiv\mathcal{L}_- &=&\frac{2r(r^3-3r+4b)+a^2(r+1)^2}{(r-1)^2}\label{eq18}.
\end{eqnarray}
The character of the extrema follows from the sign of  $\partial^2\mathcal{L}_{max}/\partial r^2$. One finds that

\begin{eqnarray}
\frac{\partial^2 \mathcal{L}_{max}}{\partial r^2} &=& \frac{8r^2}{\Delta},\quad\textrm{for}\quad \lambda = \lambda_+,\label{eq19}\\
\frac{\partial^2 \mathcal{L}_{max}}{\partial r^2} &=&\frac{8r^2}{\Delta} - \frac{8r}{(r-1)^2},\quad\textrm{for}\quad \lambda=\lambda_-.\label{eq20}
\end{eqnarray}
Clearly, there are only minima of $\mathcal{L}_{max}$ along for $\lambda=\lambda_{+}$, corresponding to unstable
spherical orbits.
\begin{figure}[ht]
\begin{tabular}{cc}
  \includegraphics[width=6cm]{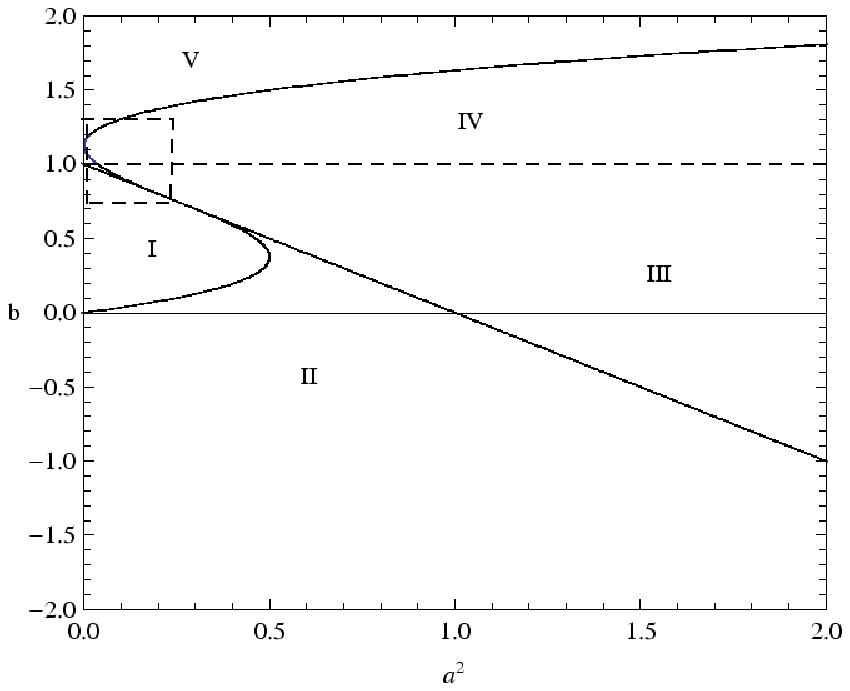}&\includegraphics[width=6cm]{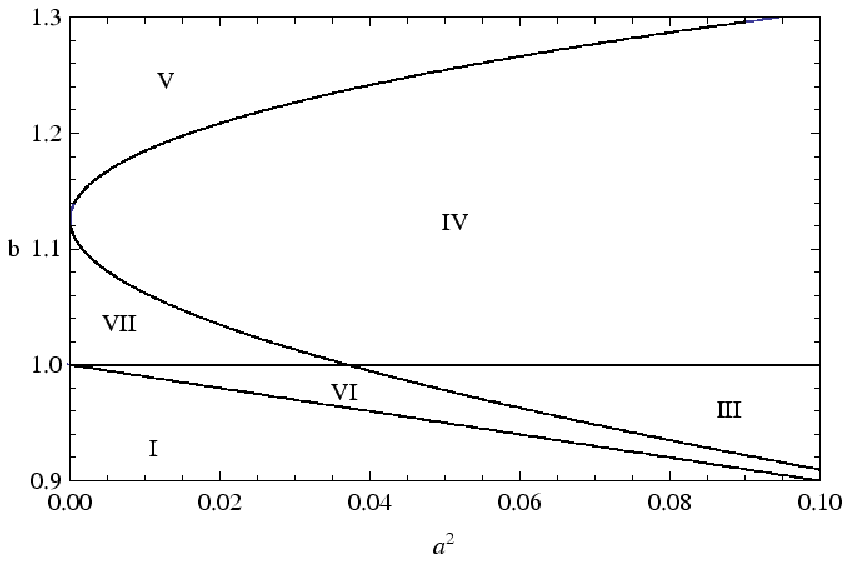} 
\end{tabular}
\caption{\label{fig2_a_b}Left: classification of Kerr spacetime in braneworld universe according to the values of $a^2+b$, $b$ and $n_{ext}$ (the number of local extrema of the curves $\tilde\lambda_\pm$, which is also the number of circular photon orbits in the equatorial plane). 
The classification regions are: I) for $a^2+b\leq 1$ and $n_{ext}=2$, II) for $a^2+b\leq 1$ and $n_{ext}=4$, III) $a^2+b>1$ and $b<1$ and $n_{ext}=2$, IV) for $a^2+b>1$ and $b>1$ and $n_{ext}=2$, V) for $a^2+b>1$ and $n_{ext}=0$, VI) for $a^2+b>1$ and $b<1$ and $n_{ext}=4$, VII) for $a^2+b>1$ and $b>1$  and $n_{ext}=4$. 
Right: zoom of the area in the dashed rectangle of the left plot, to cover regions VI and VII.} 
\end{figure}

Further, we have to determine where the restrictions given by the latitudinal motion $\mathcal{L}_{min}$ meet the restrictions on the radial motion $\mathcal{L}_{max}$. We find that $\mathcal{L}_{max}=\lambda^2$ (for $|\lambda|\ge a$) is fullfilled where

\begin{equation}
 \lambda=\tilde\lambda_\pm\equiv\frac{a(b-2r\pm r^2\sqrt{\Delta})}{r^2-2r+b},\label{eq23}
\end{equation}
while $ \quad\mathcal{L}_{max}= 2a|\lambda| - a^2$ (for $|\lambda|<a$) is fullfilled where

\begin{equation}
 \lambda=\bar\lambda \equiv \frac{1}{\Delta}[4r-r^2-2b-a^2+2\sqrt{\Delta(b - 2r)}].\label{eq24}
\end{equation}

The extreme points of curves $\tilde\lambda_\pm$, which are also the intersection points of these curves  with $\lambda_-$, are determined by the equation

\begin{equation}
	f(r;a,b)\equiv r^4-6r^3+(9+4b)r^2-4(3b+a^2)r+4b(b+a^2)=0.\label{eq25}
\end{equation}
The equation $f(r;a,b)=0$ determines loci of the photon equatorial circular orbits; in an implicit form the radii are given by the condition

\begin{equation}
 a^2=a^2_{ph\pm}(r;b)=\frac{r^2(r-3)^2+4b(r^2-3r+b)}{4(r-b)}.
\end{equation}
The maxima of the curve $\bar\lambda$, which also determine the intersections of curves $\bar\lambda$ and $\lambda_-$ are located on $r$ satisfying the equation

\begin{equation}
 2r^3-(3+b)r^2+2br+a^2 = 0.\label{eq25a}
\end{equation}

\begin{figure}[!ht]
  \includegraphics[width=12.0cm]{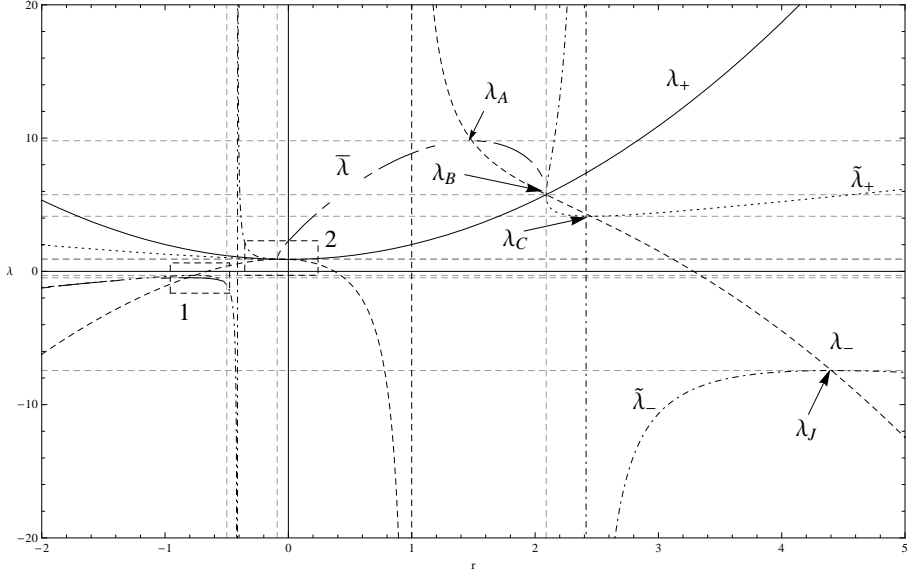}
\caption{\label{fig3}The graphs of the $\lambda_\pm$, $\tilde\lambda_\pm$ and $\bar\lambda$ functions are plotted for representative values of the parameters  $a=0.9$ and $b=-1.0$. The two dashed rectangle areas labeled with numbers $1$ and $2$ are zoomed in the following figures. The horizontal gray dashed lines represent special values of the impact parameter $\lambda$, denoted according to the text as  $\lambda_A$...$\lambda_J$.}
\end{figure}

\begin{figure}[ht]
\begin{tabular}{cc}
  \includegraphics[width=6cm]{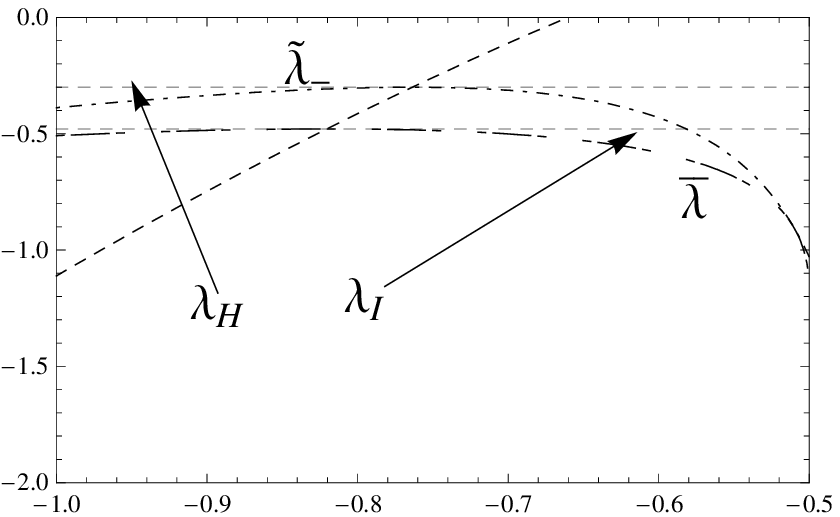}&\includegraphics[width=6cm]{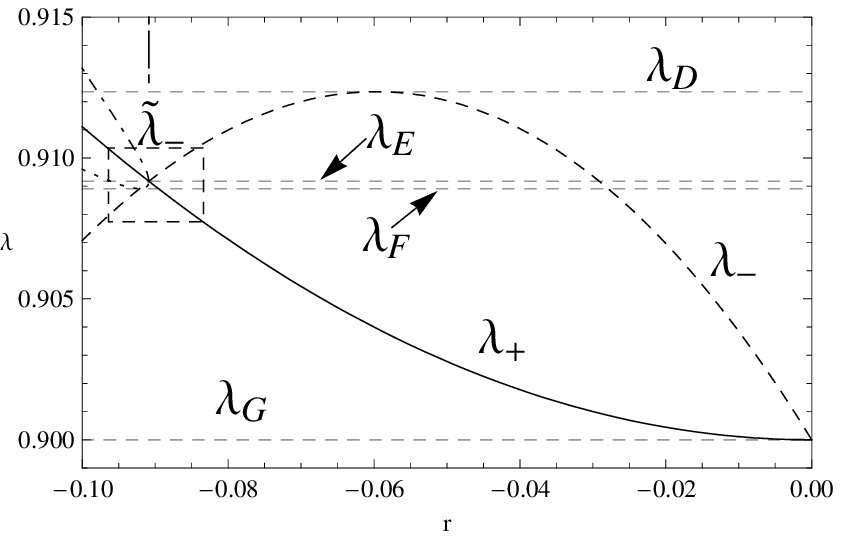}
\end{tabular}
\caption{\label{fig4}Left figure is the zoom of dashed area labelled $1$ in previous figure. Right figure is the zoom of dashed area labelled $2$ in previous figure. The dashed rectangle area here is zoomed in the next figure.}
\end{figure}

\begin{figure}[ht]
  \includegraphics[width=6.2cm]{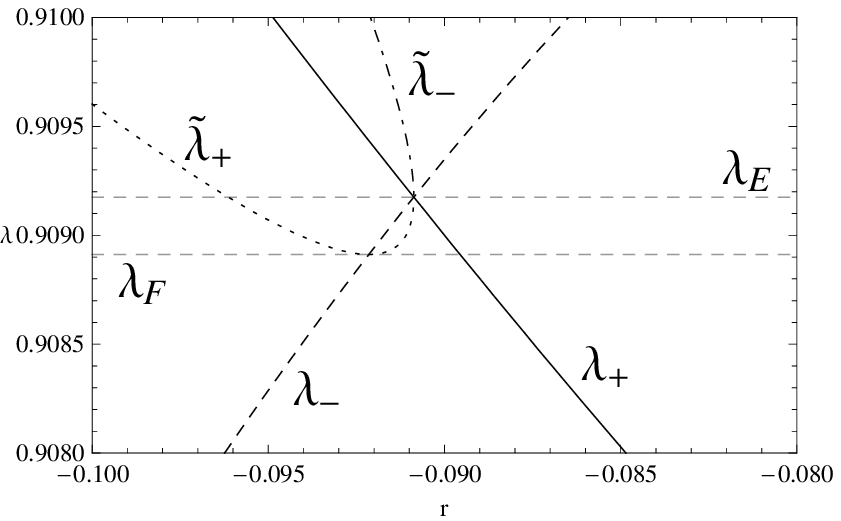}
\caption{\label{fig5}The zoom of the dashed rectangle area in previous figure.}
\end{figure}

The braneworld Kerr spacetimes can be classified due to the properties of the photon motion as determined by the behaviour of the functions $\lambda_\pm$, $\tilde\lambda_\pm$, $\bar\lambda$. The classification is governed  by their divergences (i.e., by existence of the horizons) and the number of local extrema determining equatorial photon circular orbits $n_{ex}$. There exist seven classes of the braneworld Kerr spacetimes, with the criteria of separation being $a^2+b^{\phantom{i}<}_{\phantom{i}>} 1$, $b^{\phantom{i}<}_{\phantom{i}>} 1$ and $n_{ex}$ . The classification is represented in Figure \ref{fig2_a_b}. There are two different classes of the black-hole spacetimes, differing by the presence of the photon circular orbits under the inner horizon. However, in the astrophysically relevant region outside the outer horizon, both the classes are of the same character, having two unstable  equatorial photon circular orbits, one corotating (at $r_{ph1}$) and the other counter-rotating (at $r_{ph2}>r_{ph1}$). The tidal charge $b$ introduces no qualitatively new feature into the behaviour of photon motion in the Kerr spacetimes, but the quantitative impact of $b<0$ with
high magnitude are quite relevant, as shown in next sections. All the braneworld Kerr black holes with tidal charge $b<0$ belong to the class II discussed in the case of standard Kerr-Newman spacetimes \cite{Stu:1981b:}. We illustrate in Figures \ref{fig3}-\ref{fig5} functions $\lambda_\pm$, $\tilde\lambda_\pm$ and $\bar\lambda$ for such a black hole spacetime with parameters $a=0.9$ and $b=-1.0$. In this case typical for braneworld Kerr black hole with $b<0$ there exist ten significiant values of $\lambda$ as given in Figures \ref{fig3} - \ref{fig5}.

For each interval of $\lambda$ as determined by the sequence of $\lambda_A$ - $\lambda_J$ introduced in Figure \ref{fig3}, there exists a characteristic type of behaviour of the restricting "radial" function $\mathcal{L}_{max}$ and its relation to the "latitudinal" restricting function $\mathcal{L}_{min}$. They can be found in \cite{Stu:1981b:} and will not be repeated here.

The allowed values of the impact parameter $\mathcal{L}$ lie between the limiting functions  $\mathcal{L}_{min}$ and $\mathcal{L}_{max}$. If the minimum $\mathcal{L}_{max}^{min}\equiv\mathcal{L}_{max}(r_{min},\lambda_0)$ of the limiting function $\mathcal{L}_{max}$ is less than the value of the limiting function $\mathcal{L}_{min}$, an incoming photon ($k^r < 0$) travelling from infinity will return back for all values of $\mathcal{L}_0\in[\mathcal{L}_{min};\mathcal{L}_{max}]$. If $\mathcal{L}_{max}^{min}>\mathcal{L}_{min}$, 
the incoming photon ($k^r < 0$) travelling from infinity returns back if its impact parameter $\mathcal{L}_0$  
satisfies the condition  $\mathcal{L}_{0}\ge\mathcal{L}_{max}^{min}$ and is captured by the black hole 
if $\mathcal{L}_0<\mathcal{L}^{min}_{max}$.  
The minimum $\mathcal{L}_{max}^{min}$ determines (with the particular value of $\lambda$) a photon spherical orbit, 
i.e., a sphere where photons move with $r=const$ but with varying latitude $\theta$ (and, of course, varying $\varphi$). 
When the condition  $\mathcal{L}_0 = \mathcal{L}_{min}$ is satisfied simultaneously, the spherical photon orbit is transformed 
to an equatorial photon circular orbit. Photons with $\mathcal{L}_0=\mathcal{L}_{max}^{min}$ coming from distant regions or 
regions close to the black hole horizon will wind up around the photon sphere. 
 
\clearpage

\section{\label{sec:LEC}Light escape cones}
The optical phenomena related to accretion processes in  the field of rotating black holes could be efficiently studied by using the notion of light escape cones of local observers (sources) that determine which portion of radiation emitted by a source could escape to infinity and, complementary, which portion is trapped by the black hole \cite{SSJ:RAGTime:2005:Proceedings}. Here we focus our attention to four families of observers (sources) that are of direct physical relevance.

\subsection{Local frames of stationary and free-falling observers}
We consider three families of stationary frames, namely  $LNRF$ (Locally Nonrotatig Frame), $SF$ (Static Frame) and $GF_\pm$(Circular Geodesic Frame) and one non-stationary frame, namely $RFF$ (Radially Falling Frame). 
The $LNRF$ are of highest physical importance since the physical phenomena take the simplest form when expressed in such frames, because the rotational spacetime effects are maximally suppressed there \cite{Bardeen:1973:,MTW}. The $GF_\pm$ are directly related to Keplerian accretion discs in the equatorial plane of the spacetime, both corotating and counterrotating, while $RFF$ are related to free-falling spherical accretion. The $SF$ are fixed relative to distant observers. The $GF_\pm$ and $RFF$ are geodetical frames, while $SF$ and $LNRF$ are generally accelerated frames.

The radial and latitudinal 1-forms of the three stationary frame tetrads are common for all three stationary cases and read

\begin{eqnarray}
	\omega^{(r)}&=&\left\{0,\sqrt{\Sigma/\Delta},0,0 \right\},\label{LC9}\\
	\omega^{(\theta)}&=&\left\{0,0,\sqrt{\Sigma},0 \right\}.\label{LC10}
\end{eqnarray}
$LNRF$ correspond to observers with $\Phi=0$ (zero angular momentum observers). Their time and azimuthal 1-forms read

\begin{eqnarray}
	\omega^{(t)}&=&\left\{\sqrt{\frac{\Delta\Sigma}{A}},0,0,0 \right\},\label{LC11}\\
	\omega^{(\varphi)}&=&\left\{-\Omega_{LNRF}\sqrt{\frac{A}{\Sigma}}\sin\theta,0,0,\sqrt{\frac{A}{\Sigma}}\sin\theta\right\}.\label{LC12}
\end{eqnarray}
where 

\begin{equation}
	\Omega_{LNRF}=\frac{a(2Mr-b)}{A}\label{LC13}
\end{equation}
is the angular velocity of $LNRF$ as seen by observers at infinity. 
\par
The tetrad of $SF$ corresponding to observers with $\Omega=0$ ,i.e. static relative to observers at infinity, is given by the formulae

\begin{eqnarray}
	\omega^{(t)}&=&\left\{ \sqrt{1-\frac{2r-b}{\Sigma}},0,0,\frac{a(2r-b)\sin^2\theta}{\sqrt{\Sigma^2-(2r-b)\Sigma)}}  \right\},\\
	\omega^{(\varphi)}&=&\left\{ 0,0,0,\sqrt{\frac{\Delta\Sigma}{\Sigma-(2r-b)}}\sin\theta  \right\}.
\end{eqnarray}

The $GF_\pm$ observers move along $\varphi$-direction in the equatorial plane with velocity $V_{GF\pm}$(+...corotating, -...counterrotating) relative to the $LNRF$ and with angular velocity $\Omega$ relative to the static observers at infinity given by \cite{}[SK]
\begin{equation}
\Omega_\pm=\pm\frac{\sqrt{r-b}}{r^2 \pm a\sqrt{r-b}}. \label{ang_vel_gf}
\end{equation}

The velocity $V_{GF\pm}$ is given by

\begin{equation}
	V_{GF\pm}=\pm\frac{(r^2+a^2)Y\mp a(2r-b)}{\sqrt{\Delta}(r^2\pm aY)}.\label{VGF}
\end{equation}
where $Y=\sqrt{r-b}$. The standard Lorentz transformation of the $LNRF$ tetrad gives the tetrad of $GF_\pm$ in the form
\begin{eqnarray}
	\omega^{(t)}_\pm&=&\left\{ \frac{r^2-2r+b\pm a Y}{Z_\pm},0,0,\mp\frac{(r^2+a^2)Y\mp a(2r-b)}{Z_\pm} \right\},\\
\omega^{(\varphi)}_\pm&=&\left\{\mp \frac{\sqrt{\Delta}Y}{Z_\pm},0,0,\frac{\sqrt{\Delta(r^2\pm a Y)}}{Z_\pm}, \right\}
\end{eqnarray}
where 

\begin{equation}
	Z_\pm = r\sqrt{r^2-3r+2b\pm2aY}.
\end{equation}
Note that the $GF_\pm$ family is restricted to the equatorial plane, while $LNRF$ are defined at any $\theta$.

The $RFF$ observers have velocity

\begin{equation}
	V_{RFF}=\{V^{(r)},\,V^{(\theta)},\,V^{(\varphi)}\}
\end{equation} 
as measured in $LNRF$. The radially free-falling (or free-escaping) observers starting (finishing) at infinity move with $\theta = const$. Using the results of \cite{Stu-Bic-Bal:1999:}, we find the velocity components of the free-falling frames in  the $LNRF$ frames

\begin{eqnarray}
	V^{(r)}&=&\pm\sqrt{1-\frac{\Sigma\Delta}{A}},\\
	V^{(\theta)}&=&0,\\
	V^{(\varphi)}&=& 0.
\end{eqnarray}
Clearly, the free-falling (free-escaping) observers move only radially in the $LNRF$, in analogy to particles radially moving in static frames of the Schwarzchild spacetimes.
 For the radially free-falling sources, the tetrad components $\omega^{(\theta)}$ and $\omega^{(\varphi)}$ coincide with those of the LNRF tetrad, while $\omega^{(t)}$ and $\omega^{(r)}$ are transformed. The local Lorentz transformation of the $LNRF$ to the $RFF_\pm$ tetrad yields

\begin{eqnarray}
\omega_\pm^{(t)}&=&\left\{ \gamma\frac{\Delta\Sigma}{A}, \mp\sqrt{\frac{\Sigma}{\Delta}}V,0,0 \right\},\\
\omega_\pm^{(r)}&=&\left\{\mp\gamma\sqrt{\frac{\Delta\Sigma}{A}}V,\sqrt{\frac{\Sigma}{\Delta}}\gamma,0,0\right\},\\
\omega_\pm^{(\theta)}&=&\{0,0,\sqrt{\Sigma},0\},\\
\omega_\pm^{(\varphi)}&=&\left\{-\Omega_{LNRF}\sqrt{\frac{A}{\Sigma}}\sin\theta,0,0,\sqrt{\frac{A}{\Sigma}}\sin\theta \right\}.
\end{eqnarray}

\subsection{Construction of escape cones}

\begin{figure}[ht]
	\includegraphics[width=10cm]{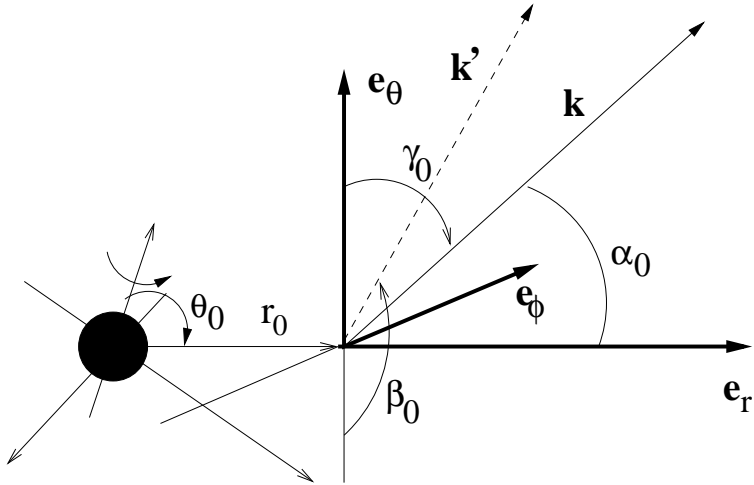}
\caption{\label{fig8}Definition of directional angles $\alpha_0$, $\beta_0$ and $\gamma_0$ in a local frame. Vectors $\vec{e}_r$, $\vec{e}_\theta$, $\vec{e}_\varphi$ are the basic tetrad vectors. Position of the observer (source) is given by the coordinates $(r_0,\theta_0)$. Vector $\vec{k}$ represents a photon as observed by the observer in the given tetrad and vector $\vec{k}^\prime$ is its projection into the plane ( $\vec{e}_\theta$, $\vec{e}_\varphi$). }
\end{figure}

 For each direction of emission in the local frame of a source, there is a corresponding pair of values of the impact parameters $\lambda$ and $\mathcal{L}$ which can be related to the directional cosines of the photon trajectory in the local frame at the position of the source. Of course, the analysis of the turning points of the radial motion of photons, presented in the previous section, is crucial in determining the local escape cones as the boundary of the escape cone is given by directional angles related to spherical photon orbits.

Projection of a photon 4-momentum $\vec{k}$ onto the local tetrad of an observer is given by the formulae

\begin{eqnarray}
k^{(t)}&=&-k_{(t)}=1,\label{LC1}\\
k^{(r)}&=&k_{(r)}=\cos\alpha_0,\label{LC2}\\
k^{(\theta)}&=&k_{(\theta)}=\sin\alpha_0\cos\beta_0,\label{LC3}\\
k^{(\varphi)}&=&k_{(\varphi)}=\sin\alpha_0\sin\beta_0,\label{LC4}
\end{eqnarray} 
where $\alpha_0$, $\beta_0$ are directional angles of the photon in the local
frame (see Figure \ref{fig8}) and $\cos\gamma_0=\sin\alpha_0\sin\beta_0$. 
In terms of the local tetrad components of the photon 4-momentum and the related directional angles, the conserved quantities, namely, the azimutal momentum $\Phi$, energy $E$ and $K$ read

\begin{eqnarray}
	\Phi&=&k_\varphi=-\omega^{(t)}_{\phantom{(t)}\varphi}k^{(t)} + \omega^{(r)}_{\phantom{(r)}\varphi}k^{(r)}+\omega^{(\theta)}_{\phantom{(\theta)}\varphi}k^{(\theta)}+\omega^{(\varphi)}_{\phantom{(\varphi)}\varphi}k^{(\varphi)},\label{LC6}\\
	E&=&-k_t=\omega^{(t)}_{\phantom{(t)}t}k^{(t)} - \omega^{(r)}_{\phantom{(r)}t}k^{(r)}-\omega^{(\theta)}_{\phantom{(\theta)}t}k^{(\theta)}-\omega^{(t)}_{\phantom{(\varphi)}\varphi}k^{(\varphi)},\label{LC7}\\
   K&=&\frac{1}{\Delta}\left\{ [E(r^2+a^2)-a\Phi]^2-(\Sigma k^r)^2\right\}.\label{LC8}
\end{eqnarray}
The impact parameters $\lambda$ and $\mathcal{L}$ defined by relations (\ref{eq9}) and (\ref{eq10}) are thus fully determined by any double, $D$, of angles from the set $M=[\alpha_0,\beta_0,\gamma_0]$.

 Having defined the source frame, we can construct light escape cones assuming fixed coordinates of the source $r_0$, $\theta_0$. Their construction proceedes in the following steps:

\begin{itemize}
\item for given $D$, say $D=[\alpha_0,\beta_0]$, we calculate $\lambda=\lambda(\alpha_0,\beta_0)$,
\item $\lambda$ determines the behaviour of $\mathcal{L}_{max}=\mathcal{L}_{max}(r;\lambda)$,
\item from the analysis presented in the previous section we calculate minimum of $\mathcal{L}_{max}$, which reads $\mathcal{L}_{max}^{min}=\mathcal{L}_{max}(r_{min};\lambda)$,
\item we search for such a double $D$ which satisfies equation $\mathcal{L}_0(\alpha_0,\beta_0)=\mathcal{L}_{max}(r_{min};\lambda)$.
\end{itemize}
Here, we present in detail  the construction of light escape cones in particular case of the $LNRF$. The procedure is analogous for the other stationary frames and simply modified for the free-falling frames, being radius dependent.

\begin{figure}[ht]
\begin{tabular}{ll}
\includegraphics[width=6cm]{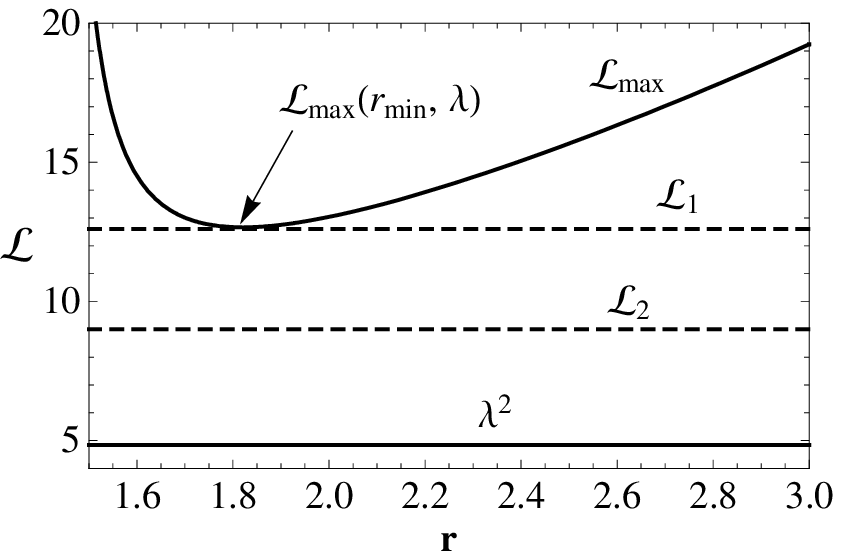} & \includegraphics[width=6cm]{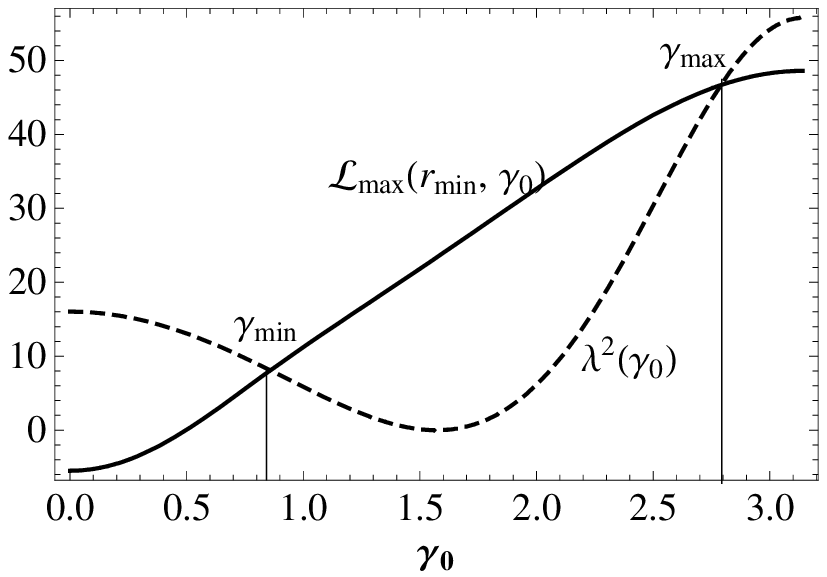}
\end{tabular}
\caption{\label{fig9_a_b}Left. The functions $\mathcal{L}_{max}$ and $\mathcal{L}_{min}=\lambda_0^2$ are plotted together with representative constant functions $\mathcal{L}_1$ and $\mathcal{L}_2$ to demonstrate the construction of the photon escape cone. Right. The intersections of $\mathcal{L}_{max}(\gamma_0)$ with $\lambda^2(\gamma_0)$ give the interval of relevant values of $\gamma_0\in[\gamma_{min};\gamma_{max}]$.}
\end{figure}

\par
The impact parameter $\lambda$  expressed in terms of the angle $\gamma_0$, related to the $LNRF$, reads

\begin{equation}
	\lambda_0=\frac{1}{\Omega_{LNRF0}+\frac{\Sigma_0\sqrt{\Delta_0}}{A_0\sin\theta_0\cos\gamma_0}},
\end{equation}
where index '$0$' refers to the frame with coordinates $[r_0,\theta_0]$. The minimum of $\mathcal{L}_{max}$ is located at

\begin{equation}
	r_{min}=\left\{ \begin{array}{lcr}
			\sqrt{a\lambda - a^2} & \textrm{for} & \lambda\geq\lambda_G = a\\
			1-\frac{k_1}{k_2}+\frac{k_2}{3} & \textrm{for} & \lambda<\lambda_G = a
			\end{array}\right.\label{eq_rmin}
\end{equation} 
where 

\begin{eqnarray}
	k_1&=&a^2+2b+a\lambda-3,\\
	k_2&=&\left\{ 27(1-a^2-b)+2\sqrt{3}\sqrt{27(1-a^2-b)^2+k_1^3}\right\}^{1/3}.
\end{eqnarray}
The relevant values of $\mathcal{L}$ lie between $\mathcal{L}_{max}$ and
$\mathcal{L}_{min}$ determined by Eqs (\ref{eq13}) and (\ref{eq14}). The
intersections of functions $\mathcal{L}_{max}=\mathcal{L}_{max}(\gamma_0)$ and
$\mathcal{L}_{min}(\gamma_0)$ give the relevant interval of angles
$\gamma\in[\gamma_{min},\gamma_{max}]$ (see Figure \ref{fig9_a_b}). For each $\gamma$ from $[\gamma_{min},\gamma_{max}]$ we calculate minimal value of the photon impact parameter $\mathcal{L}$ for which the photon reaches the turning point $r_{min}$ and escapes to infinity. This minimal value is the minimum of $\mathcal{L}_{max}$ which is located at $r_{min}$, eg. $\mathcal{L}_{max}=\mathcal{L}_{max}(r_{min};\lambda_0(\gamma_0),a,b)$, where $r_{min}$ is given by (\ref{eq_rmin}).
Now we can calculate the value of $\alpha_0$ using equation

\begin{equation}
	\cos\alpha_0=\frac{k^{(r)}}{k^{(t)}}=\frac{\omega^{(r)}_{LNRF\mu}k^\mu}{\omega^{(t)}_{LNRF\mu}k^\mu}.
\end{equation} 
We arrive to the formula

\begin{equation}
	\cos\alpha_0=\pm\sqrt{A_0}\frac{\sqrt{(r_0^2+a^2-a\lambda_0)^2-\Delta_0(\mathcal{L}_{max}^{min}-2a\lambda_0+a^2)}}{-a(a\sin^2\theta_0-\lambda_0)\Delta_0+(r_0^2+a^2)(r_0^2+a^2-a\lambda_0)},
\end{equation}
where $A_0=A(r_0,\theta_0)$, $\Delta_0=\Delta(r_0)$  and $\mathcal{L}_{max}^{min}=\mathcal{L}_{max}(r_{min};\lambda_0,a,b)$. The angle $\beta_0$ can be  calculated from the formula (\ref{LC4}).
In this way we obtain angles from the arc $\beta_0\in\langle -\pi/2; \pi/2\rangle$. The remaining arc $\beta_0\in\langle \pi/2; 3\pi/2\rangle$ can be obtained by turning the arc $\beta_0\in\langle -\pi/2; \pi/2\rangle$ around the symmetry axis determined by angles $\beta_0=-\pi/2$ and $\beta_0=\pi/2$. This procedure can be done because photons released under angles $\beta_0$ and $\pi-\beta_0$ have the same constants of motion. 
Clearly, for sources under the radius corresponding to the corotating
equatorial photon circular orbit, only outward directed photons with no
turning point of the $r$-motion can escape. With radius of the source
approaching the event horizon ($r_0\rightarrow r_+$), the escape cone shrinks
to infinitesimal extension, except the case of extreme black hole \cite{Bardeen:1973:}. For the other frames considered here, the procedure of
the related light escape cone construction can be directly repeated, but with
the relevant tetrad 1-form components being used in the procedure.

In order to reflect properly the effect of the tidal charge $b$ on the escape cone structure, we shall give the cones for black hole sequences of two kind: first we keep the spin $a$ fixed and change $b$, second we keep fixed "distance" to the extreme black hole states, i.e., $a^2+b$ is fixed, and both $a$ and $b$ are changed. The positive tidal charges have tendency to slightly increase the asymmetry of the cones as compared with $b=0$ case, keeping its character similar to the case of Kerr black holes (see next section). Therefore, we focus our attention to the influence of negative tidal charges.  

\begin{figure}[ht]
  \begin{tabular}{ccc}
    \includegraphics[width=4.0cm]{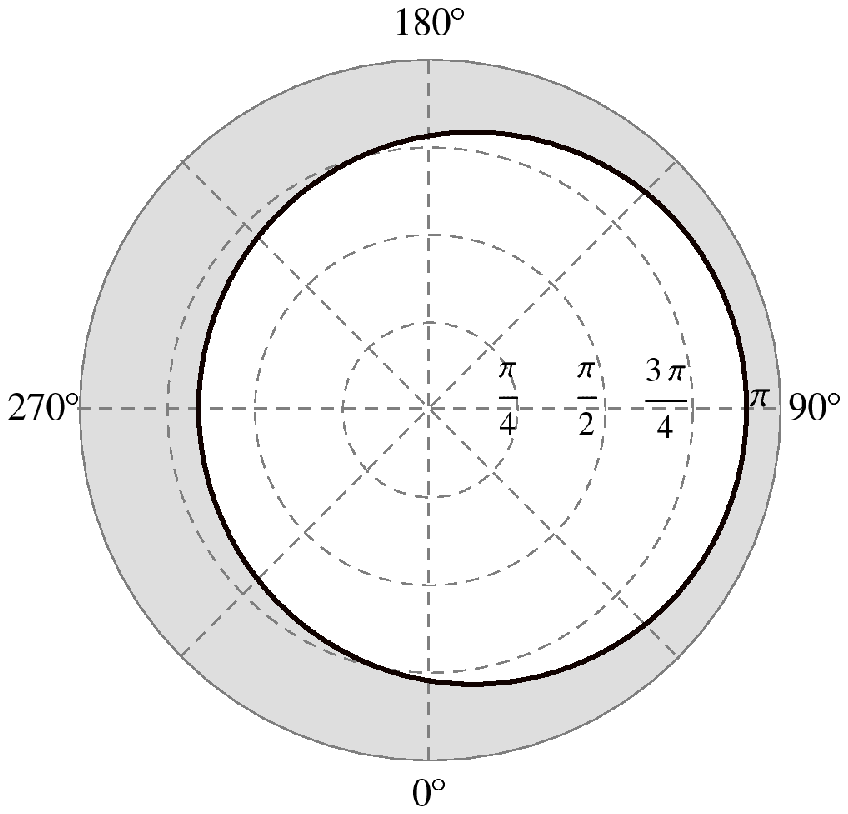}& \includegraphics[width=4.0cm]{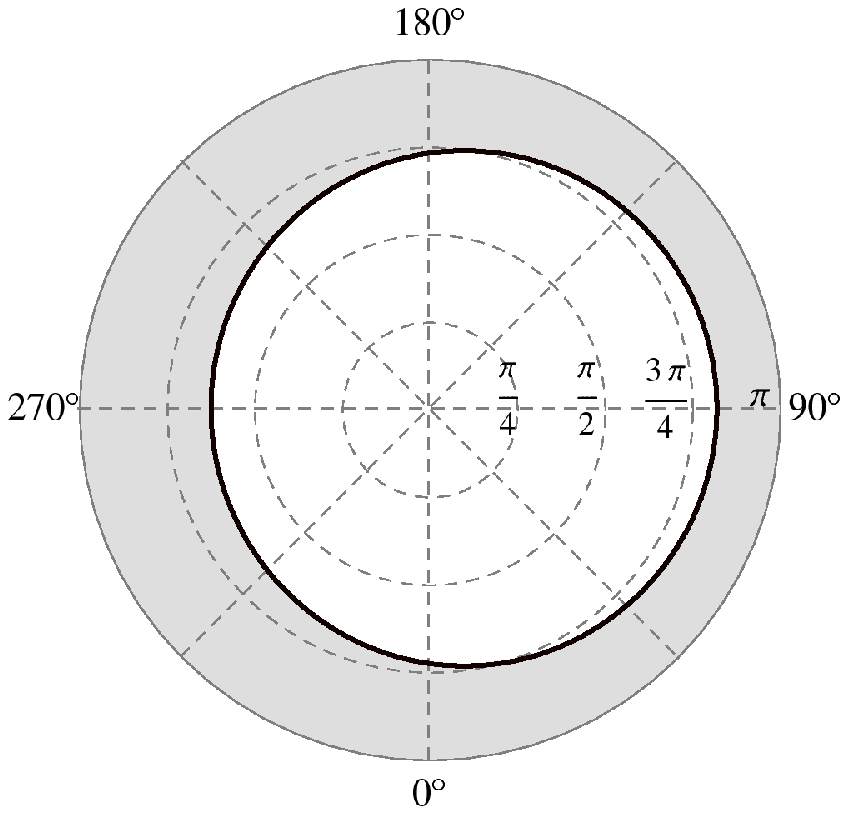}& \includegraphics[width=4.0cm]{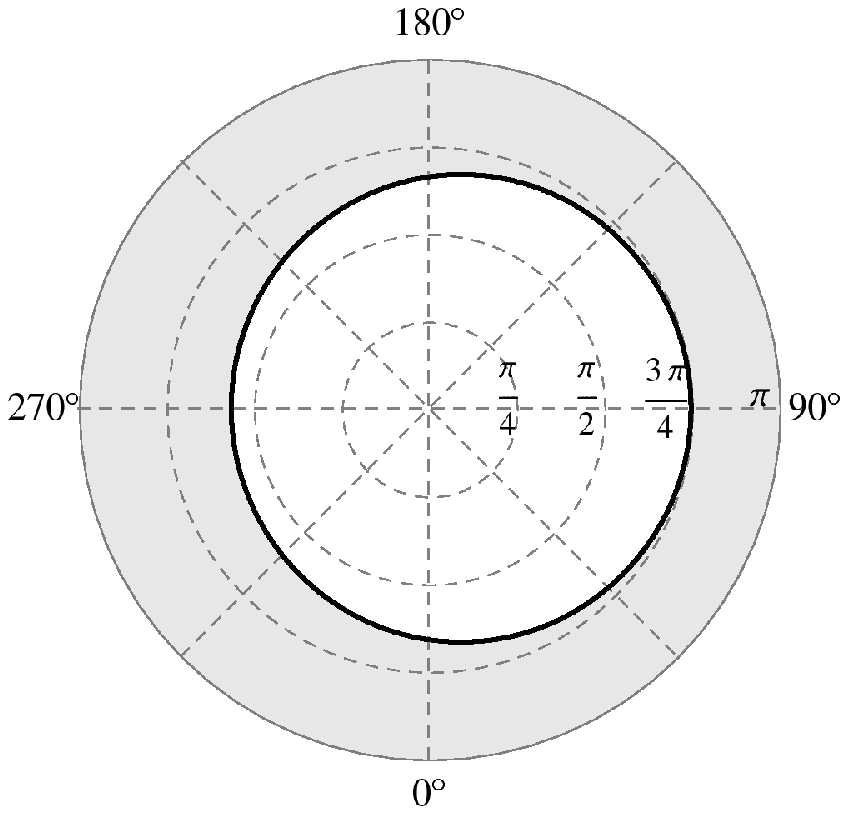}\\
  \includegraphics[width=4.0cm]{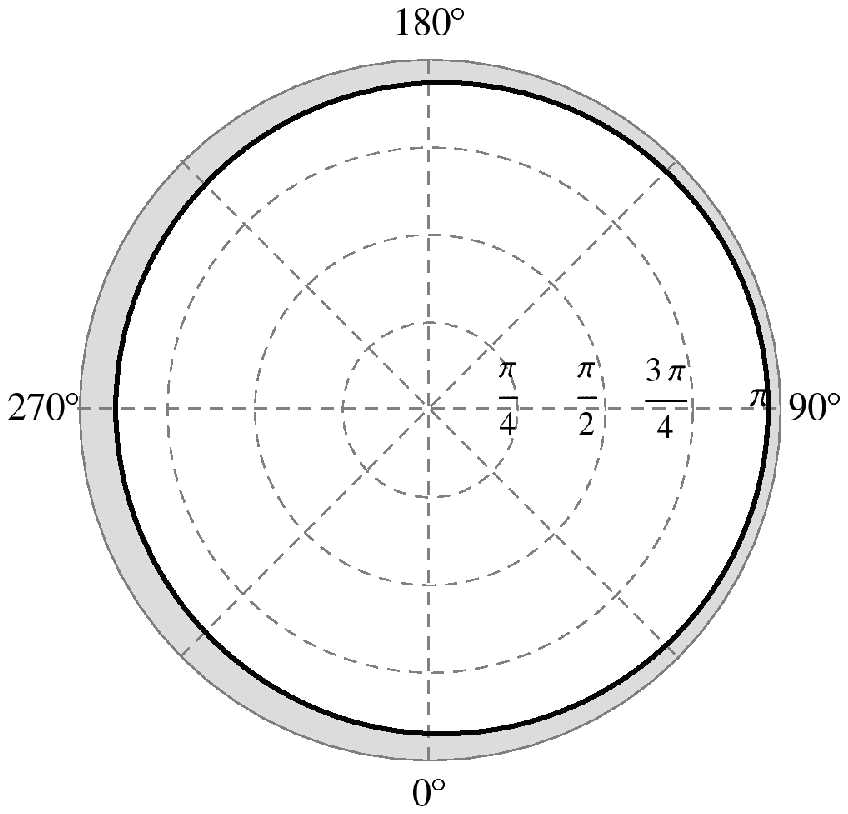}& \includegraphics[width=4.0cm]{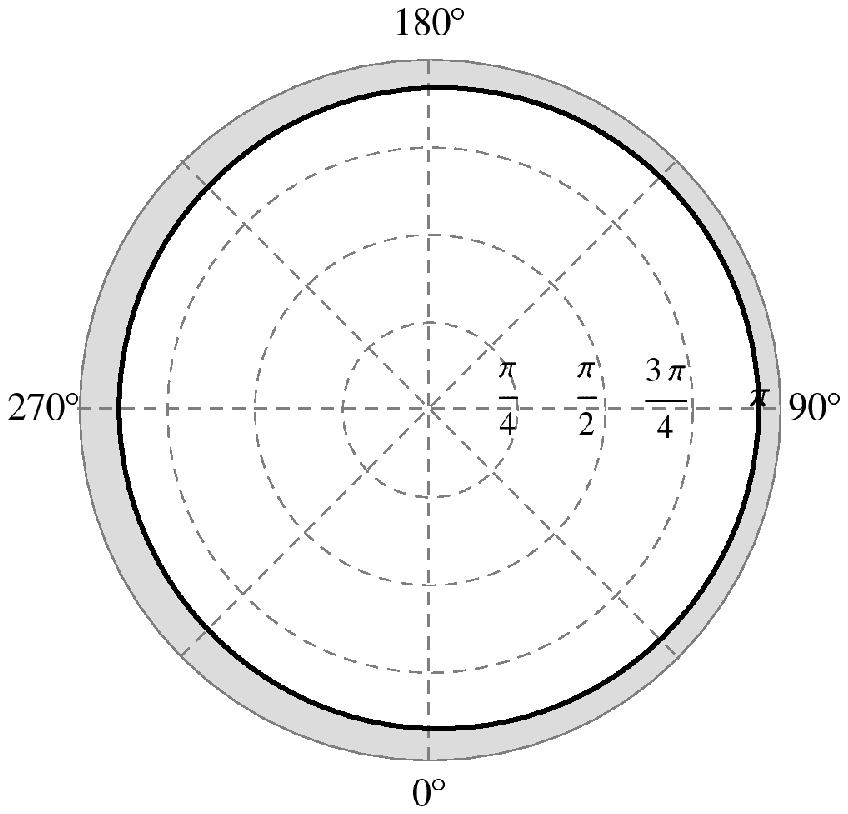}& \includegraphics[width=4.0cm]{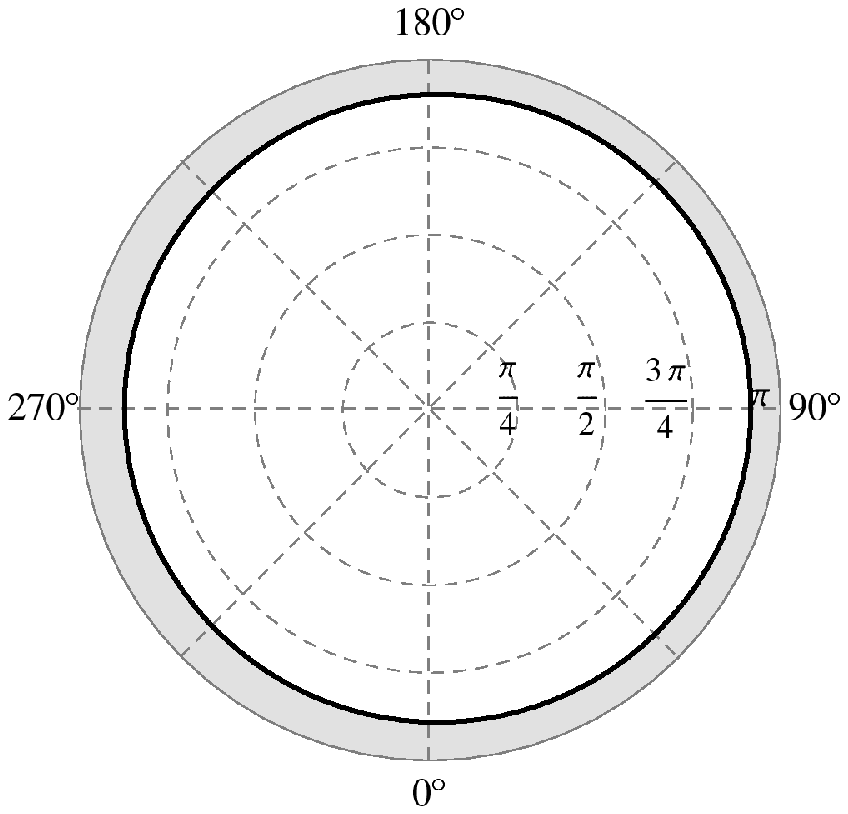}
  \end{tabular}
\caption{Light escape cones as seen by $LNRF$ in the vicinity of the  braneworld kerr black hole. 
Top set of images is plotted for radial coordinate of emitter $r_e=6M$ and bottom set for $r_e=20M$.
The rotational parameter $a=0.9981$ is fixed and the representative values of the braneworld parameter $b$ are $0$ (left), $-1$ (middle) and $-3$ (right). The shaded area represents photons captured by black hole. }\label{LNRF_fixed_a_on_b}
\end{figure}

\begin{figure}[ht]
  \begin{tabular}{ccc}
    \includegraphics[width=4.0cm]{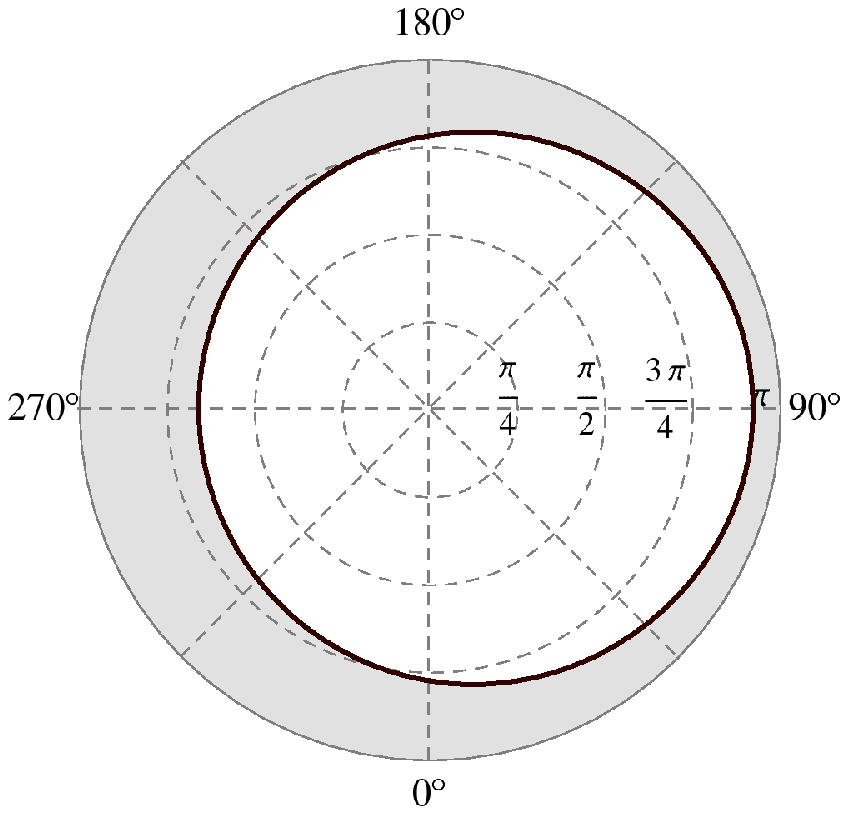}& \includegraphics[width=4.0cm]{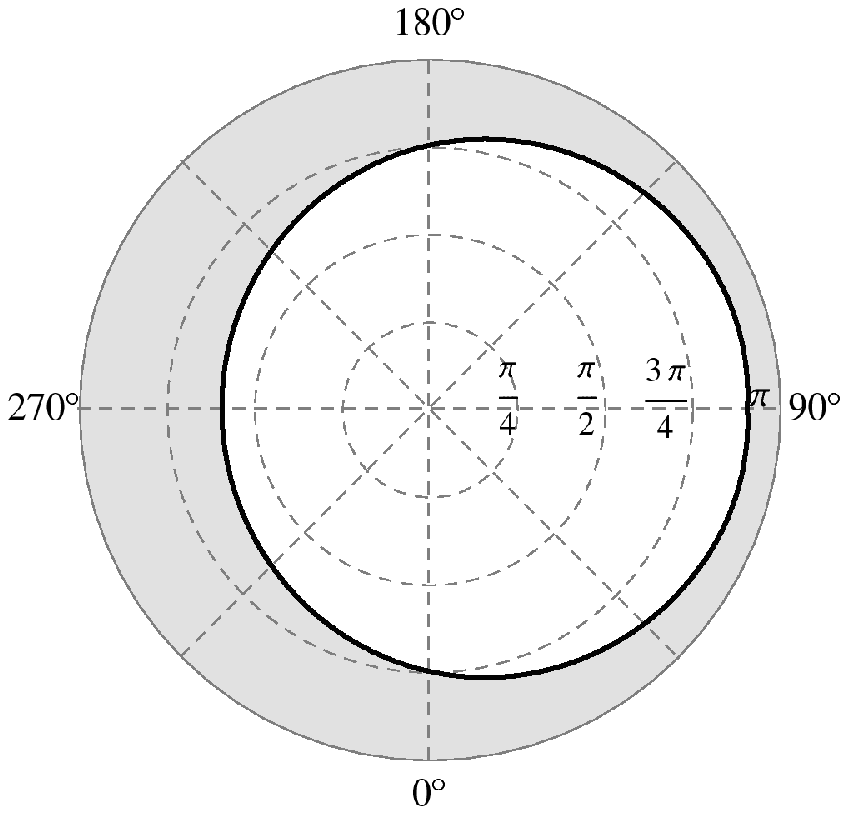}& \includegraphics[width=4.0cm]{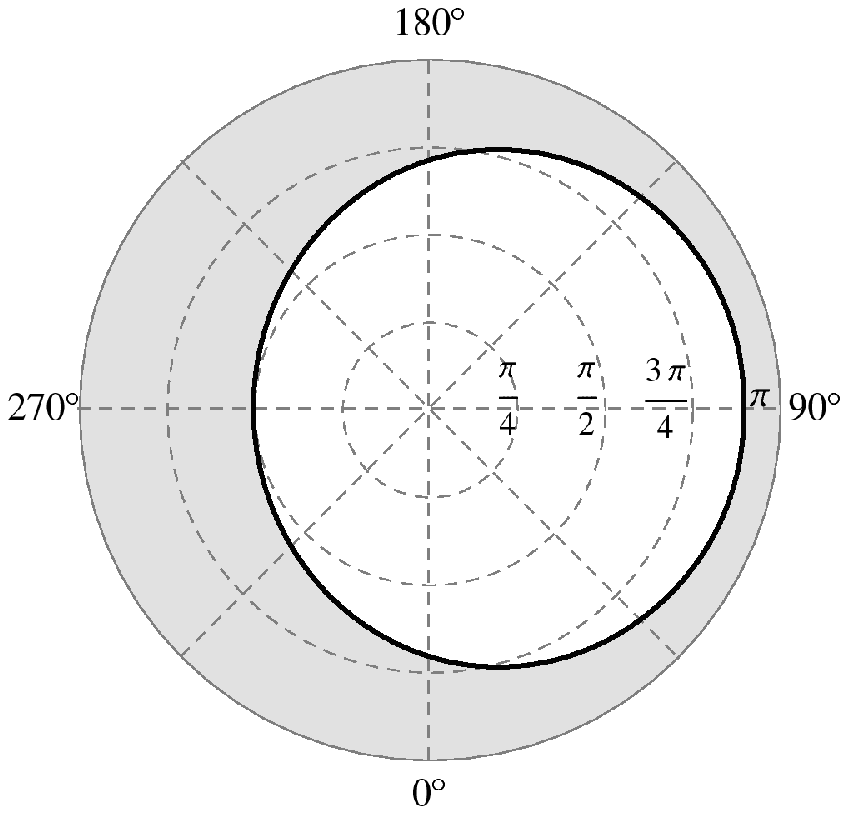}\\
   \includegraphics[width=4.0cm]{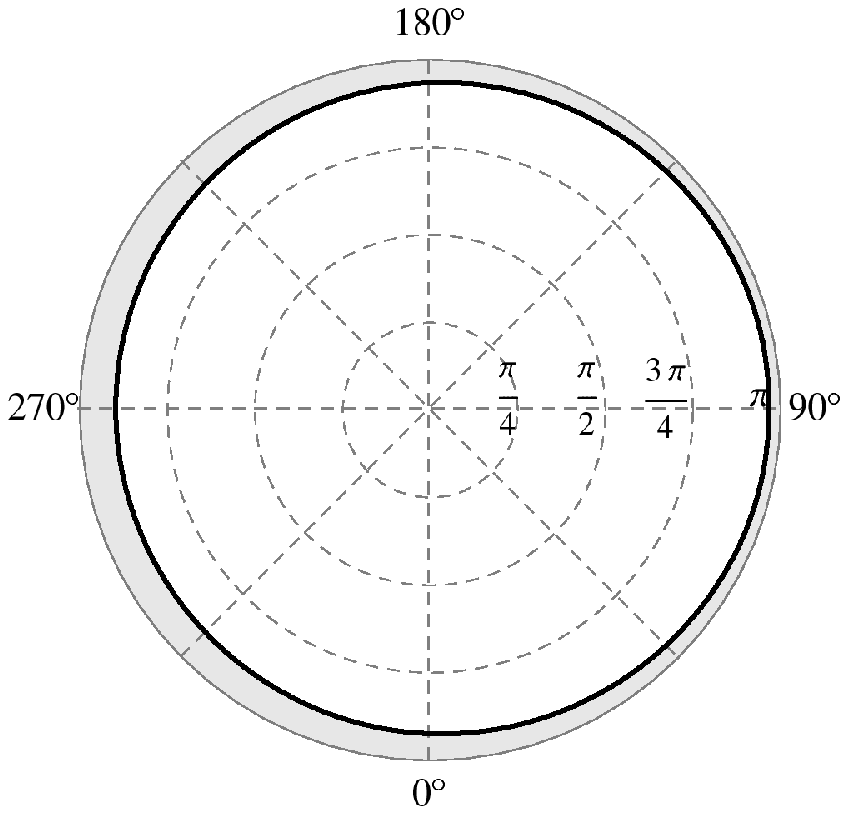}& \includegraphics[width=4.0cm]{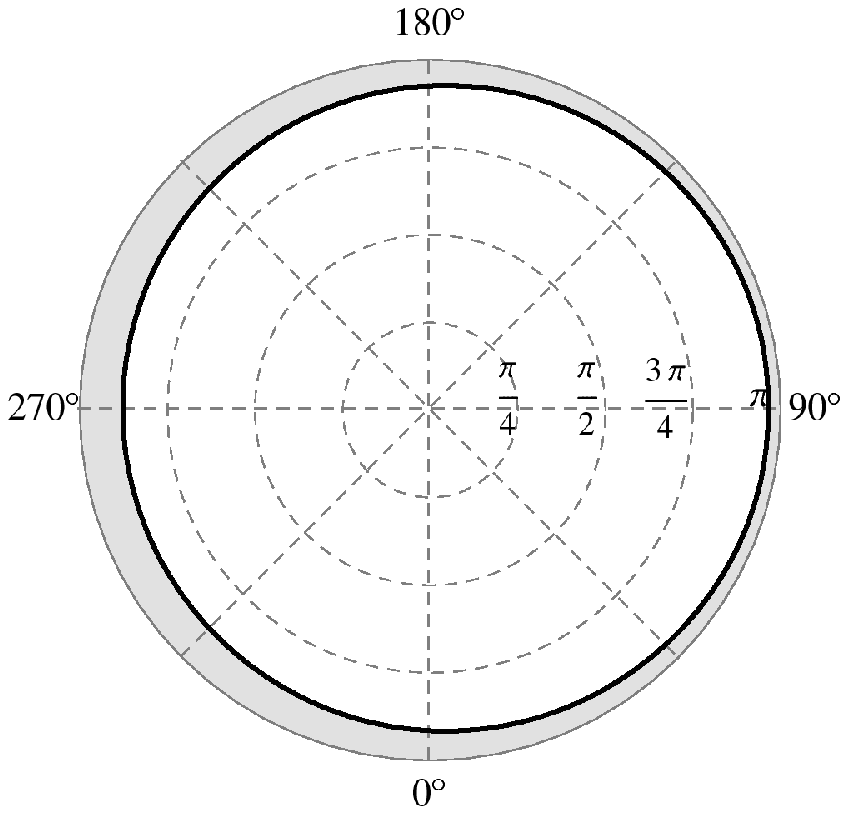}& \includegraphics[width=4.0cm]{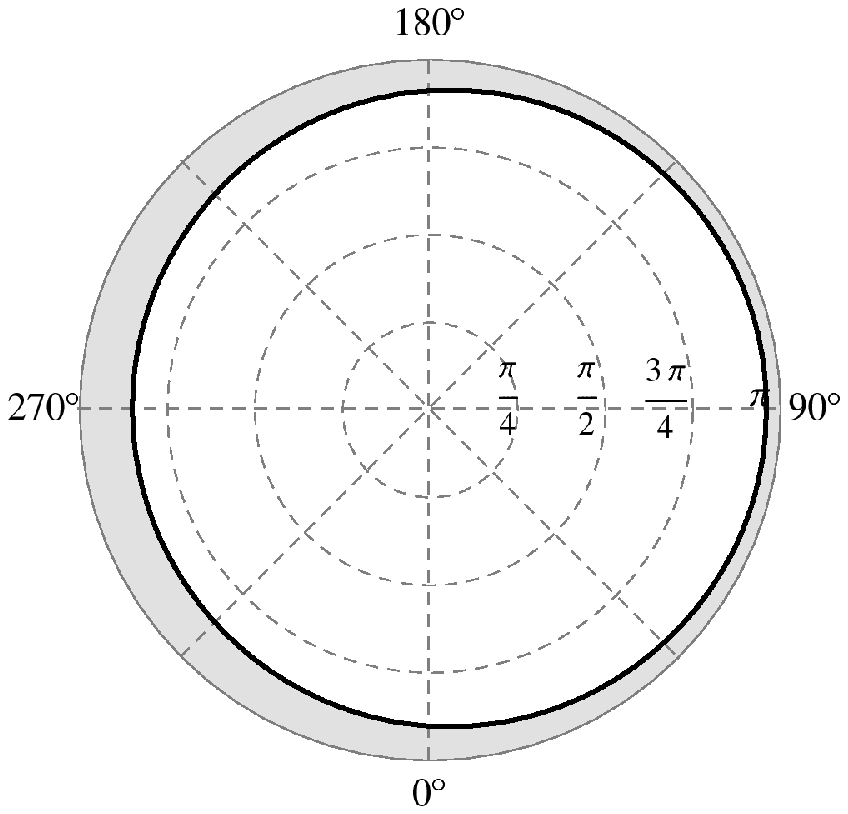}
  \end{tabular}
\caption{Light escape cones as seen by $LNRF$ in the vicinity of the extreme braneworld kerr black hole. Top set of images is plotted for radial coordinate of emitter $r_e=6M$ and bottom set for $r_e=20M$. The representative rotational  and braneworld parameters [$a^2$,$b$] are [$1.0$,$0.0$](left), [$2.0$,$-1.0$](middle) and [$4.0$,$-3.0$](right). The shaded area represents photons captured by black hole. }\label{LNRF_extreme_on_b}
\end{figure}

Behaviour of the $LNRF$ escape cones in dependence on the braneworld parameter
$b$ (and the spin $a$) is represented in Figures \ref{LNRF_fixed_a_on_b} and \ref{LNRF_extreme_on_b}.
The complementary  trapped cones, corresponding to photons captured by the black hole, are shaded. 

At a fixed radius expressed in units of $M$ the extension of the trapped cone grows with descending of $b$ to higher negative values and fixed spin $a$ and mass $M$, demonstrating  thus the growing gravitational pull of the black hole due to growing magnitude of the negative braneworld parameter. The same statement holds also in the case of extreme Kerr black holes, when $a$ grows and $b$ descends, while $M$ is fixed. Clearly, the positive braneworld parameters have tendency to increase the asymmetry of the cones, while the negative ones symmetrize the escape cones with growing of $|b|$.  On the other hand, the asymmetry of the escape cone grows with descending of $b$ for extreme black holes (Figure \ref{LNRF_extreme_on_b}).

	\begin{figure}[ht]
		\begin{tabular}{ccc}
			\includegraphics[width=4cm]{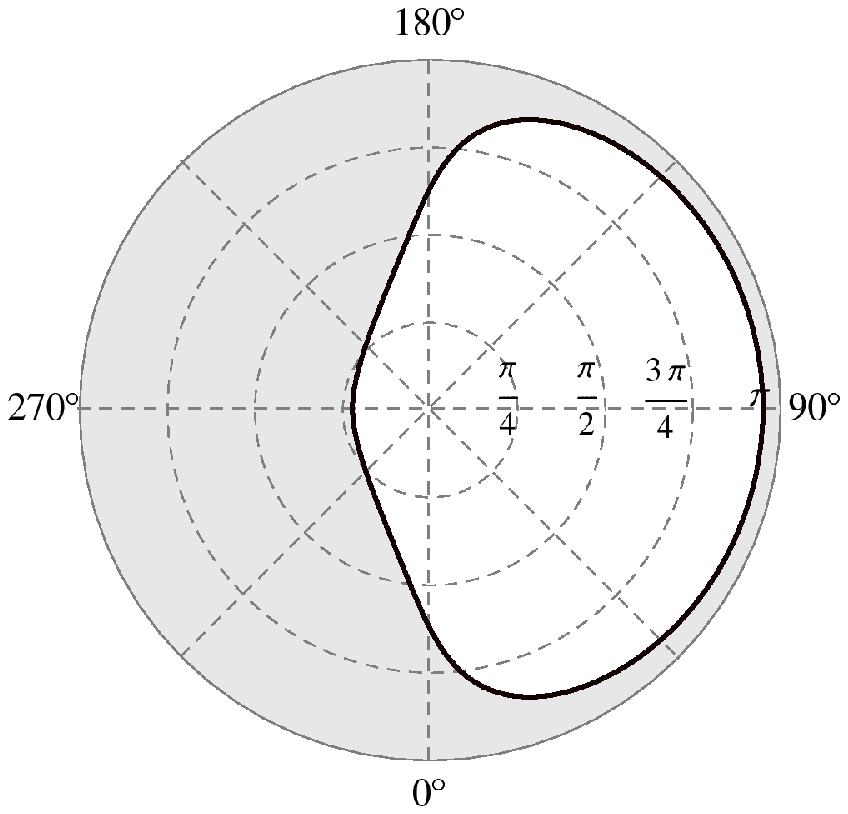}&
			\includegraphics[width=4cm]{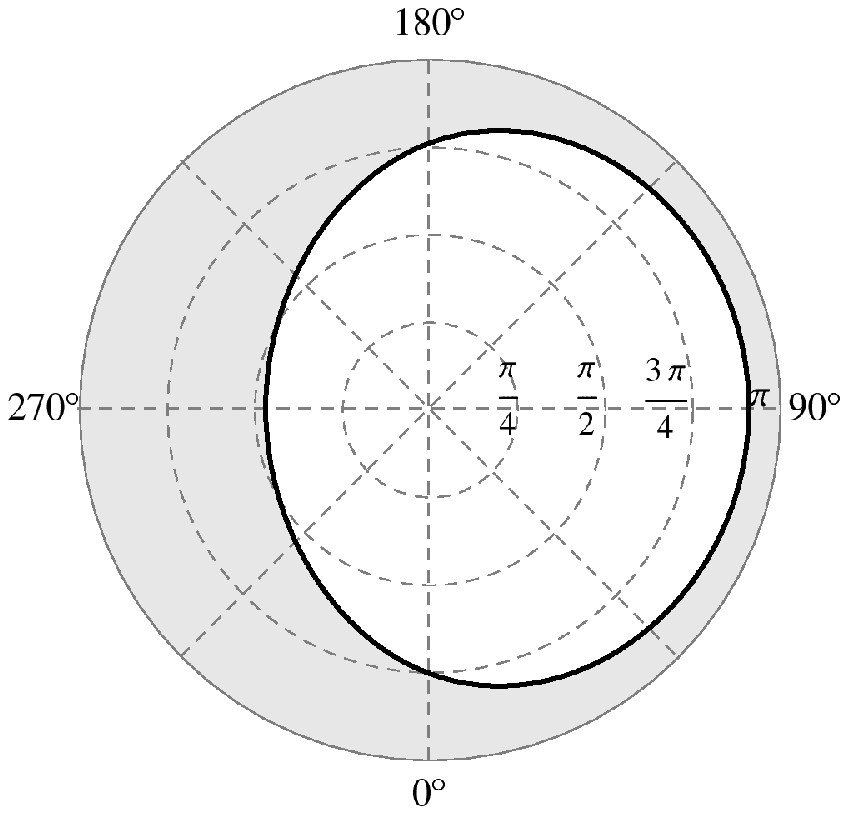}&
			\includegraphics[width=4cm]{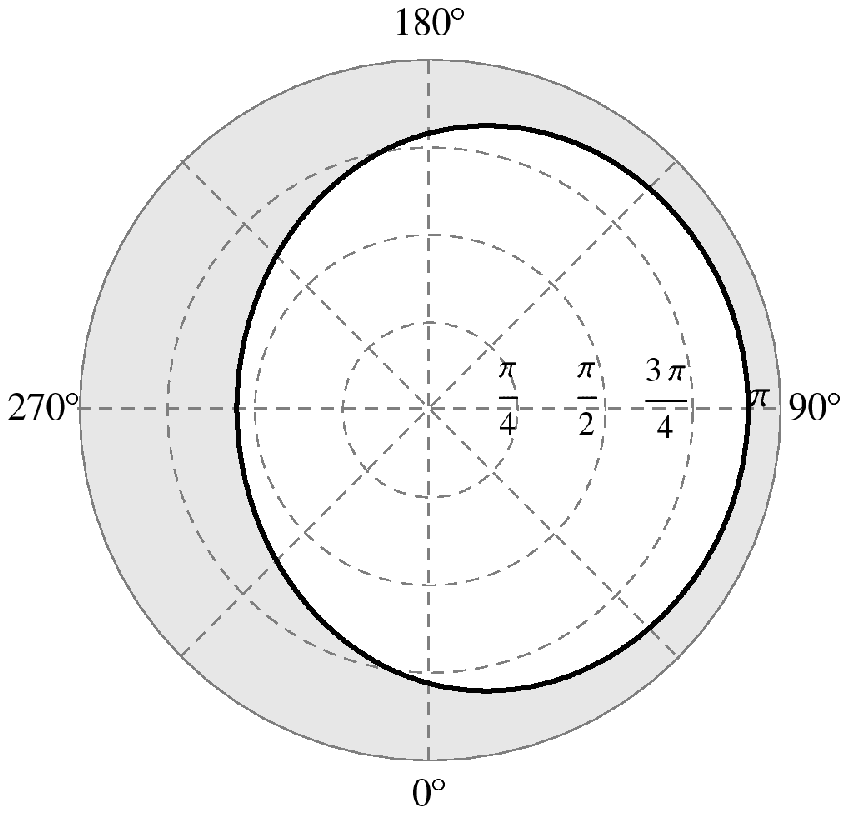}\\
			\includegraphics[width=4cm]{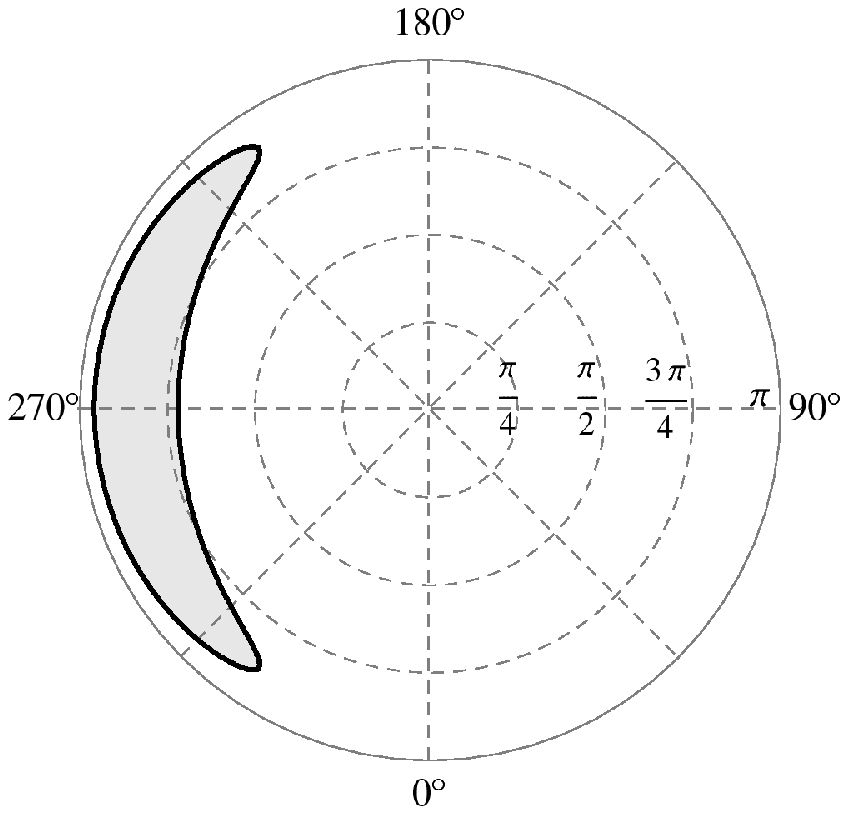}&
			\includegraphics[width=4cm]{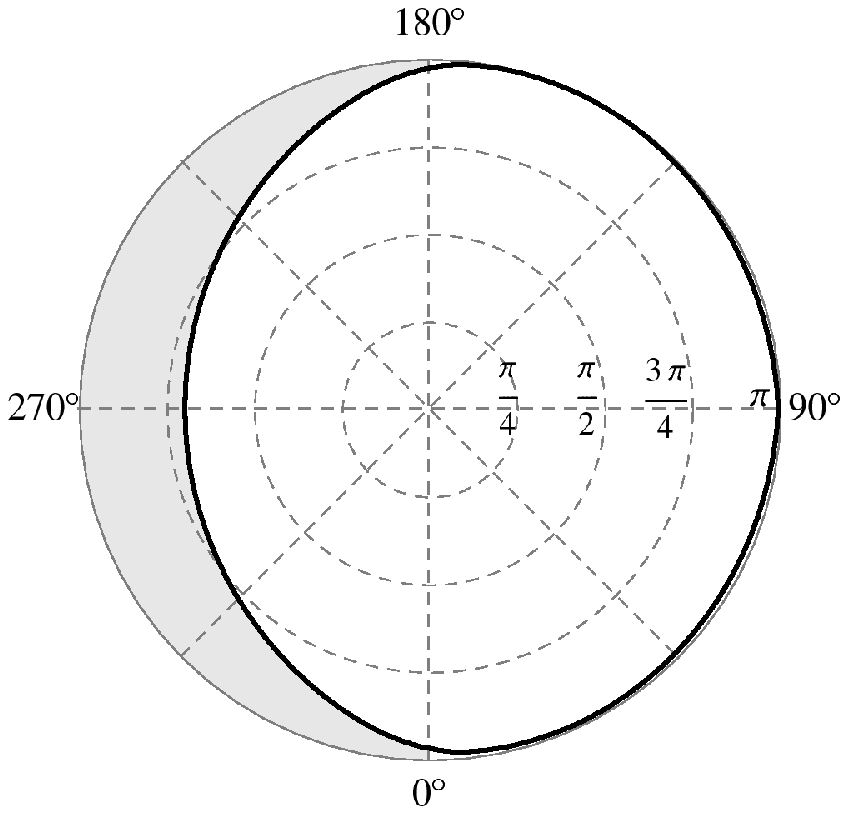}&
			\includegraphics[width=4cm]{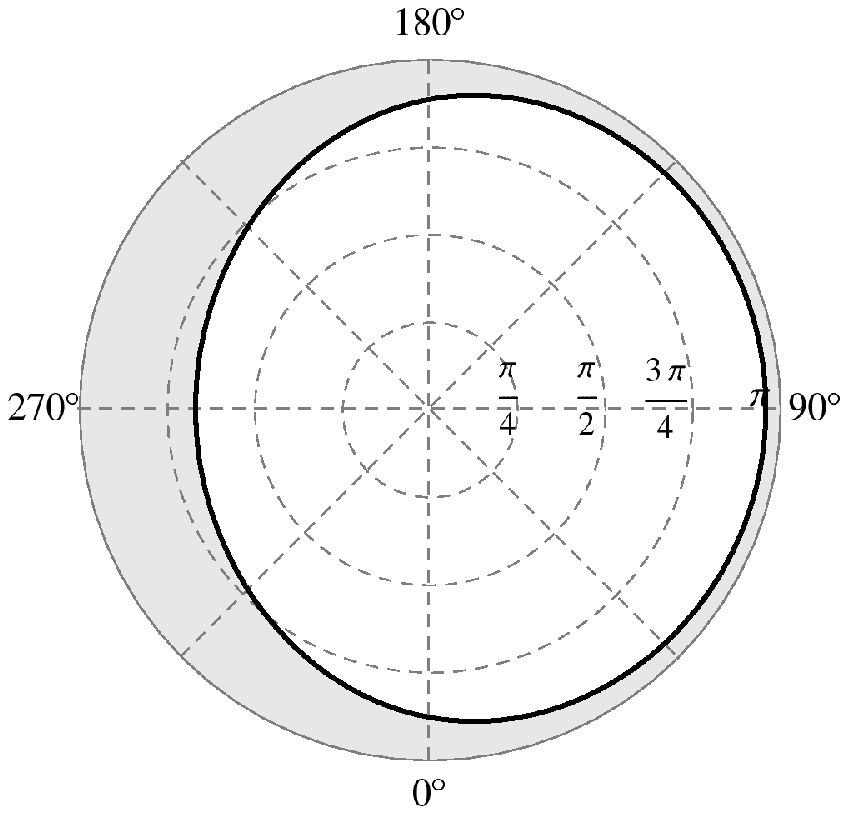}\\
			\includegraphics[width=4cm]{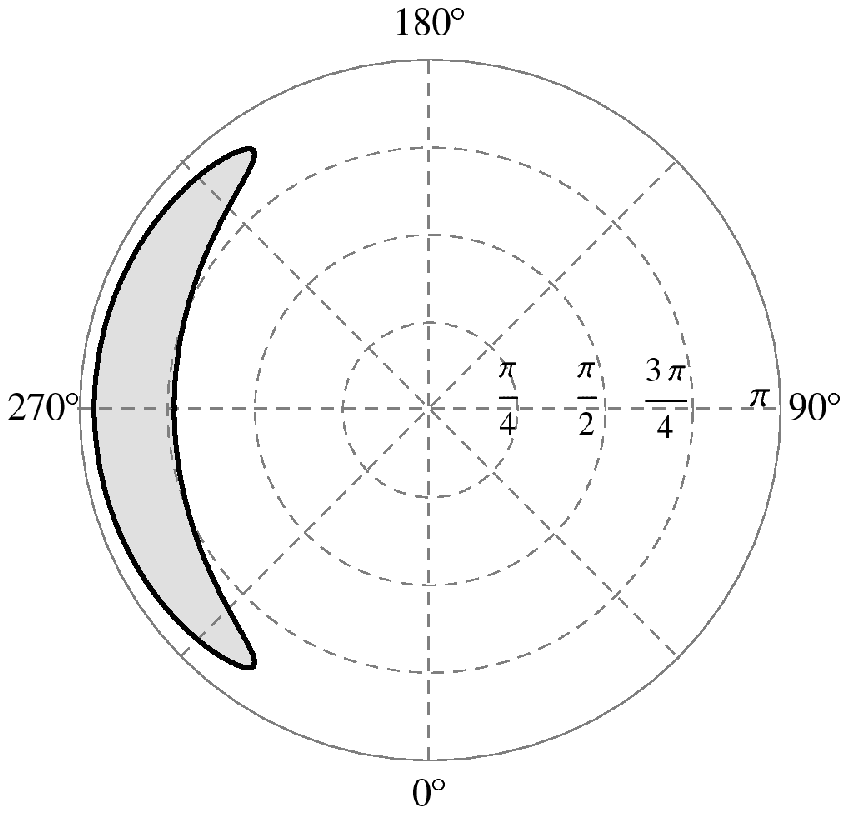}&
			\includegraphics[width=4cm]{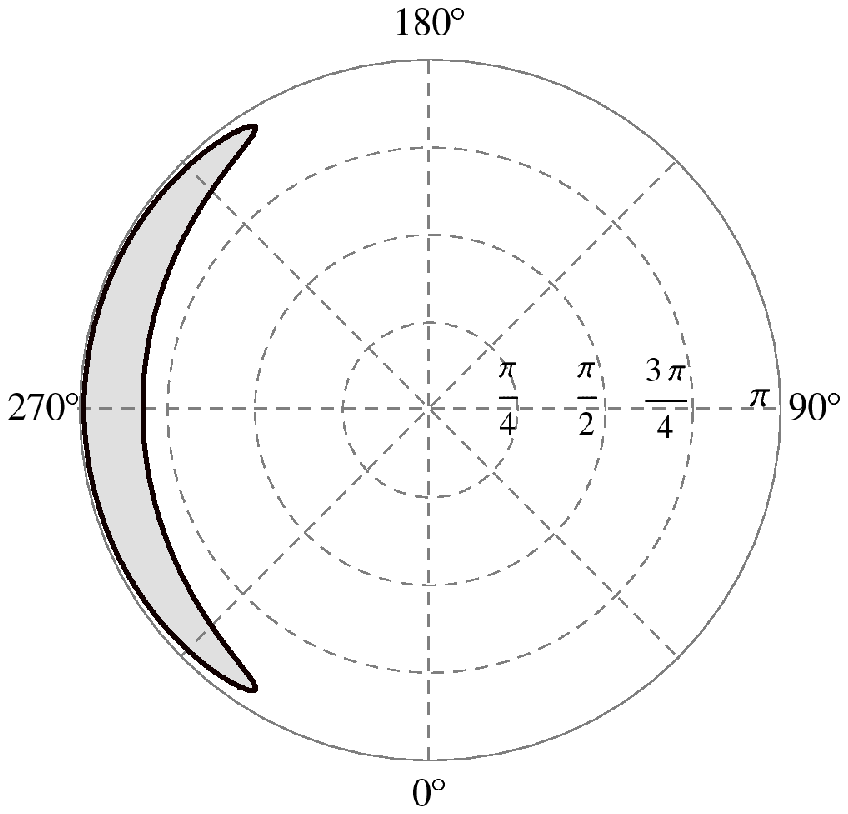}&
			\includegraphics[width=4cm]{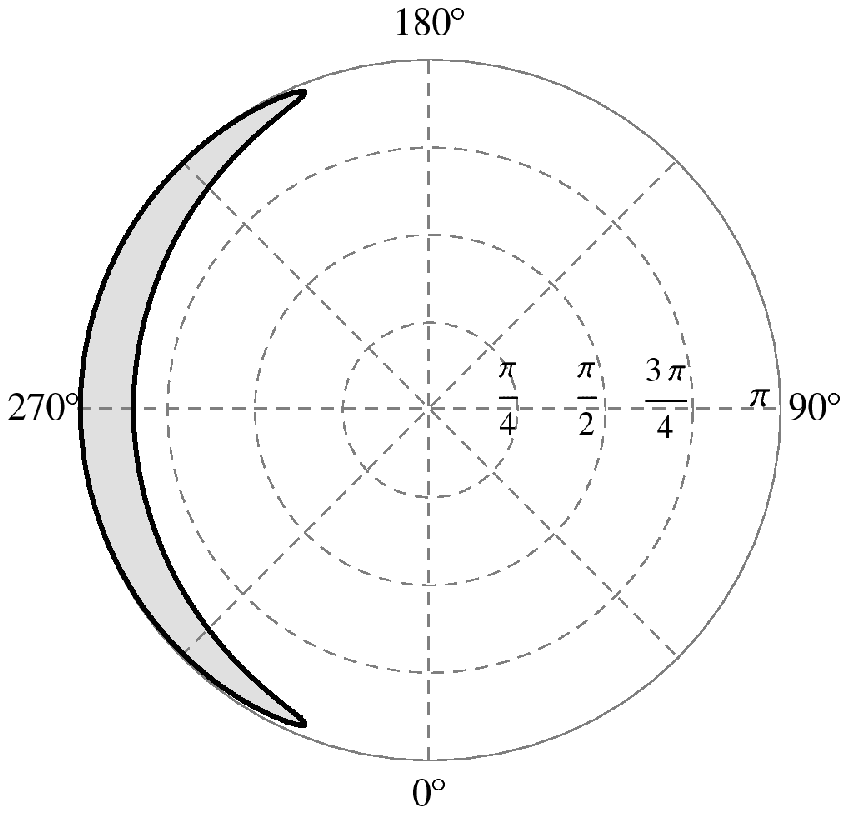}
		\end{tabular}
		\caption{Escape cones of GF+ observers. Top images are plotted for observer (emitter) at $r=r_{ms}$, middle images $r=10M$ and bottom images for 
		$r=10\cdot r_h$. The value of $a=0.9981$ is kept fixed. The representative values of $b$ are (from left to right) $0.0$, $-1.0$ and $-3$. }\label{GF_escape_cones}
	\end{figure}	

	\begin{table}[ht]
		\tbl{Table of relevant values of $r_{ms}$ and $r_{h}$ used in plots on Figs \ref{GF_escape_cones} and \ref{GF_escape_cones_extreme}.}
		{\begin{tabular}{@{}cccc@{}} 
		\toprule
		$(a^2, b)$ & (0.9981,0.0) & (0.9981,-1.0) & (0.9981,-3.0)\\ 
		\colrule
		$r_{ms}$ & 1.24M & 3.91M & 6.27M\\
		\colrule
		$r_h$ & 1.062M & 2.002M  & 2.73M\\
		\botrule
		\end{tabular}\label{tabulka1}}
	\end{table}

\begin{figure}[ht]
		\begin{tabular}{ccc}
			\includegraphics[width=4cm]{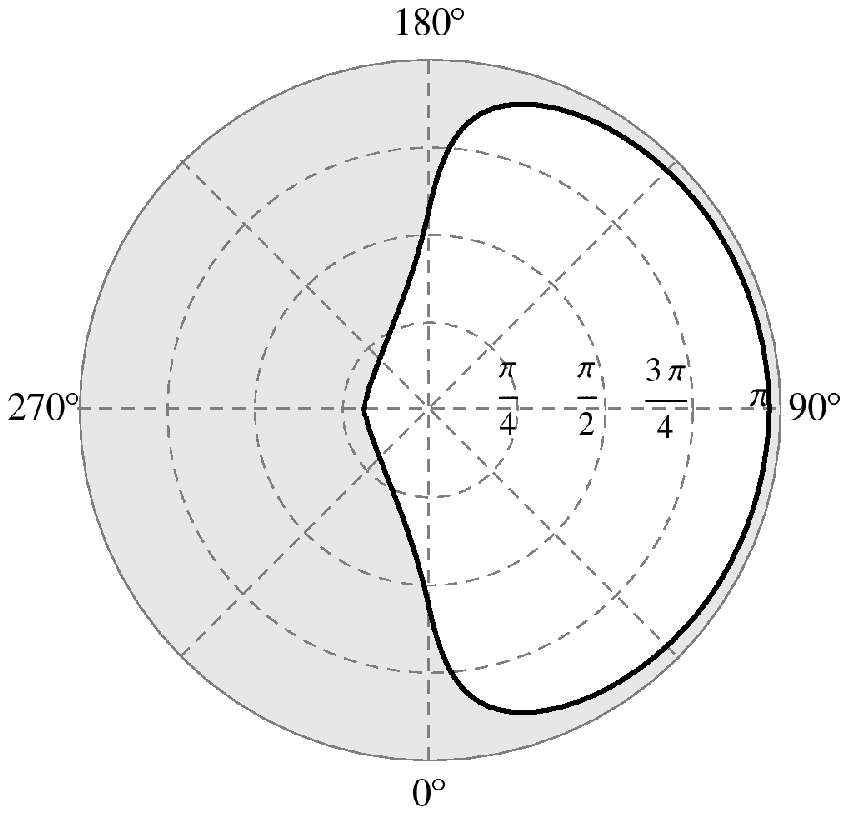}&
			\includegraphics[width=4cm]{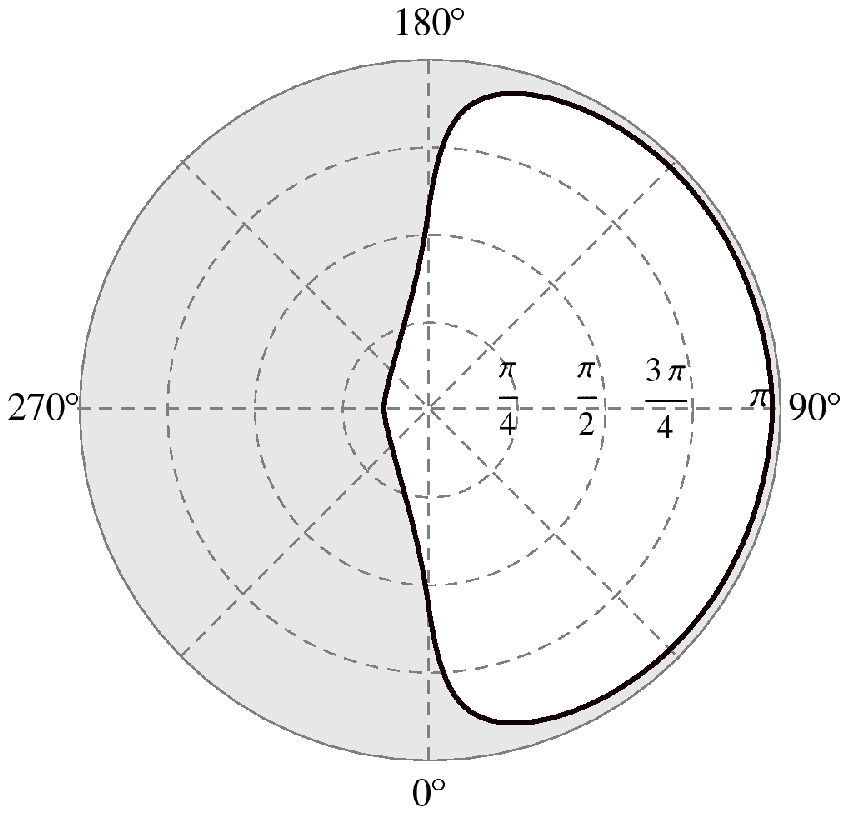}&
			\includegraphics[width=4cm]{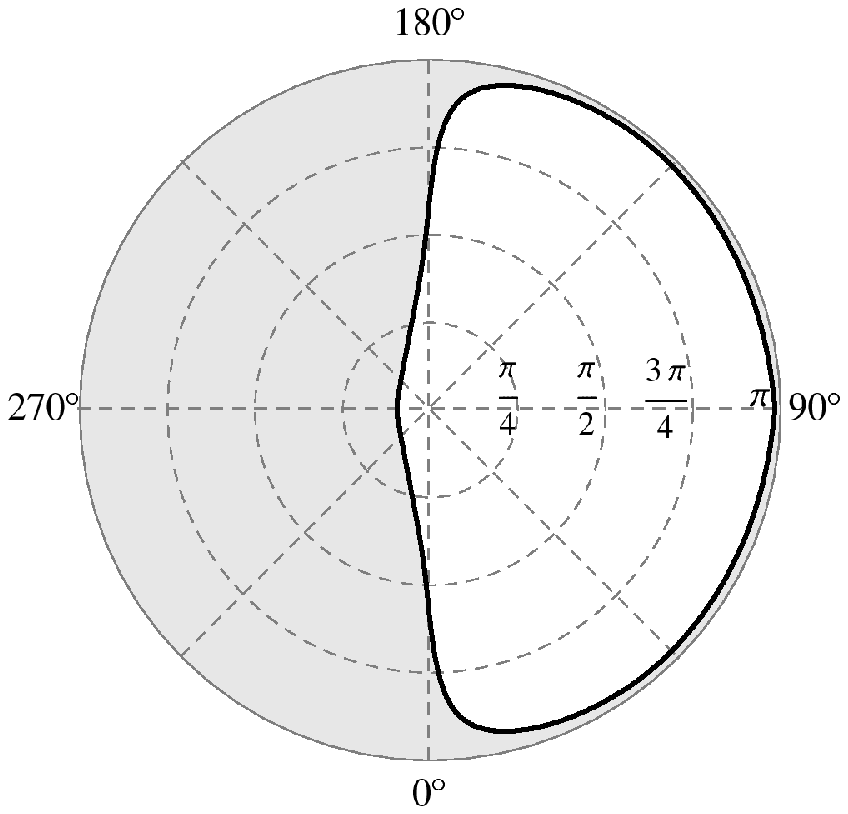}\\
			\includegraphics[width=4cm]{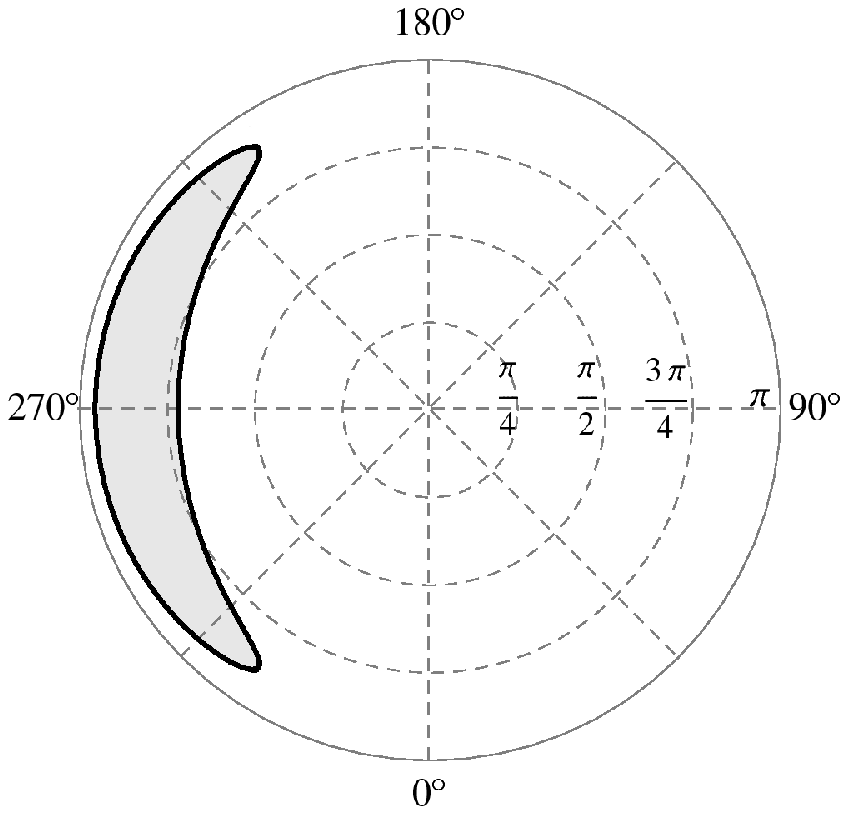}&
			\includegraphics[width=4cm]{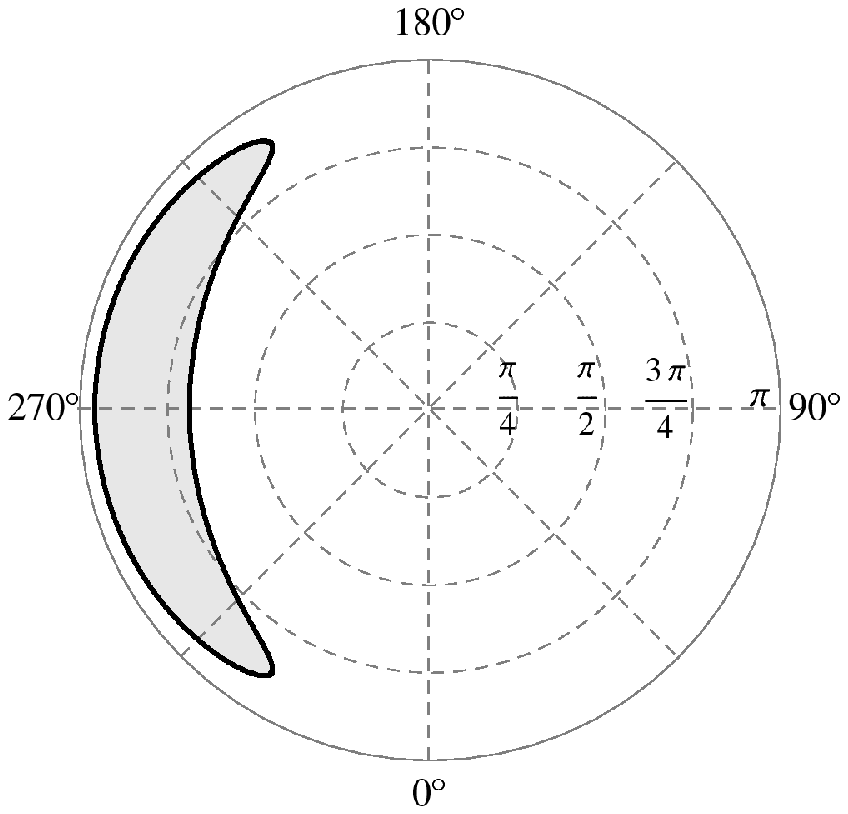}&
			\includegraphics[width=4cm]{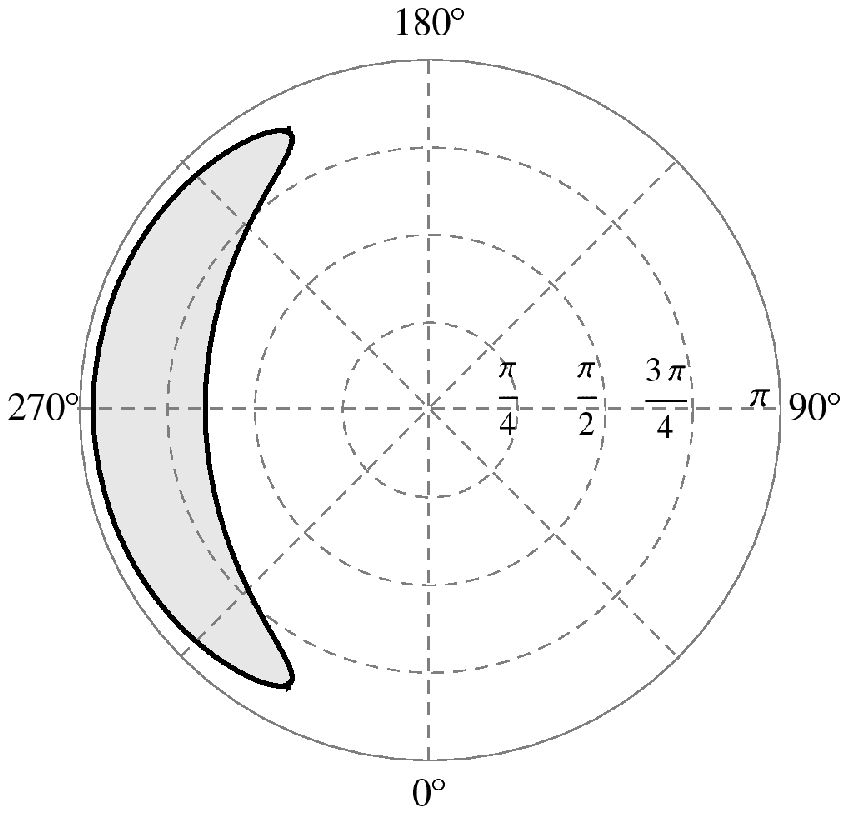}
		\end{tabular}
		\caption{Escape cones of GF+ observers. Top images are plotted for observer (emitter) at $r=r_{ms}$ and bottom images for $r=10M$. The value of $a^2+b=0.9999$ is kept fixed. The representative values of $(a^2; b)$ are (from left to right) $(0.9999;0.0)$, $(1.9999;-1.0)$ and $(3.9999;-3)$.  }\label{GF_escape_cones_extreme}
	\end{figure}	
 
 Further, we represent the influence of the braneworld parameter on the escape cones for the circular (corotating) geodesic frames in Figure \ref{GF_escape_cones}. Assuming astrophysically relevant sources in Keplerian accretion discs, their orbits must be located above the marginally stable orbit $r_{ms}$, determined implicitly by the condition \cite{Ali-Gum:2005:,Stu-Kot:2008}

\begin{equation}
 	a=a_{ms}(r;b)\equiv\frac{4(r-b)^{3/2}\mp r \sqrt{3r^2-2r(1+2b)+3b}}{3r-4b}.
\end{equation}
Therefore, we construct the escape cones for observers at $r=r_{ms}(a,b)$ and at fixed radii. In the sequence of black holes with fixed spin  $a=0.9981$ (Figure \ref{GF_escape_cones}) we include also  a subsequence of escape cones constructed at the same relative distance from the black hole horizon in order to better illustrate the role of the tidal charge $b$. In the sequence of near-extreme black holes with $a^2+b=0.9999$ (Figure \ref{GF_escape_cones_extreme}) the third sequence is not necessary as the black hole horizon is fixed at $r_h=1.01M$. Figures \ref{GF_escape_cones} and \ref{GF_escape_cones_extreme} demonstrate that the trapped cone expands as the tidal charge descends to lower negative values, both for black holes with fixed spin $a$ and for near-extreme holes. On the other hand, considering the cones at $r_{ms}$ we can conclude that the descending tidal charge ($b<0$) symmetrizes their shape for fixed $a$, but makes them strongly asymmetric for near-extreme black holes shrinking them strongly in the direction of the black hole rotation.

\begin{figure}
 \begin{tabular}{ccc}
  \includegraphics[width=4cm]{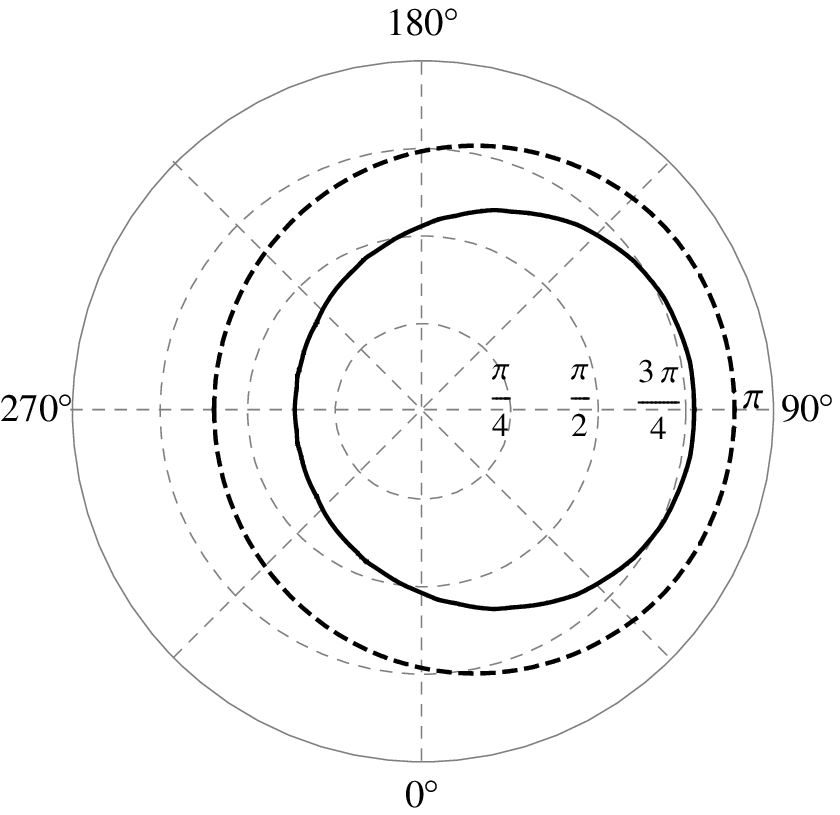}&\includegraphics[width=4cm]{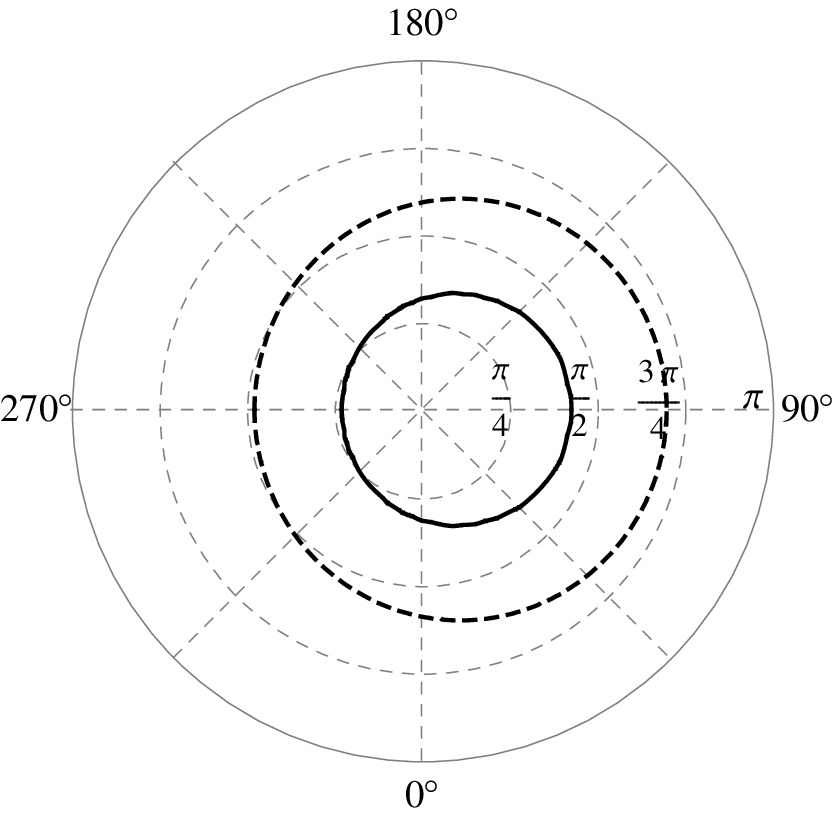}&\includegraphics[width=4cm]{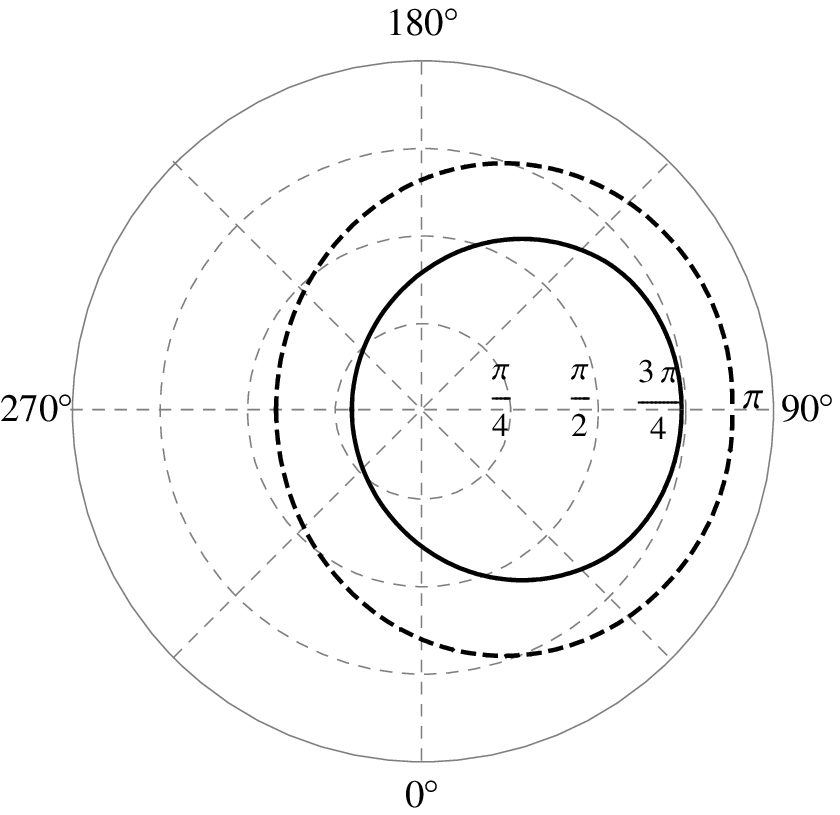}\\
  \includegraphics[width=4cm]{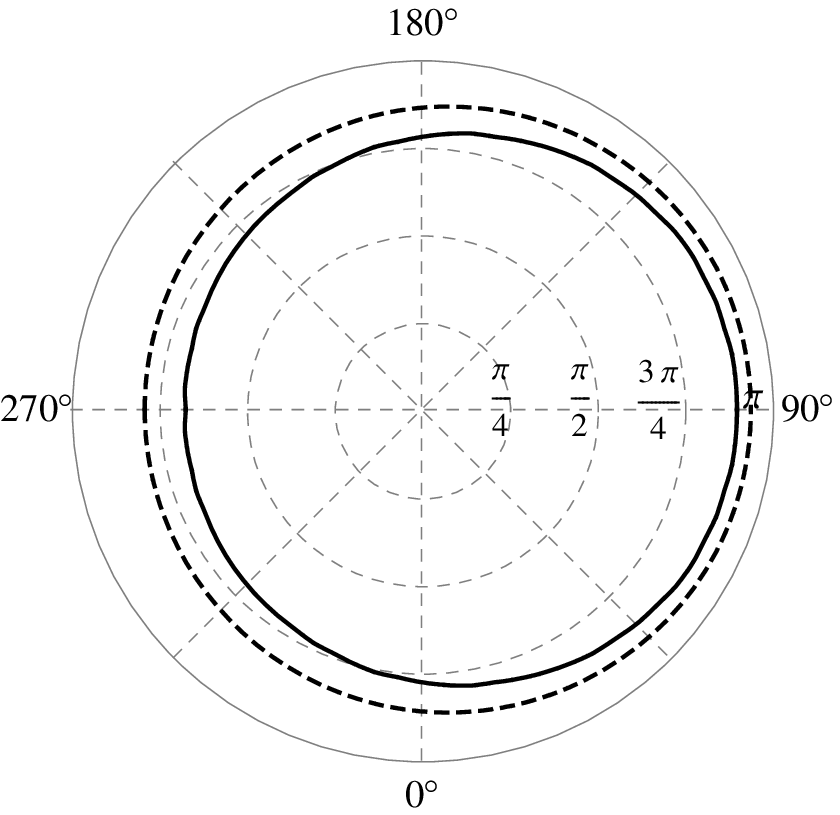}&\includegraphics[width=4cm]{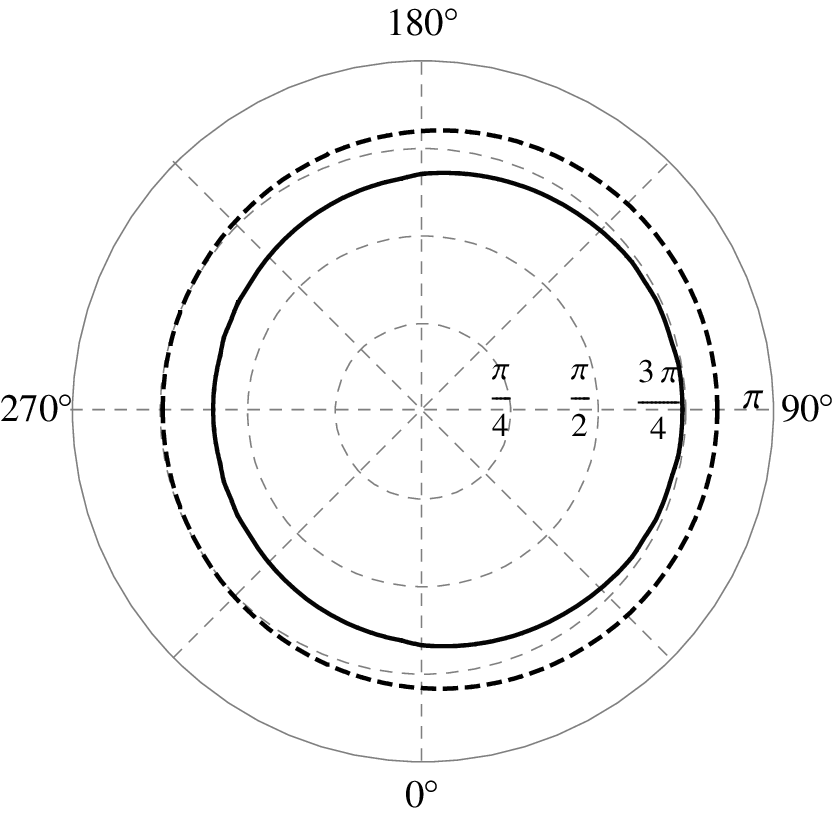}&\includegraphics[width=4cm]{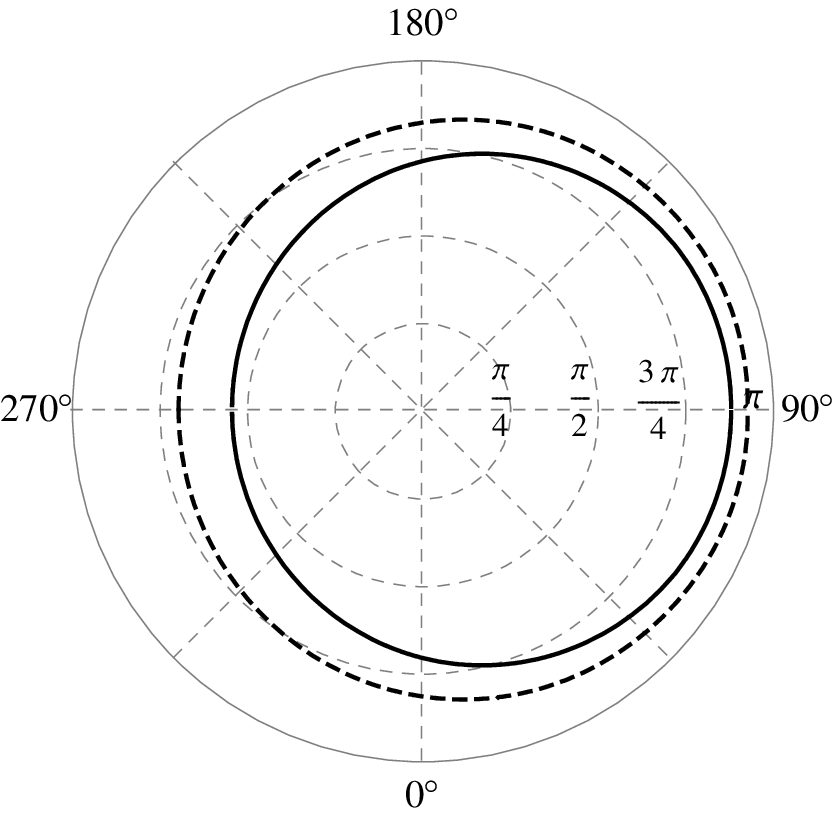}\\
  \includegraphics[width=4cm]{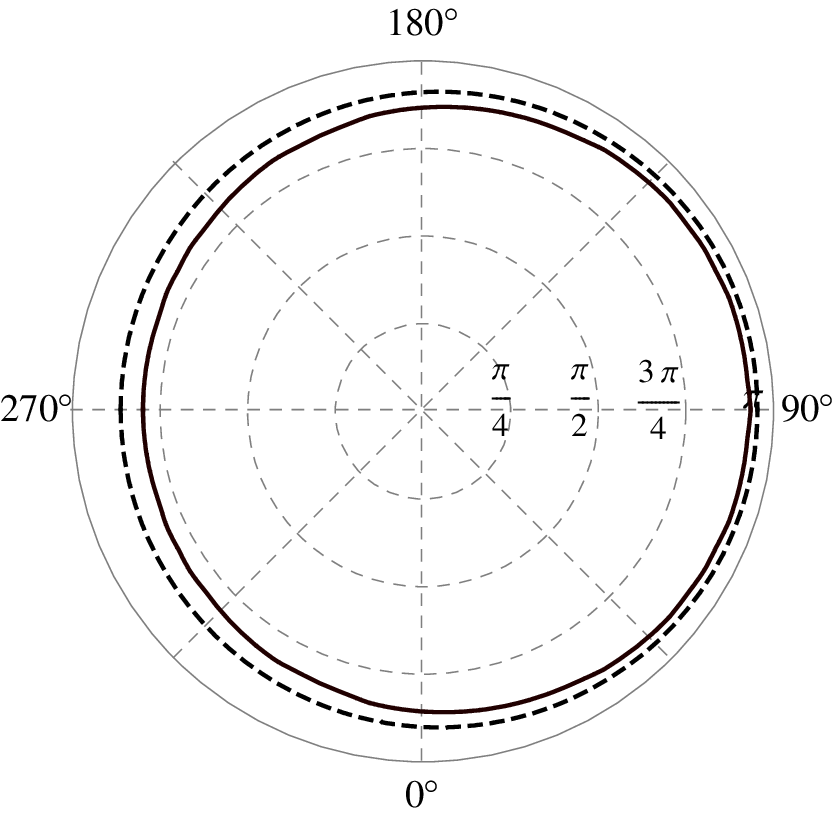}&\includegraphics[width=4cm]{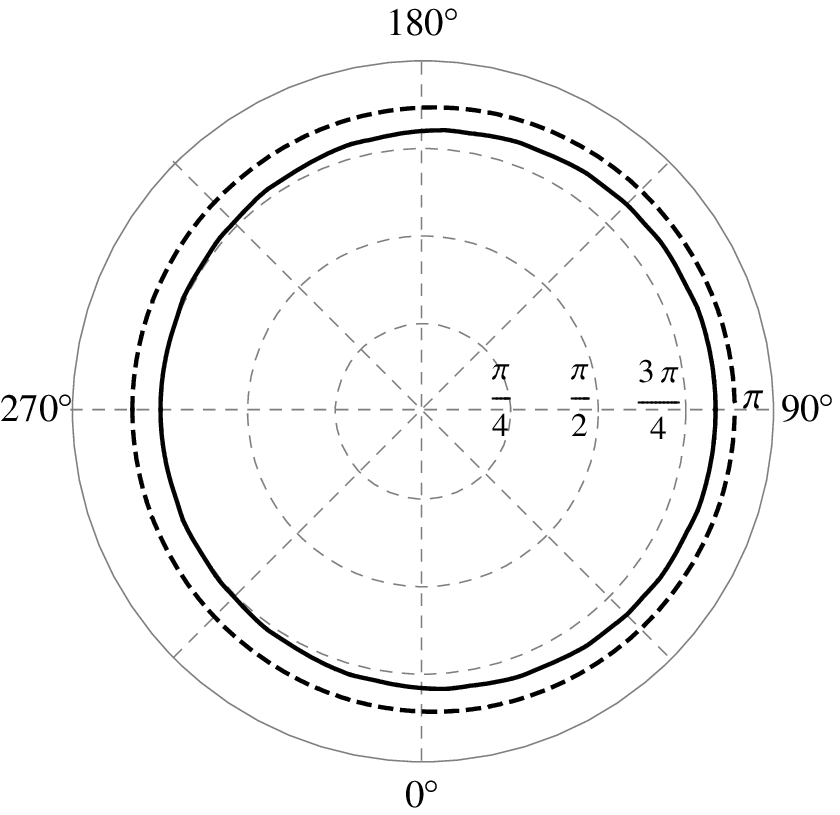}&\includegraphics[width=4cm]{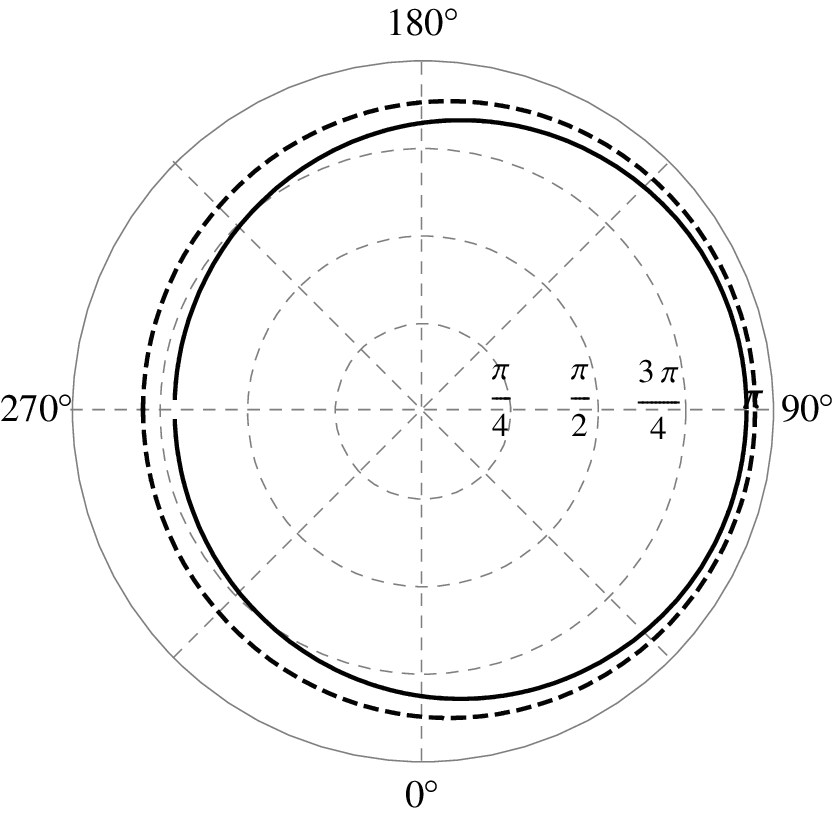}
 \end{tabular}
\caption{\label{fig13a_f}Comparison of the effect of the tidal charge $b$ on the shape of
  light escape cones of locally nonrotating (dashed curves) frames  and free falling (solid curves)
  frames. In the left column light escape cones are plotted for the tidal
  charge parameter $b=0$ and in the middle one the light escape cones are
  plotted for $b=-3$. The spin $a=0.9981$ is kept fixed in both columns. The right column gives the sequence of the escape cones for an extreme black hole with [$a^2=4;b=-3$]. Emitting sources in all plots are moving in the equatorial plane. The radial distances of emitter are $r_e=5M$ (top row), $r_e=10M$ (middle row) and $r_e=15M$ (bottom row).}
\end{figure}

Finally we demonstrate the relevance of the tidal charge $b$ in the character of escape cones of the $RFF_-$ (comparing them with those related to $LNRF$) in Figure \ref{fig13a_f}. We construct the escape cones for two typical values of the tidal charge ($b=0$, $b=-3$) in a sequence of radii where the free-falling source is radiating, demonstrating thus the combined growing influence of the black hole gravitational pull on the photon motion and the velocity of the free-falling source. In order to illustrate the phenomena in a clear way, we compare the $RFF_-$ escape cones to the corresponding $LNRF$ escape cones. Clearly, the tidal charge descending to higher negative values makes stronger squeezing of the free-falling  cones relative to the $LNRF$ escape cones at any fixed radius. Notice that both the $RFF_-$ and $LNRF$ cones are shifted due to the black hole rotational dragging. 
We again observe the tendency of negative brany parameters to symmetrize and squeeze the escape cones. At a fixed $r$, the escape cones  become smaller for growing $|b|$ due to stronger gravity. For completeness we present sequence of both the $RFF_-$ and $LNRF$ escape cones at the three fixed radii for an extreme black hole with $b=-3$ and $a^2=4$. We observe that both the $RFF_-$ and $LNRF$ cones are strongly shifted in the sense of the black hole rotation in vicinity of black hole horizon due to growing influence of the spin. The symmetrizing effect of descending values of negative tidal charge is canceled by strong influence of the rotational effects due to growing black hole spin.

\clearpage

\section{\label{sec:Silhuette}Silhuette of braneworld Kerr black hole}

In principle, it is of astrophysical importance to consider a black hole in front of a source of illumination whose angular size is large compared with the angular size of the black hole \cite{Bardeen:1973:}. A distant observer will see a silhuette of the black hole, i.e., a black hole in the larger bright source. The rim of the black hole silhuette corresponds to photon trajectories spiralling around the black hole many times before they reach the observer. Of course, the shape of the silhuette enables, in principle, determination of the black hole parameters. But we have to be aware of the strong dependency of the silhuette shape on the observer viewing angle; clearly, the shape will be circular for observers on the black hole rotation axis, and its deformation grows with observer approaching the equatorial plane.

Assuming that distant observers measure photon directions relative to the symmetry center of the gravitational field, the component of the angular displacement perpendicular to the symmetry axis is given by $-p^{(\varphi)}/p^{(t)}$ (for black hole rotating anticlockwise relative to distant observers), while for angular displacement parallel to the axis it is given by $p^{(\theta)}/p^{(t)}$. These angles are proportional to $1/r_0$, therefore, it is convenient to use the impact parameters in the form independent of $r_0$ \cite{Bardeen:1973:}

\begin{equation}
 \tilde{\alpha}=-r_0\frac{p^{(\varphi)}}{p^{(t)}}=-\frac{\lambda}{\sin\theta_0},\label{silalpha}
\end{equation}

and

\begin{eqnarray}
 \tilde{\beta}&=&r_0\frac{p^{(\theta)}}{p^{(t)}}=\left[q+a^2\cos^2\theta_0-\lambda^2\cot^2\theta_0\right]^{1/2}\nonumber\\
&&=\left[\mathcal{L}+a^2\cos^2\theta-\frac{\lambda^2}{\sin^2\theta_0}\right]^{1/2}.\label{silbeta}
\end{eqnarray}
Photon trajectories reaching the observer are represented by points in the $(\tilde{\alpha}-\tilde{\beta})$ plane representing a small portion of the celestial sphere of the observer.

The shape of the black hole silhuette is the boundary of the no-turning-point region, i.e., it is the curve $\mathcal{L}=\mathcal{L}^{min}_{max}(\lambda)$ expressed in the $(\tilde{\alpha}-\tilde{\beta})$ plane of the impact parameters. For observers in the equatorial plane $(\theta_0 = \pi/2)$, $\tilde{\alpha}=-\lambda$, $\tilde{\beta}=(\mathcal{L}-\lambda^2)^{1/2}=q^{1/2}$.

\begin{figure}[ht]
\begin{tabular}{cc}
  \includegraphics[width=5.5cm]{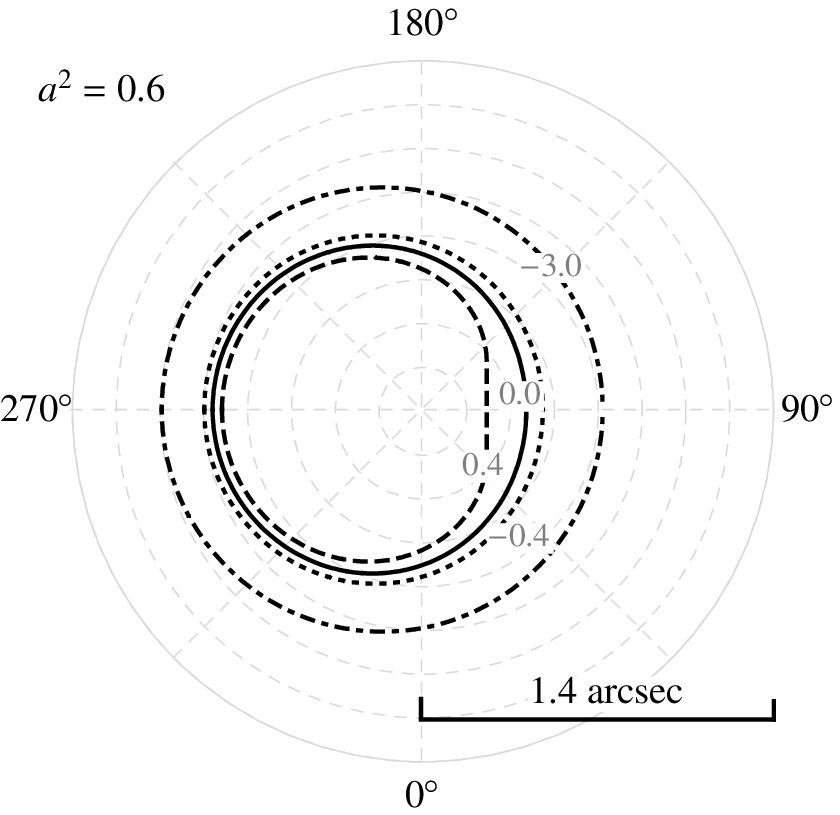}&\includegraphics[width=5.5cm]{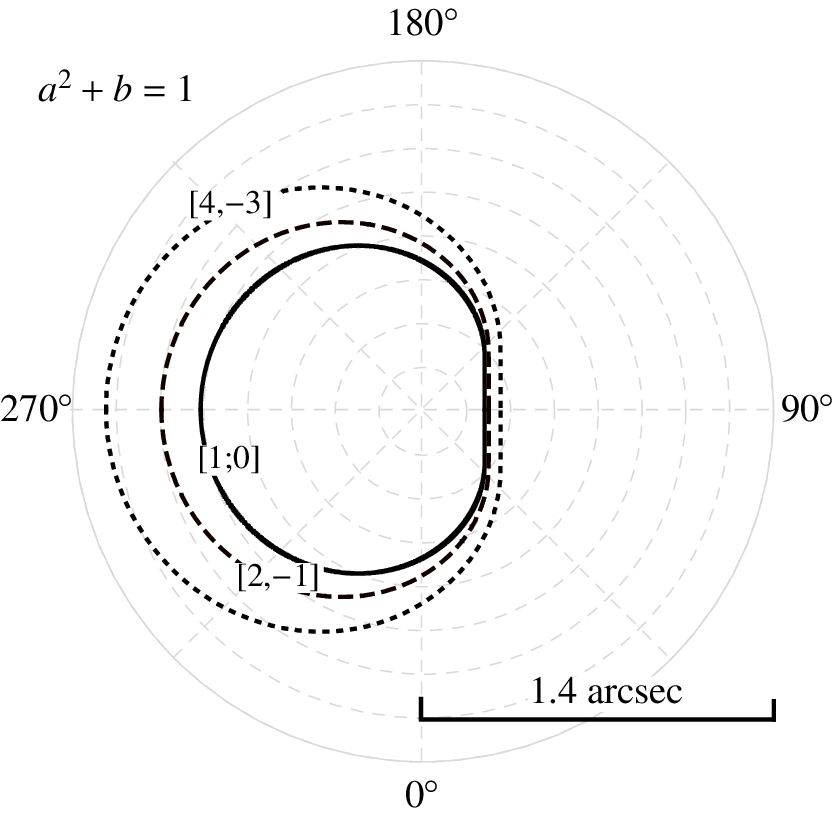}
\end{tabular}
\caption{\label{fig15}Left figure. The $(\bar\alpha_0,\bar\beta_0)$ plots of the silhuettes of braneworld Kerr black hole on a bright background for rotational parameter $a^2=0.6$ and four representative values of tidal charge parameter $b=-3.0$, $b=-0.4$, $b=0.0$ and $b=0.4$. The observer is located at $r_0=10^4 M$ and $\theta_0=90^\circ$.
Right figure. The silhuettes of extreme black holes for three representative values of braneworld parameter $b=0$ (solid), $b=-1$ (dashed) and $b=-3$ (dotted). Static observer is in equatorial plane at radial distance from the centre $r_0=10^4 M$.}
\end{figure}


We consider that the black hole is observed by static distant observers. Therefore, we shall use the static frames introduced above. The silhuette of the black hole is quite naturally related to their trapped (escape) light cones.

The marginal values of impact parameters $\lambda_0$ and $\mathcal{L}_0$(resp $q_0$) are obtained from the  light escape cone. Using the stationarity of the braneworld Kerr spacetime we ``shoot out`` virtual photons from observer (static frame at very large distance $r_0$) and we are looking for the light escape cone of this virtual source (using the results of the previous section). The trapped light cone of this virtual source is constructed from the light escape cone of the virtual source by transformations of directional angle $\alpha_0$ to $\bar{\alpha}_0=\pi - \alpha_0$ and directional angle $\beta_0$ to $\bar{\beta}_0=\beta_0$. In this way we get marginal directions for received photons from bright background behind the black hole.  Then we can use the formulas (\ref{LC6}), (\ref{LC7}) and (\ref{LC8}) to calculate the marginal values of $\lambda_0$ and $q_0$($\mathcal{L}_0$) in order to obtain the silhuette of the braneworld Kerr black hole in the plane $(\tilde{\alpha}-\tilde{\beta})$, i.e., the set of doubles $(\tilde{\alpha}_0,\tilde{\beta}_0)$ from equations (\ref{silalpha}) and (\ref{silbeta}). Here we plotted the silhuette directly from the trapped light cone $(\bar{\alpha}_0,\bar{\beta}_0)$ on the observer's  sky $(\bar{\alpha}_0\sin\bar{\beta}_0,\bar{\alpha}_0\cos\bar{\beta}_0)$. Note that the angle $\bar\alpha_0$ is the radial coordinate and the angle $\bar\beta_0$ is the polar coordinate in the polar graph of the silhuette.

\begin{figure}[ht]
\begin{tabular}{cc}
	\includegraphics[width=6cm]{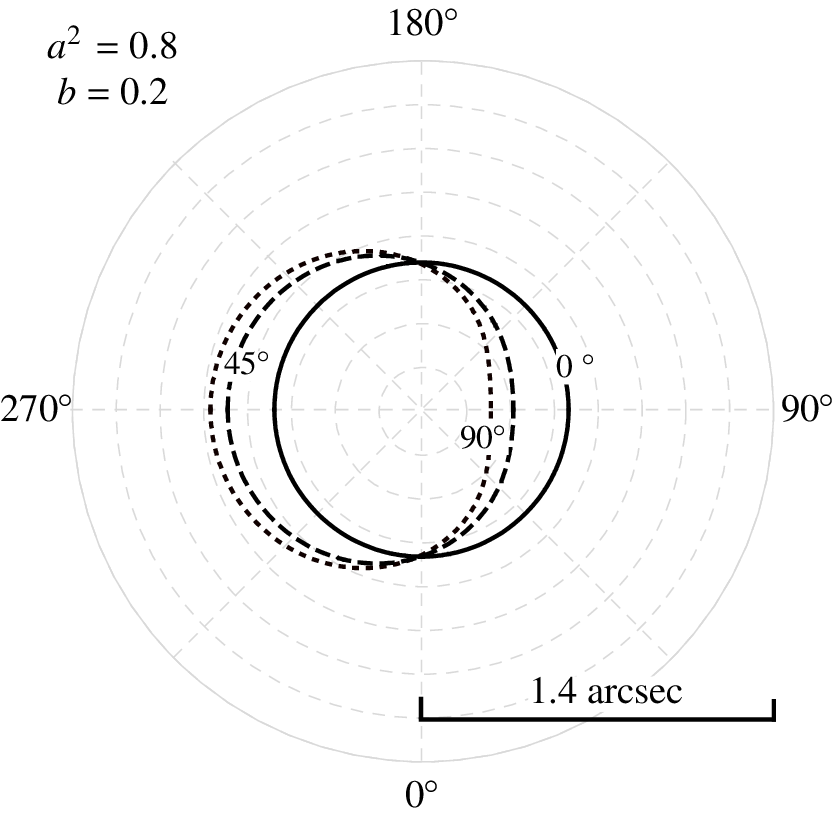}&\includegraphics[width=6cm]{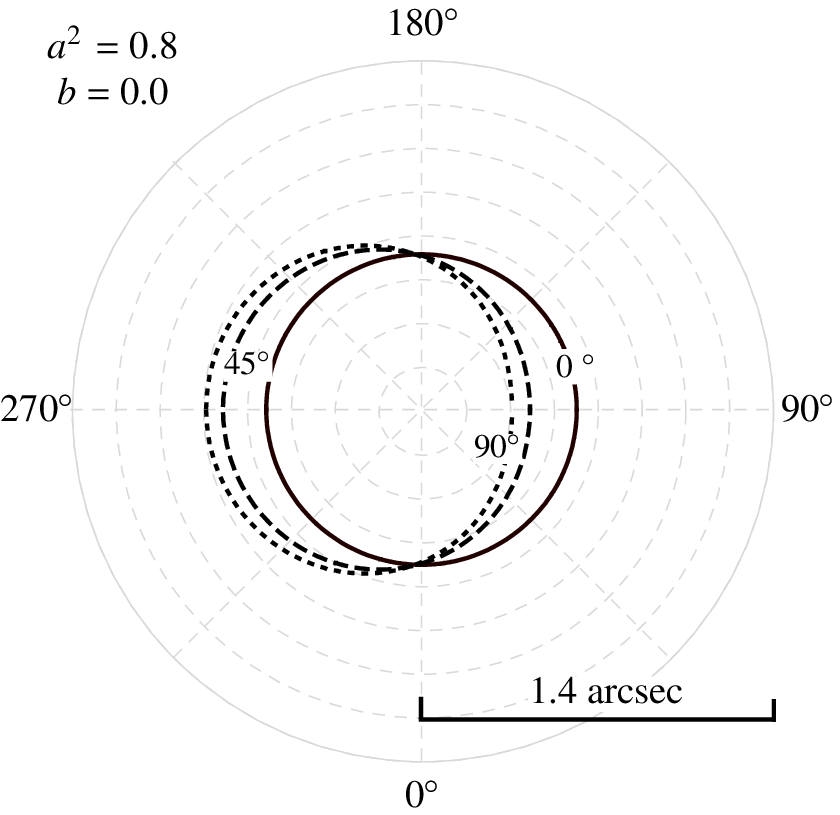}\\
\includegraphics[width=6cm]{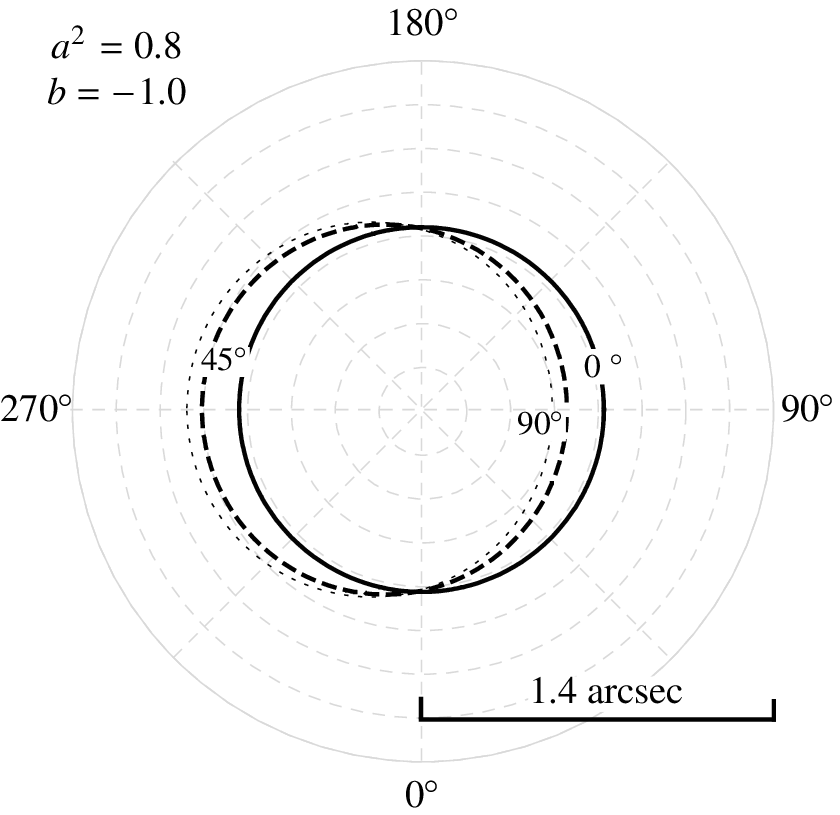}&\includegraphics[width=6cm]{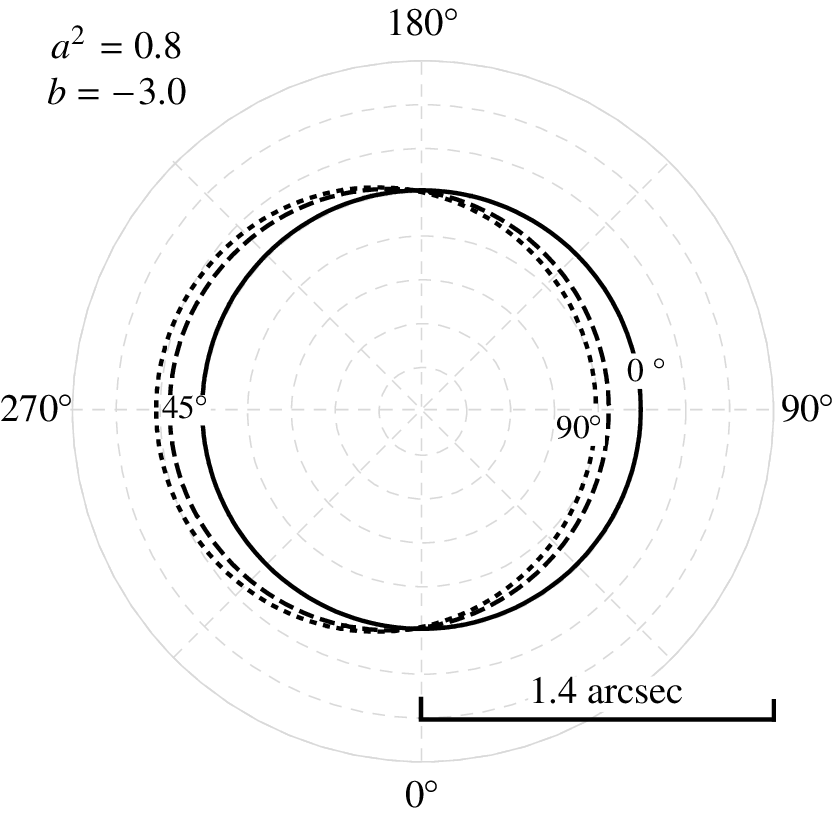}
\end{tabular}
\caption{\label{fig18_a_d}The silhuettes of rotating braneworld black hole on a bright background.  Each image contains three black hole shapes for three representative values of observer's inclination angle $\theta_0=\{0^\circ(solid),45^\circ(dashed),90^\circ(dotted)\}$, observer's radial coordinate $r_0=10^4 M$ and the rotational parameter $a^2=0.8$. Top left image: $b=0.2$. Top right image: $b=0.0$. Bottom left image: $b=-1.0$. Bottom right image: $b=-3.0$.}
\end{figure}

\begin{figure}[ht]
\begin{tabular}{cc}
	\includegraphics[width=6cm]{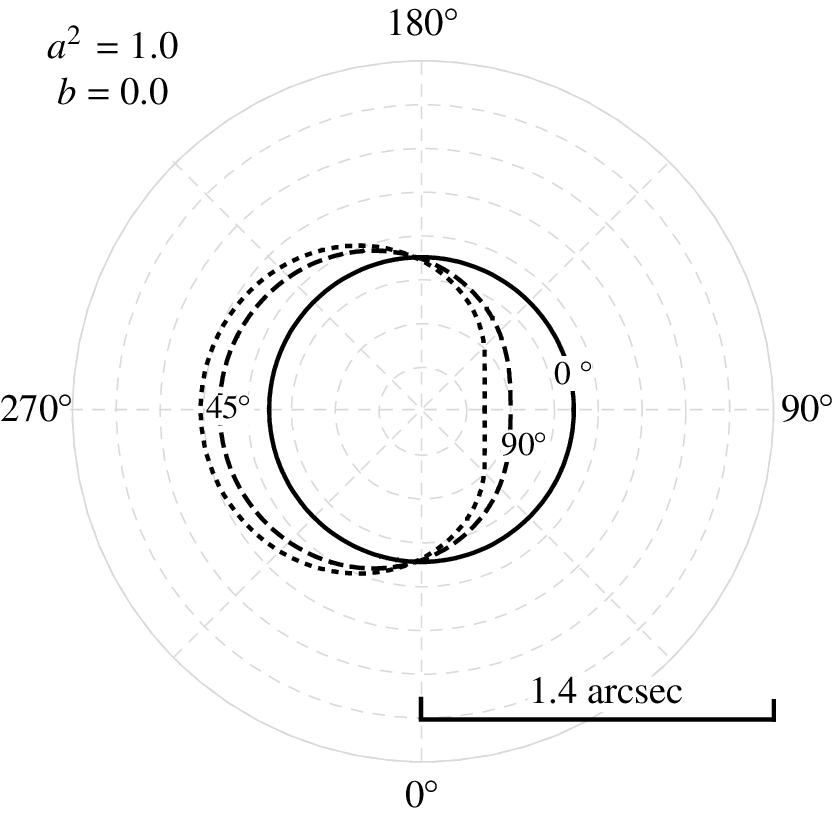}&\includegraphics[width=6cm]{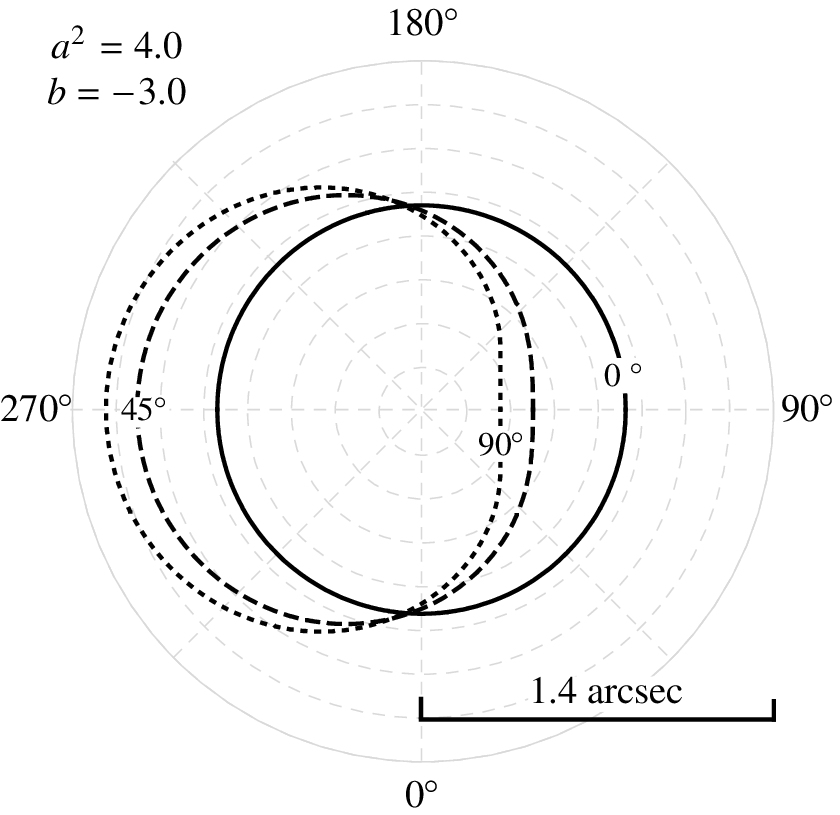}
\end{tabular}
\caption{The silhuettes of extreme rotating braneworld black holes on a bright background.  Each image contains three black hole shapes for three representative values of observer's inclination angle $\theta_0=\{0^\circ(solid),45^\circ(dashed),90^\circ(dotted)\}$, observer's radial coordinate $r_0=10^4 M$. Silhuettes on the left figure are plotted for extreme black holes with $a^2=1$ and $b=0$ and on the right side for $a^2=4$ and $b=-3$.}\label{fig19_a_b}
\end{figure}

\par 
	We shall give the silhuette of the black hole for observers located at fixed radius $r_0=10^4$M that corresponds to the angular size of $\alpha\sim 1.4$arcsec; for higher distances the angular size falls accordingly to the $1/r_0$ dependence. 

First, we give an illustrative picture of the tidal charge influence on the silhuette properties for maximal inclination angle $\theta_0=90^\circ$ when the black hole rotational effects are strongest (Figure \ref{fig15}). We present a sequence of silhuettes for fixed black hole spin and varying $b$ (left) and for extreme black holes with $a^2+b=1$ and both $a$, $b$ varying (right). We clearly see that the positive tidal charge squeezes  magnitude of the silhuette making its shape more asymmetric, while negative tidal charge enlarges silhuette's diameter symmetrizing its shape when $a$ is fixed. For extreme black holes the silhuette asymmetry is kept but its extension grows with $b$ descending to higher negative values.

	Second, there is a crucial effect of the viewing angle $\theta_0$ onto the shape of the black hole silhuette, demonstrated in Figure \ref{fig18_a_d} for representative values of $b$ and fixed spin $a$, and in Figure \ref{fig19_a_b} for extreme black holes with parameters [$a^2=1$;$b=0$] and [$a^2=4$;$b=-3$].

The rotational effect on the shape of the silhuette grows with inclination angle growing and becomes strongest when $\theta_0=\pi/2$; then the suppressing effect of the braneworld parameter is given in the most explicit form as demonstrated in Figure \ref{fig15}. 


The negative values of the braneworld parameter have the tendency to make the silhuette of a Kerr black hole (with $a^2$ fixed and for $r_0$, $\theta_0$ fixed) spherical, suppressing thus the rotational effects. However the symmetrizing effect of the tidal charge could be masked by symmetrizing effect of the viewing angle $\theta_0$. Therefore, it is very important for black hole parameter estimates to have observational limits on the value of $\theta_0$.

\begin{figure}[ht]
	\includegraphics[width=8cm]{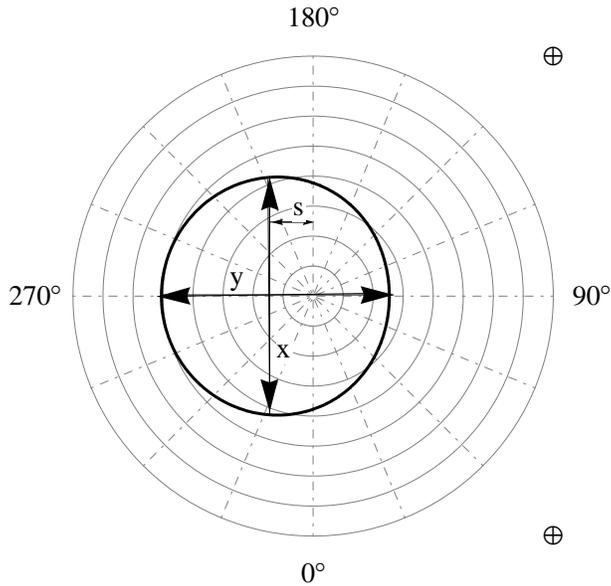} 
\caption{\label{fig14}We define shift $s$ and ellipticity $\epsilon=x/y$ as parameters enabling us to characterize the magnitude of distorsion of Kerr black hole silhuette in braneworld universe.}
\end{figure}

In order to characterize the influence of the tidal charge on the silhuette of a Kerr black hole we define two quantities in principle measurable by distant observers. The \emph{shift} $s$ of the silhuette   
\begin{equation}
 	s=\tilde\alpha(\beta_m)\sin(\beta_m - \pi),\label{eqA}
\end{equation}
and its \emph{ellipticity} $\epsilon$
\begin{equation}
 	\epsilon=\frac{\tilde\alpha(\beta=90^\circ)+\tilde\alpha(\beta=270^\circ)}{2\tilde\alpha(\beta_m)\cos(\beta_m - \pi)},\label{eqB}
\end{equation}
where $\beta_m$ is defined by $\tilde\alpha(\beta_m)\sin(\beta_m - \pi)\ge \tilde\alpha(\beta)\sin(\beta - \pi),\quad \forall \beta\in[\pi/2,3/2\pi]$ i.e., it defines maximal extension of the silhuette in the $x$-direction. The definition of \emph{shift} $s$ and \emph{elipticity} $\epsilon$ is illustrated in Figure \ref{fig14}.

\begin{figure}[ht]
\begin{tabular}{cc}
	\includegraphics[width=6cm]{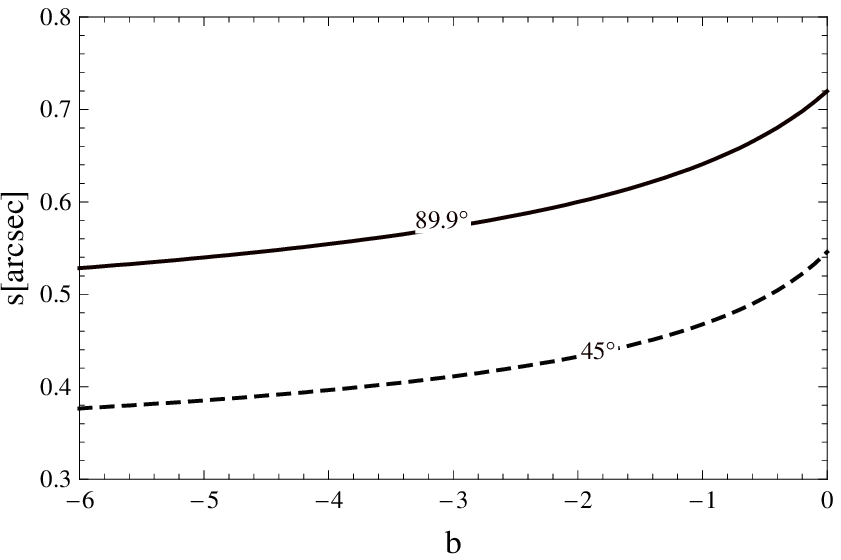} &\includegraphics[width=6cm]{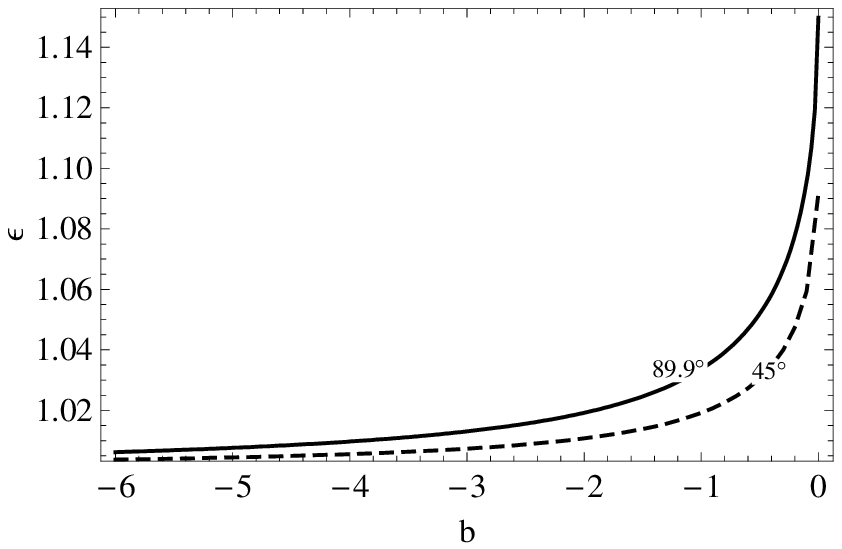} \\
	\includegraphics[width=6cm]{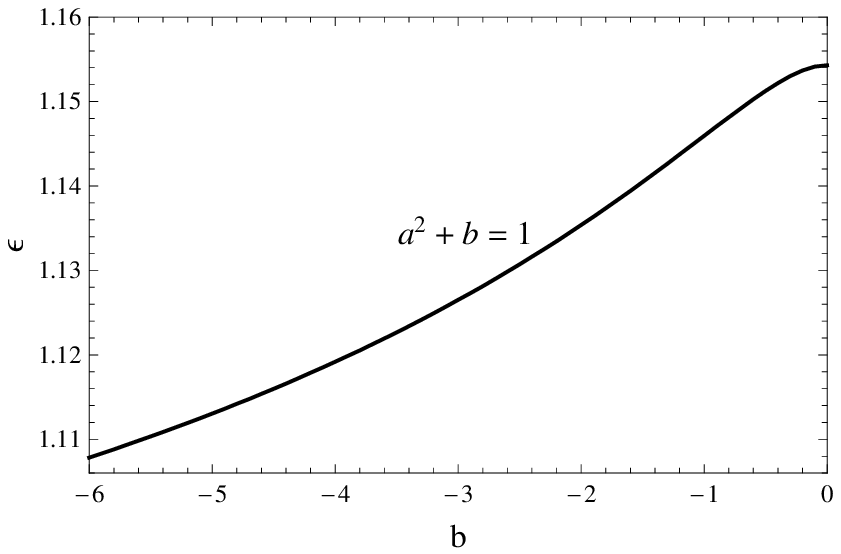}&\includegraphics[width=6cm]{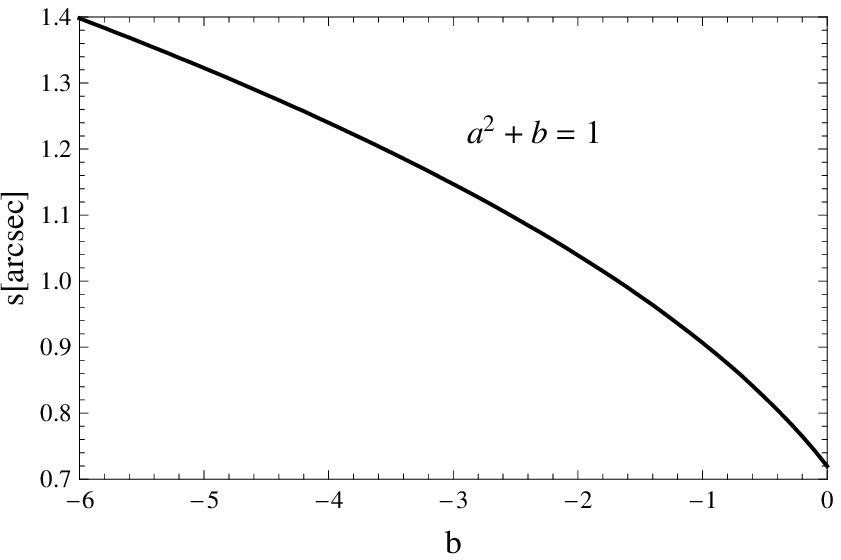}	
\end{tabular}
\caption{\label{fig16_a_b}
 Top row. Left figure: the shift $s=s(b)$ as a function of braneworld parameter $b$. Right figure: the ellipticity $\epsilon=\epsilon(b)$ as a function of $b$. There are two curves on each image, one for observer inclination angle $\theta_0=45^\circ$ and second for $\theta_0=89.9^\circ$. The rotational parameter of black hole is fixed to value $a=0.9995$ and the radial coordinate of observer if $r_0=10^4 M$.
Bottom row. The ellipticity $\epsilon$ (left) and shift $s$ (right) of the extreme black hole silhuette as functions of braneworld parameter $b$. Observer's coordinates are $\theta_0=\pi/2$ and $r_0=10^4 M$. }
\end{figure}

We calculated shift $s$ and ellipticity $\epsilon$ as functions of tidal
parameter $b$ for the Kerr black hole with rotational parameter $a^2=0.9995$
(see Figure \ref{fig16_a_b}).
Clearly, these are quantities that could be measured and used for a black hole parameters estimates, if observational techniques could be developed to the level enabling the silhuette detailed measuring. We shall discuss such a possibility for the case of the supermassive black hole predicted in the Galaxy Centre (Sgr $A^*$).
 

\clearpage
\section{\label{sec:DirAndIndirImages}Direct and indirect images of radiating disc}

Modelling of spectral line profiles of a thin radiating ring rotating in the equatorial plane of a braneworld Kerr black hole or light curve of an isotropically emitting point source orbiting such a black hole will give us information about the influence of the braneworld parameter $b$ on the optical phenomena in the strong field regime \cite{SS:b:RAGTime:2007:Proceedings}. Here we restrict our attention to images of radiating discs. We can then, at least in principle, obtain estimates on the astrophysically acceptable values of the  braneworld parameter $b$. 

\subsection{Images of isoradial geodesics}
	
\begin{figure}[ht]
  \includegraphics[width=10cm]{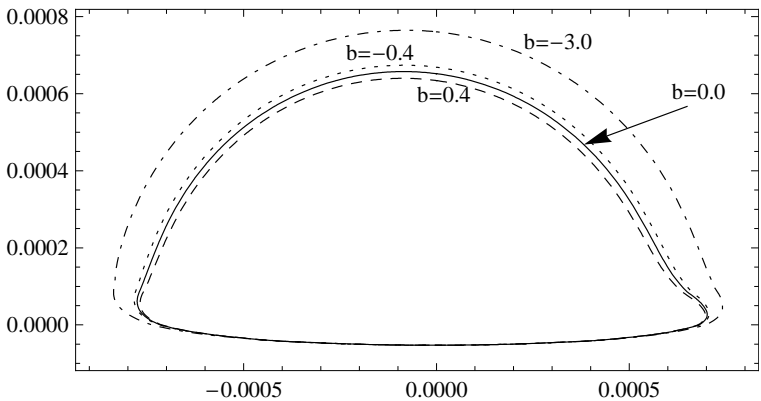}
\caption{\label{fig19}\emph{Direct} image of the rotating ring in the equatorial plane at $r_e=6M$ around braneworld Kerr black hole with rotational parameter $a^2=0.5$ for four representative values of tidal charge parameter $b=-3.0$, $b=-0.4$, $b=0.0$ and $b=0.4$. The observer is located at $r_0=10^4 M$ and $\theta_0=85^\circ$. }
\end{figure}

\begin{figure}[ht]
  \includegraphics[width=8cm]{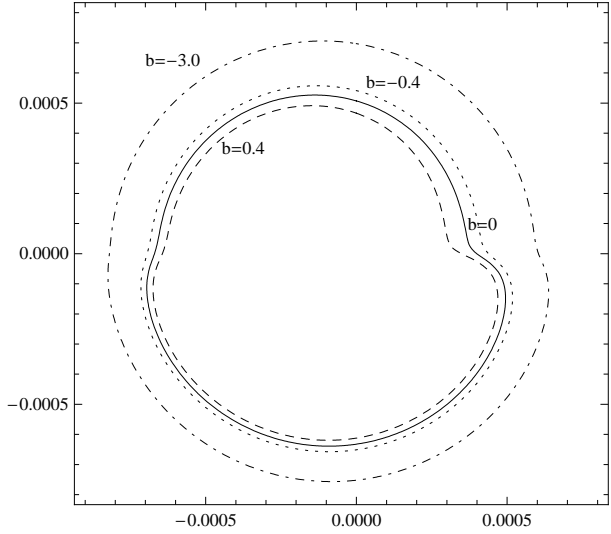}
\caption{\label{fig20}\emph{Indirect} image of the rotating ring in the equatorial plane at $r_e=6M$ around braneworld Kerr black hole with rotational parameter $a^2=0.5$ for four representative values of tidal charge parameter $b=-3.0$, $b=-0.4$, $b=0.0$ and $b=0.4$. The observer is located at $r_0=10^4 M$ and $\theta_0=85^\circ$. }\label{indirect_a_05_th85_on_b}
\end{figure}

Calculating images of an accretion disc (ring) in the equatorial plane of a braneworld Kerr black hole is the first step to calculate the optical phenomena. Generally one could obtain a direct and an indirect image (see Figures \ref{fig19} and \ref{fig20}), but in special cases the situation can be much more complicated due to complex character of the latitudinal and azimuthal photon motion. Here we focus our attention to the direct and indirect images of isoradial geodesics.

In order to find all relevant positions of points forming the rotating ring on observer's sky, we have to find photon trajectories between the ring particles and the observer, i.e., we seek for such  doubles of local observational angles $[\alpha_0,\beta_0]$ that satisfy the condition

\begin{equation}
 	I_U(\alpha_0,\beta_0;n_u,u_{sgn}) - I_M(\alpha_0,\beta_0;n,p,s)=0.\label{bvp}
\end{equation}
Here we introduced the modified radial coordinate $u=1/r$ and cosine of latitudinal coordinate $\mu=\cos\theta$ \cite{Rau-Bla:1994:}. In the condition (\ref{bvp}) $n_u$ is the number of turning points in $u$ coordinate, $n$ is the number of turning points passed in $\mu$ coordinate, $p=mod(n,2)$, $s=(1-\mu_{sgn})/2$. In terms of $u$ and $\mu$ we define the functions $I_U$ and $I_M$ by

\begin{equation}
	I_U(\alpha_0,\beta_0;n_u,u_{sgn})\equiv\left\{\begin{array}{lcr}
					-u_{sgn}\left(\int^{u_0}_{u_t} +\int^{u_e}_{u_t}\right) & \textrm{for} & n_u=1\\
					u_{sgn}\int^{u_e}_{u_0} & \textrm{for} & n_u=0
					\end{array}\right.
\end{equation}
and
\begin{eqnarray}
	I_M(\alpha_0,\beta_0;n,p,s)&\equiv&\mu_{sgn}\left[\int^{\mu_+}_{\mu_0} + (-1)^{n+1}\int^{\mu_+}_{\mu_e}+\right.\\ \nonumber
&+&\left.(-1)^s[(1-p)n+p[(1-s)(n-1)+s(n+1)]]\int^{\mu_+}_{\mu_-} \right]
\end{eqnarray}
 with
\begin{eqnarray}
	\int^{u_2}_{u_1}&\equiv&\int^{u_2}_{u_1}\frac{\diff u}{\sqrt{U(u)}},\label{u_int}\\
	U(u)&=&1+(a^2-\lambda^2-q)u^2+2[(\lambda^2-a^2)^2+q]u^3 - \nonumber\\
	&-&[q(a^2+b)+b(a-\lambda)^2]u^4
\end{eqnarray}
and

\begin{eqnarray}
	\int^{\mu_2}_{\mu_1}&\equiv&\int^{\mu_2}_{\mu_1}\frac{\diff \mu}{\sqrt{M(\mu)}},\label{mu_int}\\
	M(\mu)&=&q+(a^2-\lambda^2-q)\mu^2-a^2\mu^4.
\end{eqnarray}

\subsection{Integration of photon trajectories}

We express the integrals (\ref{u_int}) and (\ref{mu_int}) in the form of the standard elliptic integrals of the first kind. Rauch and Blandford presented the tables of reductions of $u$-integrals and $\mu$-integrals for the case of photons in Kerr geometry \cite{Rau-Bla:1994:}. Here we extended those reductions for the case of nonzero braneworld parameter $b$. Because the integration of the $\mu$-integral does not depend on braneworld parameter $b$, the transformations are the same as in the case of Kerr metric \cite{Rau-Bla:1994:}, but we include them for completeness.

There are two cases we distinguish in latitudinal integral (see table \ref{tableEIM}). In the first case there is one positive, $M_+>0$, and one negative, $M_-<0$ root of $M(m^2)$ it implies that there are two turning points located symmetrically about the equatorial plane given by $\pm\sqrt{M_+}$ (so called orbital motion \cite{Bic-Stu:1976:,Fel-Cal:1972:}. In the second case there are two positive roots, $0<M_-<M_+$ of  $M(m^2)$, which implies that the latitudinal motion is constrained to the region above or below of the equatorial plane (so called vortical motion). The relevant reductions of the integral 
$\int^m_{m_1}\diff m'/\sqrt{M(m')}=I_M$ are stored in the table \ref{tableEIM}.

 For distant observers we distinguish five relevant cases of the radial integral. These cases depend on the character of roots of the quartic equation $U(u)=0$, i.e., on the number of turning points ($n_u=0$ or $n_u=1$) in the radial motion  and the value of parameter $\tilde{q}=q(a^2+b)+b(a-\lambda)^2$. We have arranged those transformations into table \ref{tableEI}.

Denoting roots of the quartic equation $U(u)=0$ by $\beta_1$, $\beta_2$, $\beta_3$ and $\beta_4$, the meaning of each of the five cases is the following:
\begin{itemize}
\item
The \textbf{case I}: four distinct real roots of $U(u)=0$ forming the sequence $\beta_1>\beta_2>\beta_2>0$ and $\beta_4<0$. The value of modified constant of motion $\tilde{q}>0$.
\item
The \textbf{case II}: four real roots as in the case I but their values form the following order: $\beta_1>\beta_2>0$ and $\beta_4<\beta_3<0$. The value of modified constant of motion  $\tilde{q}<0$. 
\item
The \textbf{case III}: two real and two complex roots of $U(u)=0$: $\beta_1$ being a complex root, $\beta_2=\bar{\beta_1}$ and $\beta_4<\beta_3<0$. The value of modified constant of motion $\tilde{q}<0$.
\item
The \textbf{case IV}: only complex roots: $\beta_2=\bar{\beta_1}$ and $\beta_4=\bar{\beta_3}$. The value of modified constant of motion $\tilde{q}<0$. 
\item
The \textbf{case V}: two real and two complex roots of $U(u)=0$: $\beta_1>0$, $\beta_4<0$, $\beta_2$ being a complex root and $\beta_3=\bar{\beta_2}$. 
\end{itemize}

\begin{table}[!ht]
\tbl{The reductions of $\int^m_{m_1}\diff m'/\sqrt{M(m')}=I_M$} 
{\begin{tabular}{@{}lllll@{}}\toprule
  	Case & $\tan\Psi$ & $m$ & $c_1$ & $m_1$\\ \colrule
	\\
	$M_-<0$ & $\sqrt{\frac{M_+}{m^2}-1}$ & $\frac{M_+}{M_+-M_-}$ & $\frac{1}{\sqrt{a^2(M_+-M_-)}}$ & $\sqrt{M_+}$\\
	\\
	$M_->0$ & $\sqrt{\frac{M_+-m^2}{m^2-M_-}}$ & $\frac{M_+-M_-}{M_+}$ &
        $\frac{1}{a^2}$ & $\sqrt{M_+}$\\ \botrule
 \end{tabular}\label{tableEIM}}
\end{table} 

\begin{table}[!ht]
\tbl{The reductions of $\int^u_{u_1}\diff u'/\sqrt{U(u')}=I_U$}
{\begin{tabular}{@{}lllll@{}}\toprule

  	Case & $\tan\Psi$ & $m$ & $c_1$ & $u_1$\\ \colrule
	
	I & $\sqrt{\frac{(\beta_1-\beta_3)(u-\beta_4)}{(\beta_1-\beta_4)(\beta_3-u)}}$ & $\frac{(\beta_1-\beta_2)(\beta_3-\beta_4)}{(\beta_1-\beta_3)(\beta_2-\beta_4)}$ & $\frac{2}{\sqrt{\tilde{q}(b1-b3)(b2-b4)}}$ & $\beta_4$\\
	\\
	II & $\sqrt{\frac{(\beta_1-\beta_2)(u-\beta_3)}{(\beta_1-\beta_3)(\beta_2-u)}}$ & $\frac{(\beta_2-\beta_3)(\beta_1-\beta_4)}{(\beta_1-\beta_2)(\beta_4-\beta_3)}$ & $\frac{2}{\sqrt{-\tilde{q}(b1-b2)(b3-b4)]}}$ & $\beta_3$\\
	\\
	III & $\frac{2c_2(u)}{|1-c^2_2(u)|}$ & $\frac{4c_4 c_5 - (\beta_3 - \beta_4)^2 - c_4 c_5}{4c_4 c_5}$ & $\frac{1}{\sqrt{-\tilde{q}c_4 c_5}}$ & $\beta_3$\\
	\\
	IV & $\frac{u-c_3}{\Im(\beta_1)(1+c_2^2)+c_2(u-c_3)}$ & $1-\left(\frac{c_4-c_5}{c_4+c_5}\right)^2$ & $\frac{2}{(c_4+c_5)\sqrt{-\tilde{q}}}$ & $c_3$\\
	\\
	V & $\frac{2c_2(u)}{|1-c^2_2(u)|}$ & $1-\frac{(c_4+c_5)^2-(\beta_1 -
          \beta_4)^2}{4c_4 c_5}$ & $\frac{1}{\sqrt{\tilde{q}c_4 c_5}}$ &
        $\beta_4$\\ \botrule
 \end{tabular}\label{tableEI}}
\end{table}

\begin{table}[!th]
\tbl{Definitions for Table \ref{tableEI}.}
{\begin{tabular}{@{}lll@{}}\toprule
  
  	Case & $^1 c_2$ & $^1 c_3$\\ \colrule
	
	III &  $\left[\frac{c5(u-\beta_3)}{c_4(u-\beta_4)}\right]^{1/2} $ & -\\
	\\
	IV & $\left\{\frac{4[\Im(\beta_1)]^2-(c_4-c_5)^2}{(c_4+c_5)^2-4[\Im(\beta_1)]^2}\right\}^{1/2}$ & $\mbox{\fontsize{8}{10}\selectfont $\Re(\beta_1)+c_2\Im(\beta_1)$}$\\
	\\
	V & $\left[\frac{c4(u-\beta_4)}{c_5(\beta_1-u)}\right]^{1/2} $& -\\ \botrule
\end{tabular}\label{tableEI2}}
\end{table}

\begin{table}[!th]
\tbl{Definitions for Table \ref{tableEI} and Table \ref{tableEI2}.}
 {\begin{tabular}{@{}lll@{}}\toprule
  
  	Case & $^1 c_4$ & $^1 c_5$\\ \colrule
	
	III &   $\mbox{\fontsize{8}{10}\selectfont$\left\{\left[\Re(\beta_1)-\beta_3\right]^2+[\Im(\beta_1)]^2\right\}^{1/2}$}$ & $\mbox{\fontsize{8}{10}\selectfont $\left\{\left[\Re(\beta_1)-\beta_4\right]^2+[\Im(\beta_1)]^2\right\}^{1/2}$}$\\
	\\
	IV & $\mbox{\fontsize{8}{10}\selectfont $\left\{\left[\Re(\beta_1)-\Re(\beta_3)\right]^2+[\Im(\beta_1)+\Im(\beta_3)]^2\right\}^{1/2}$}$ & $\mbox{\fontsize{8}{10}\selectfont $\left\{\left[\Re(\beta_1)-\Re(\beta_3)\right]^2+[\Im(\beta_1)-\Im(\beta_3)]^2\right\}^{1/2}$}$\\
	\\
	V & $\mbox{\fontsize{8}{10}\selectfont $\left\{\left[\Re(\beta_2)-\beta_1\right]^2+[\Im(\beta_2)]^2\right\}^{1/2}$}$ & $\mbox{\fontsize{8}{10}\selectfont $\left\{\left[\Re(\beta_2)-\beta_4\right]^2+[\Im(\beta_2)]^2\right\}^{1/2}$}$\\ \botrule
\end{tabular}}
\begin{tabular}{c}
	$^1$\textit{The symbols $\Re(x)$ and $\Im(x)$  refer to real and imaginary part of $x$ here.}
\end{tabular}
\end{table}

\par

Using presented transformations we can write the integrals (\ref{u_int}) and (\ref{mu_int})  in the form

\begin{equation}
 \int^{u}_{u_1}\frac{1}{\sqrt{U(\tilde{u})}}\diff \tilde{u} = c_1\mathcal{F}(\Psi;m)\label{ellint}
\end{equation}
and 
\begin{equation}
 \int^{\mu}_{\mu_1}\frac{1}{\sqrt{M(\tilde{\mu})}}\diff \tilde{\mu} = c_1\mathcal{F}(\Psi;m)\label{ellintM}
\end{equation}
where $\mathcal{F}$ is the elliptic integral of the first kind and $u_1$(resp $\mu_1$) depends on the case of root distribution of quartic equation $U(u)=0$ (resp. $M(\mu)=0$) as given in Table \ref{tableEI} (resp \ref{tableEIM}). If, in the cases III and V, the value of $1-c_2^2(u)<0$, we have to take instead of  (\ref{ellint}) the form

\begin{equation}
 \int^{u}_{u_1}\frac{1}{\sqrt{U(\tilde{u})}}\diff \tilde{u} =  c_1(2\mathcal{K}(m)-\mathcal{F}(\Psi;m)),\label{ellint1}
\end{equation}
where $\mathcal{K}$ is the complete elliptic integral of the first kind. In the case that sign$(\mu1\cdot\mu)<0$ we have to take instead of (\ref{ellintM}) the form
\begin{equation}
 \int^{\mu}_{\mu_1}\frac{1}{\sqrt{M(\tilde{\mu})}}\diff \tilde{\mu} =  c_1(2\mathcal{K}(m)-\mathcal{F}(\Psi;m)),\label{ellintM1}
\end{equation}
where $\Psi$, $m$ and $c_1$ are taken from table \ref{tableEIM}.
 We consider two basic possibilities of trajectories, namely those corresponding to direct and indirect images (Figures \ref{fig19} and \ref{fig20}).

\subsection{Disc images}

	It is very important to demostrate the influence of the braneworld parameter on the shape of images of rings in the equatorial plane representing parts of Keplerian accretion discs. Of course, as well known from the Kerr (and even Schwarzchild) black holes, the images strongly depend on the latitude of the observer. We calculate the direct and indirect images of flat discs and combined, full image of the disc for two representative values of viewing angle $\theta_0$ and appropriatelly chosen extension of radiating disc area.


	We include the effect of frequency shift into the calculated images of part of the Keplerian discs assumed to be radiating at a given fixed frequency. The frequency shift $g$ is determined by the ratio of observed ($E_0$) to emitted ($E_e$) photon energy
\begin{equation}
 g=\frac{E_0}{E_e}=\frac{k_{0\mu} u_0^\mu}{k_{e\mu} u_e^\mu},
\end{equation}
where $u_0^\mu$($u_e^\mu$) are components of the observer (emitter) 4-velocity and $k_{0\mu}(k_{e\mu})$ are components of the photon 4-momentum taken at the moment of emission (observation). For distant observers $u^\mu_0=(1,0,0,0)$. The emitter follows an equatorial circular geodesics at $r=r_e$, $\theta_e=\pi/2$. Therefore, $u_e^\mu=(u^t_e,0,0,u_e^\varphi)$, with components given by

\begin{eqnarray}
 u_e^t&=&\left[1-\frac{2}{r_e}(1-a\Omega)^2-(r_e^2+a^2)\Omega^2+\frac{b}{r_e^2}(1-2a\Omega)\right]^{-1/2},\\ u_e^\varphi&=&\Omega u_e^t,
\end{eqnarray}
where $\Omega=\diff\varphi/\diff t$ is the Keplerian angular velocity of the emitter related to distant observers, given by equation (\ref{ang_vel_gf}).

The frequency shift including all relativistic effects is then given by

\begin{equation}
 g=\frac{\left[1-\frac{2}{r_e}(1-a\Omega)^2-(r_e^2+a^2)\Omega^2+\frac{b}{r_e^2}(1-2a\Omega)\right]^{1/2}}{1-\lambda\Omega}
\end{equation}
where $\lambda\equiv-k_\varphi/k_t$ is the impact parameter of the photon
being a motion constant for an individual photon radiated at a specific
position of the radiating disc; notice that $g$ is independent of the second
photon motion constant  (impact parameter) $q$. Of course, depending on the
position of the emitter along the circular orbit, the impact parameters
$\lambda$, $q$ of photons reaching a fixed distant observer will vary
periodically (see eg., \cite{Bao-Stu:1992:}). For each position of the emitter
the impact parameters are determined by the procedure of integration of photon
trajectories. 

\begin{figure}[ht]
  \begin{tabular}{ccc}
    \includegraphics[width=3.6cm]{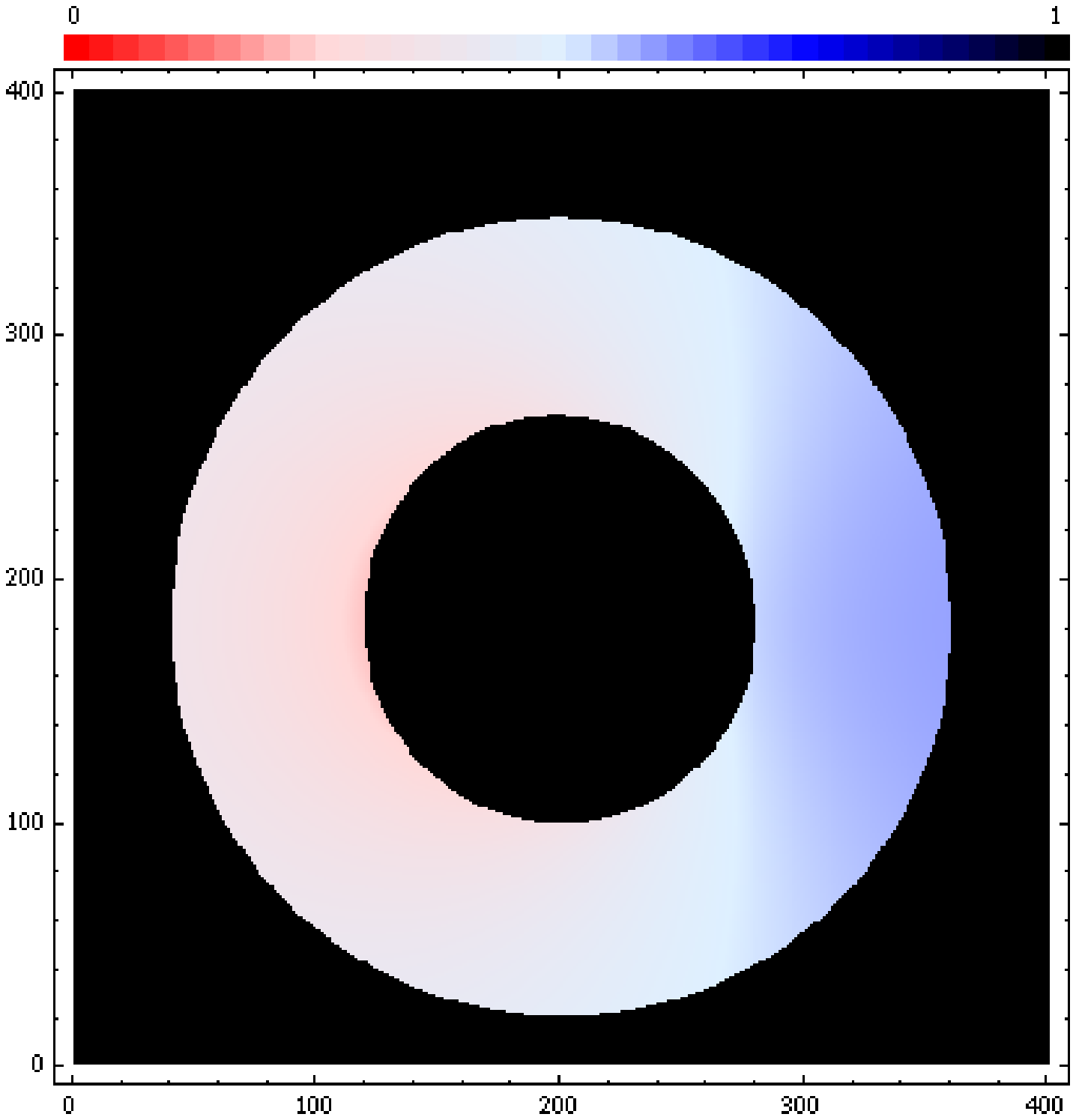}
& \includegraphics[width=3.6cm]{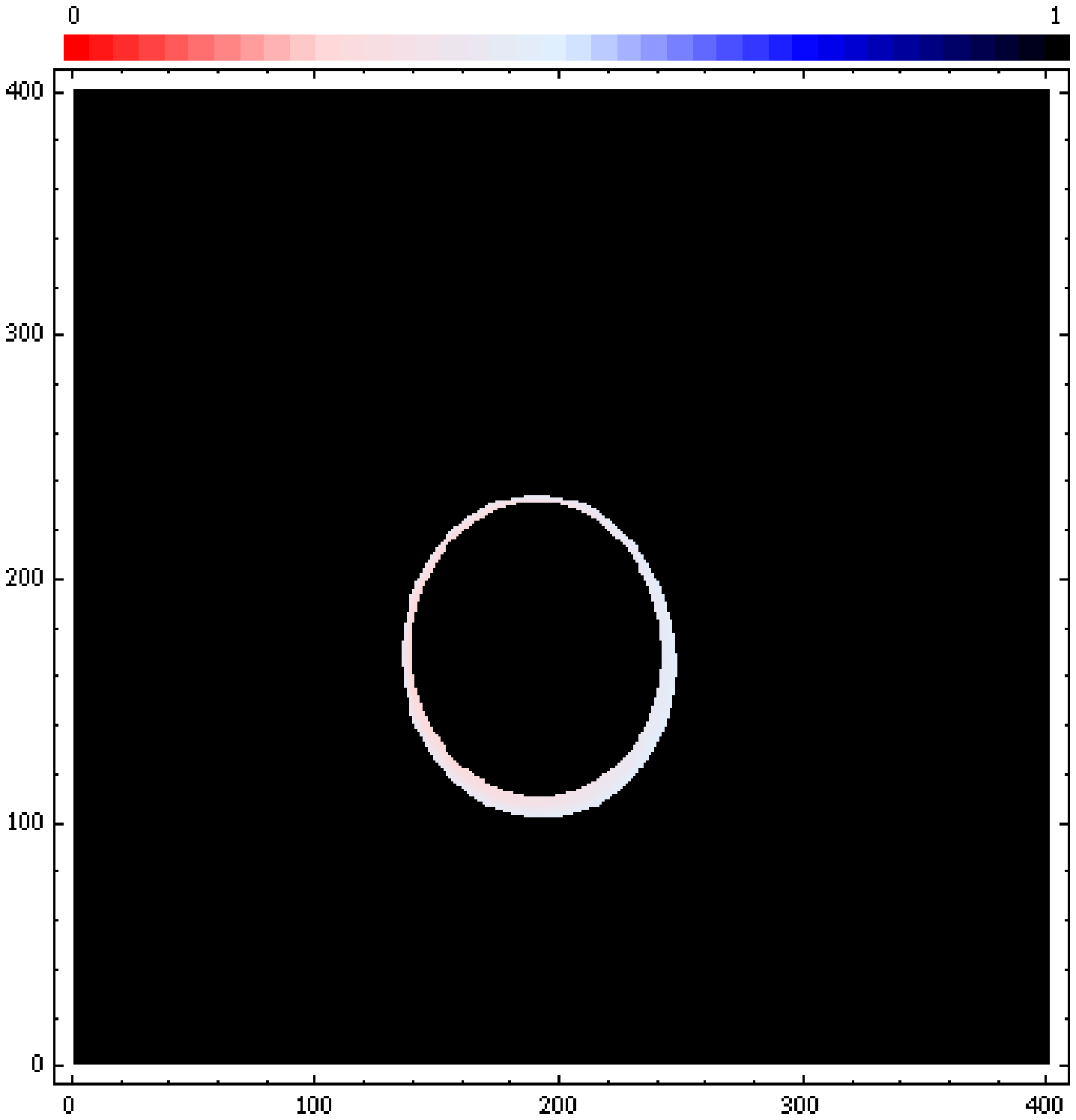}
&\includegraphics[width=3.6cm]{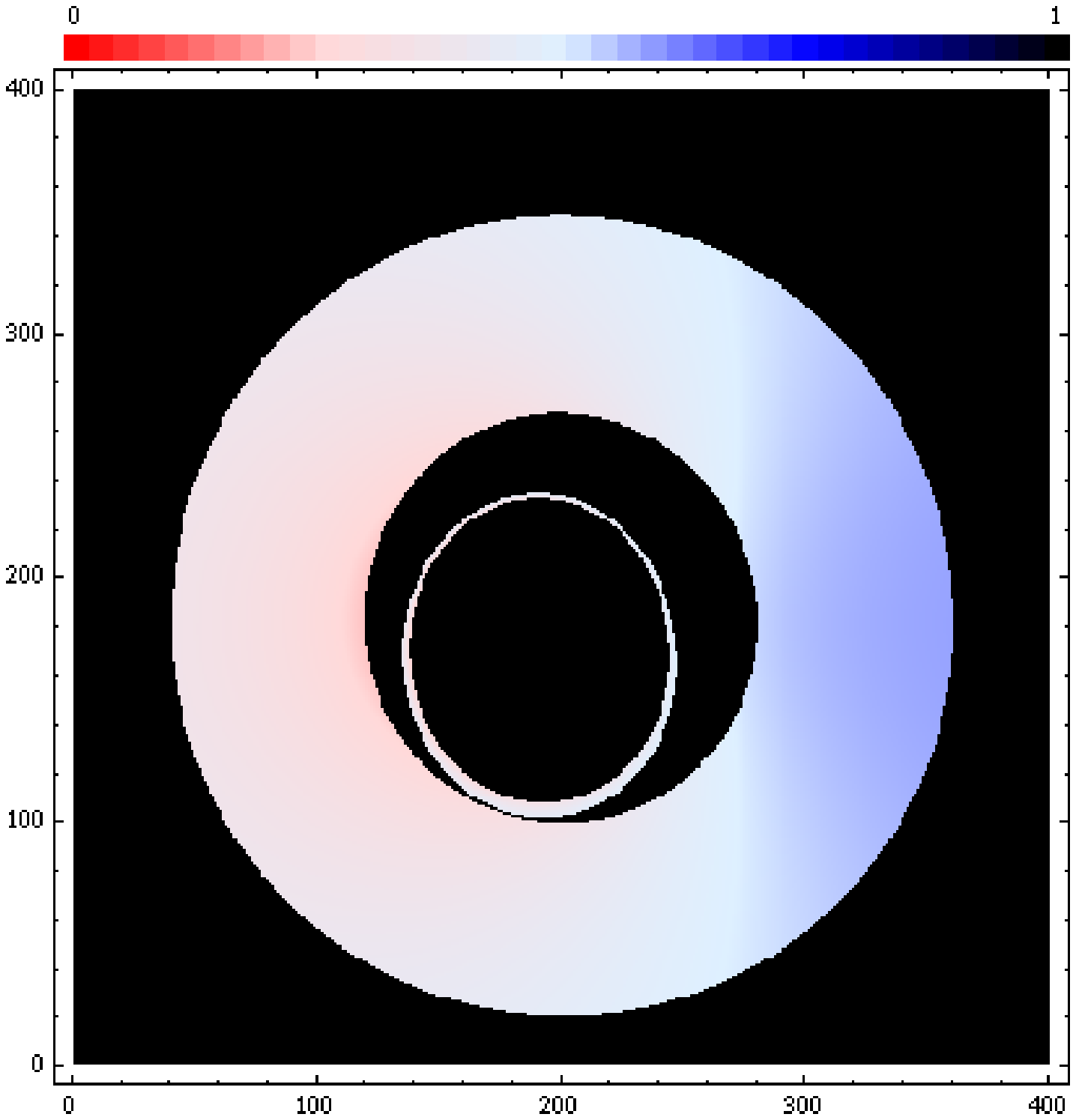}\\
 \includegraphics[width=3.6cm]{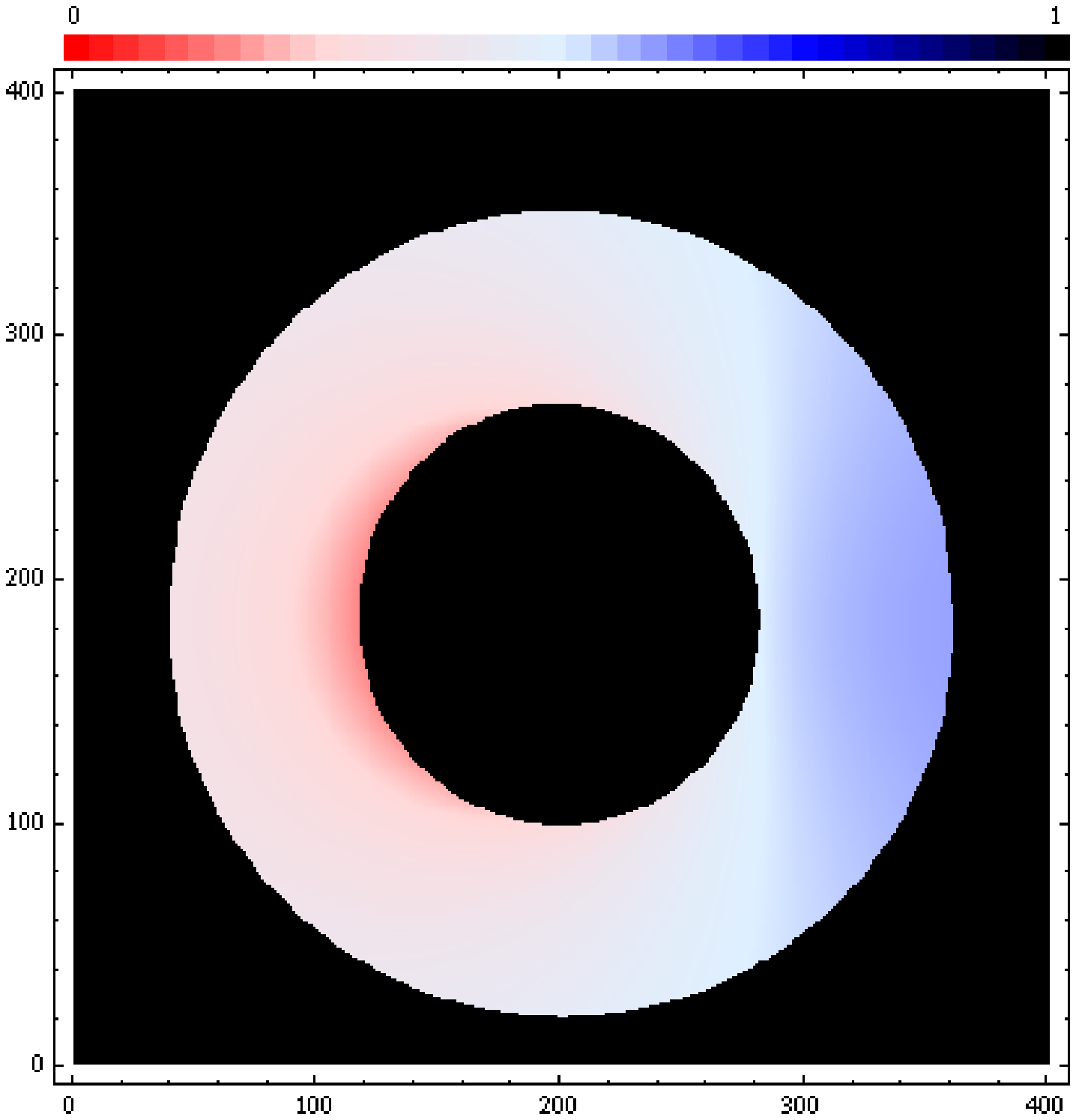}
& \includegraphics[width=3.6cm]{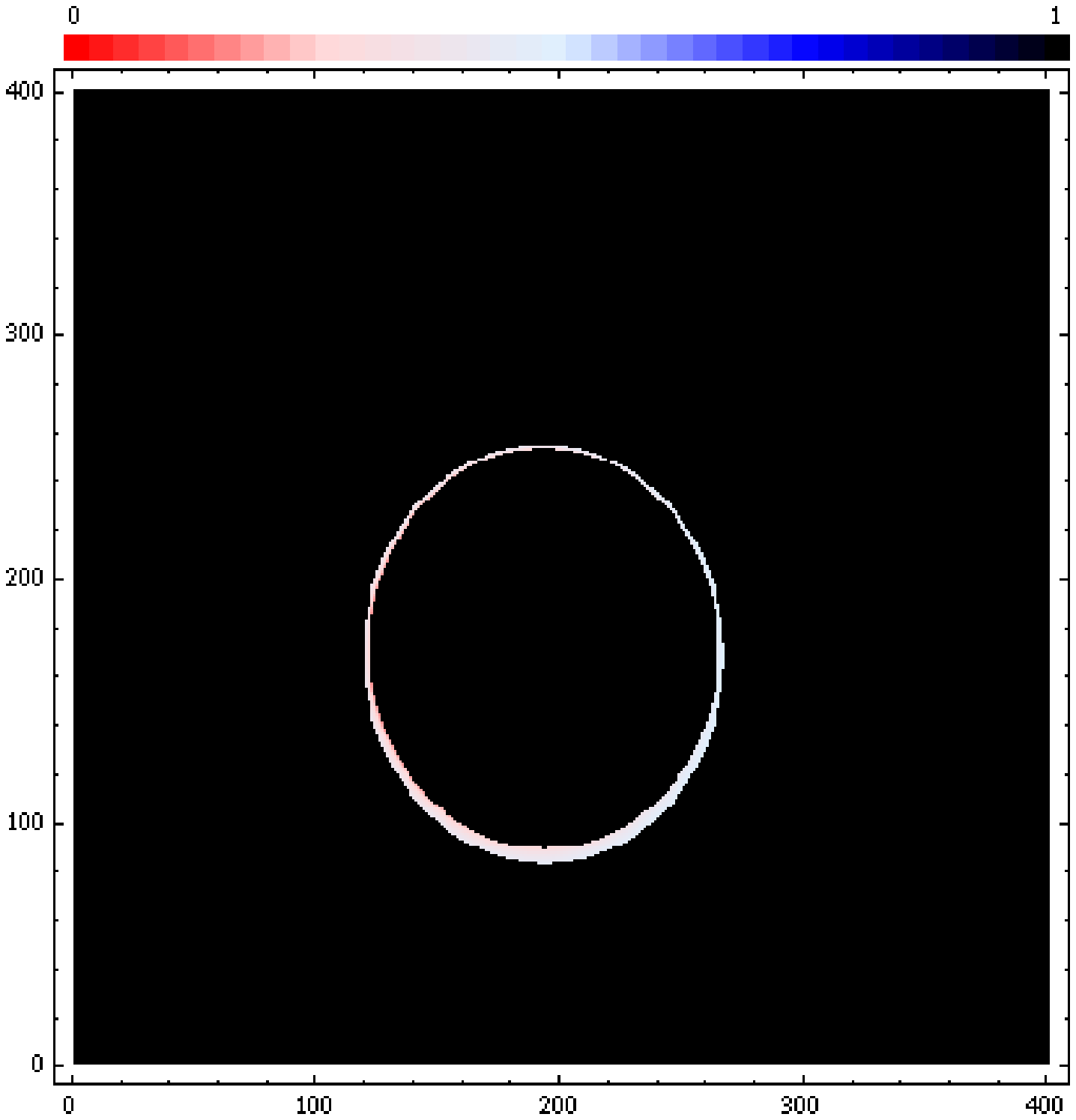}
&\includegraphics[width=3.6cm]{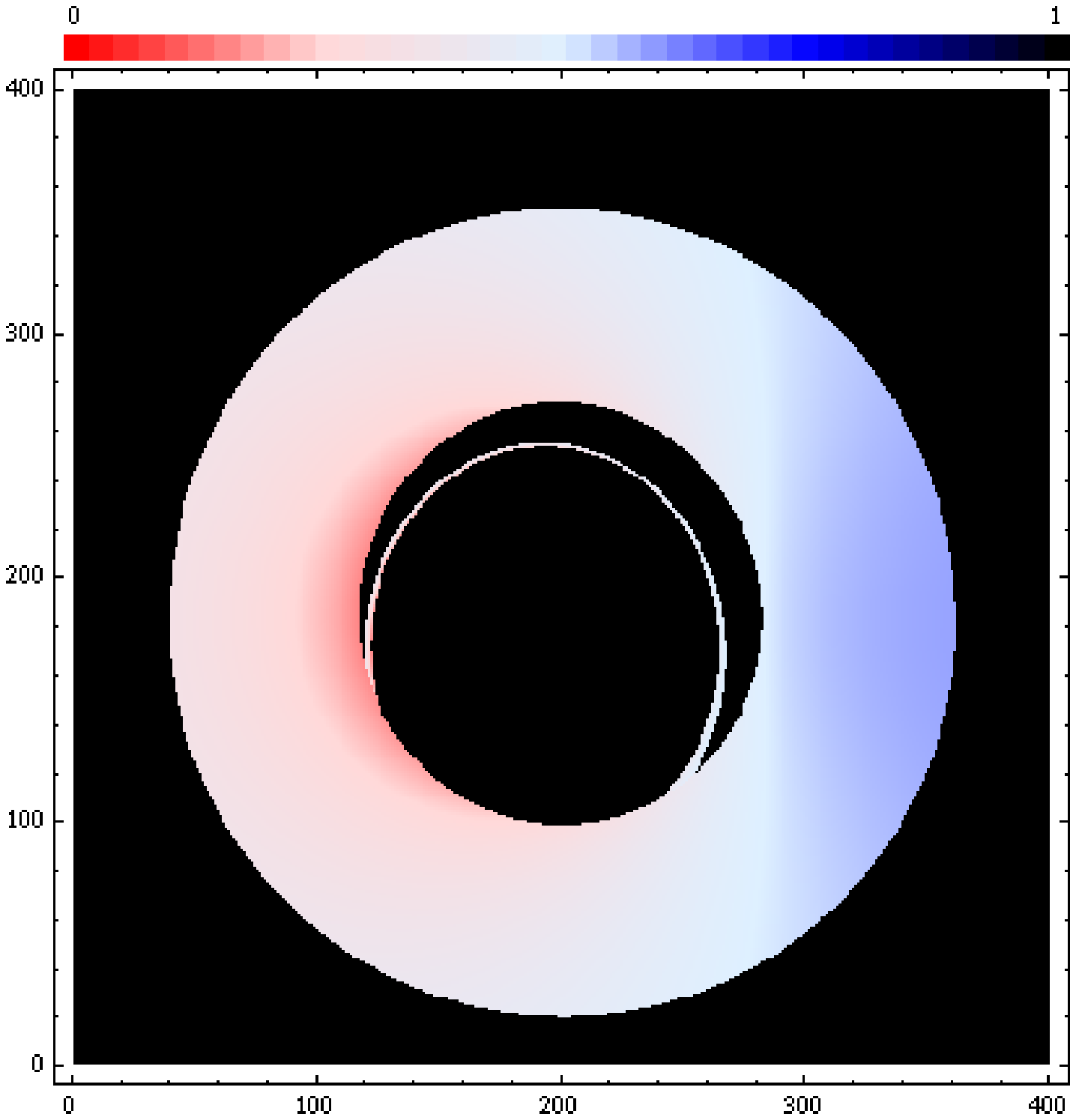}\\
 \includegraphics[width=3.6cm]{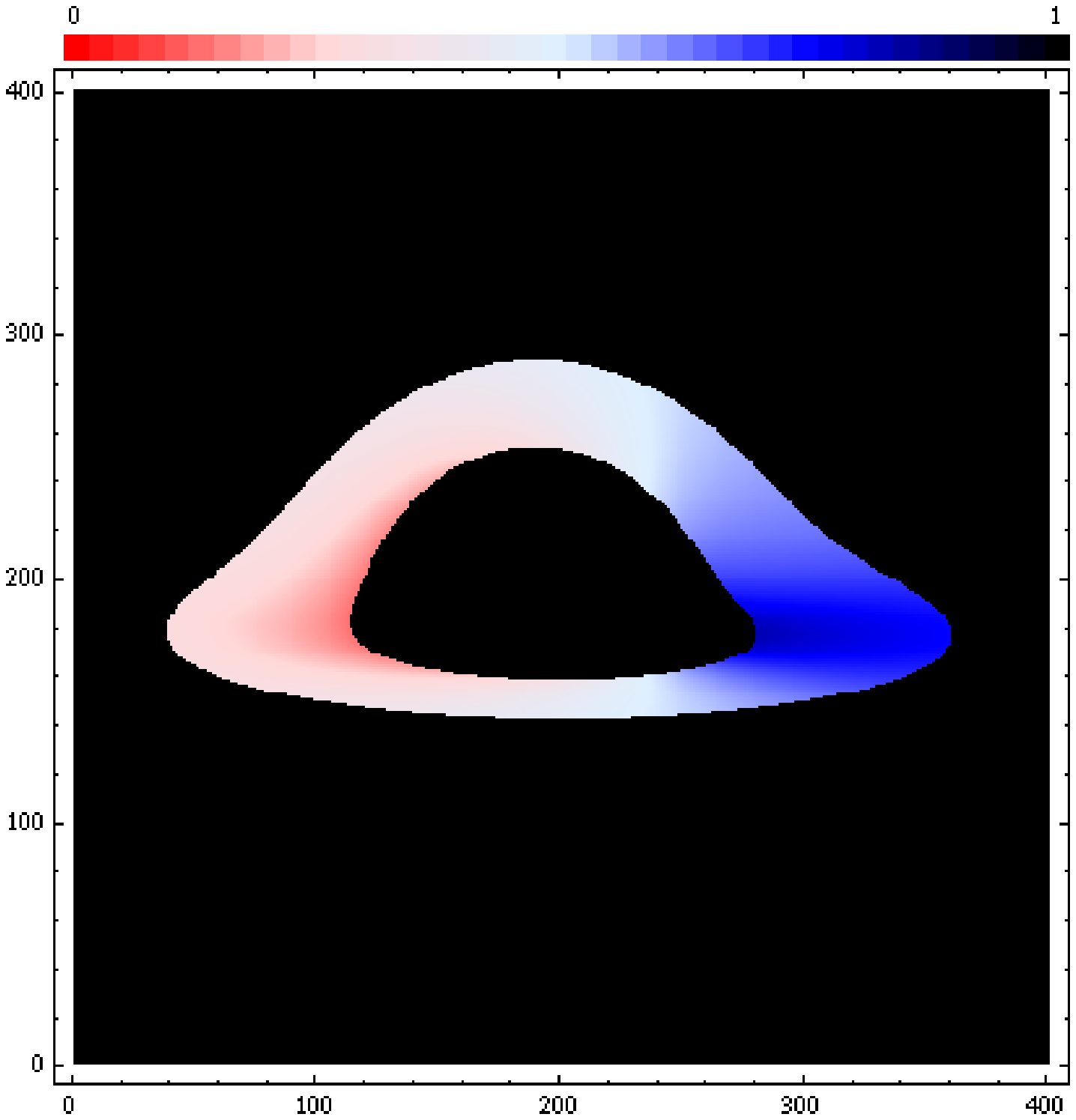}
& \includegraphics[width=3.6cm]{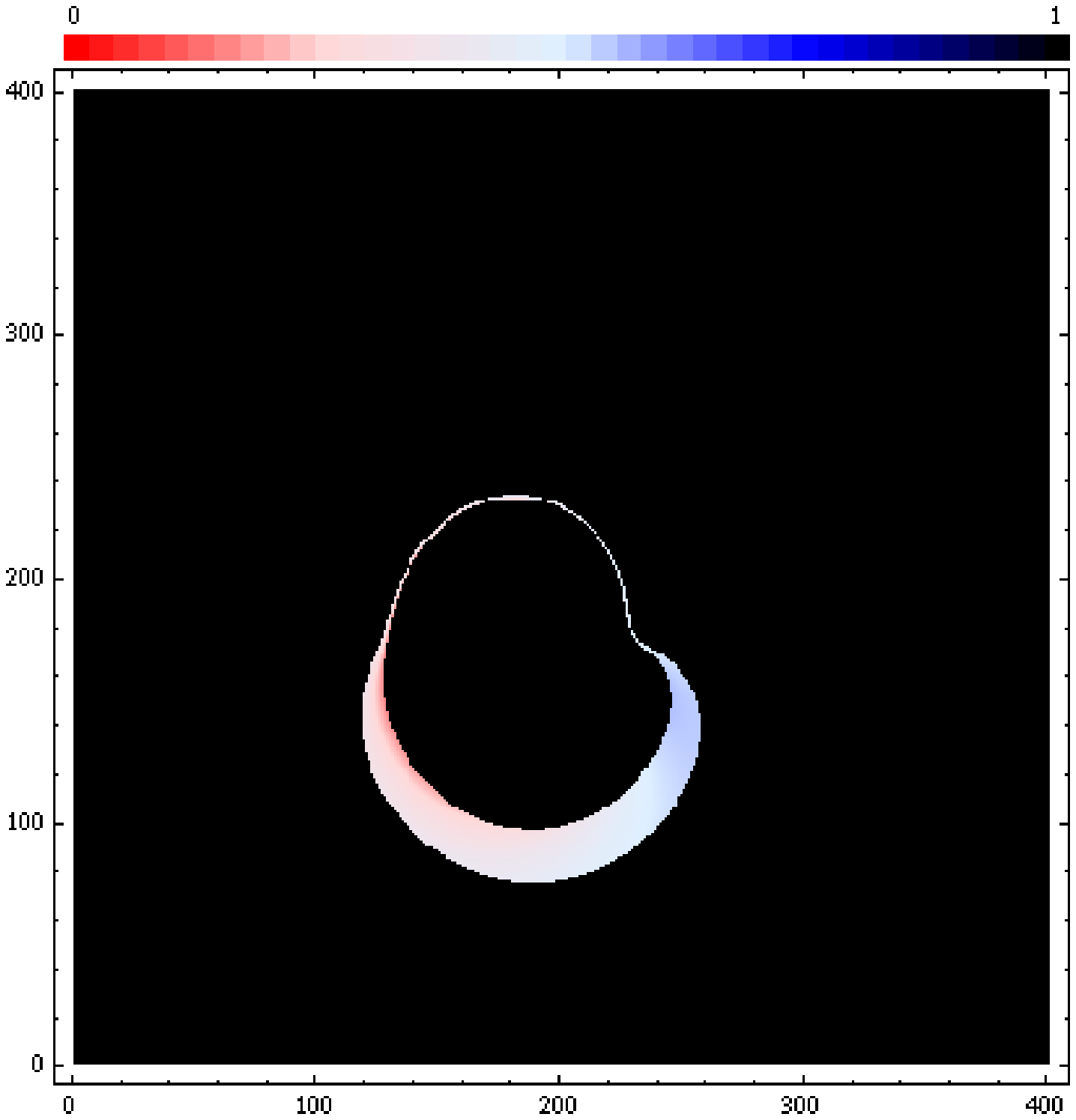}
&\includegraphics[width=3.6cm]{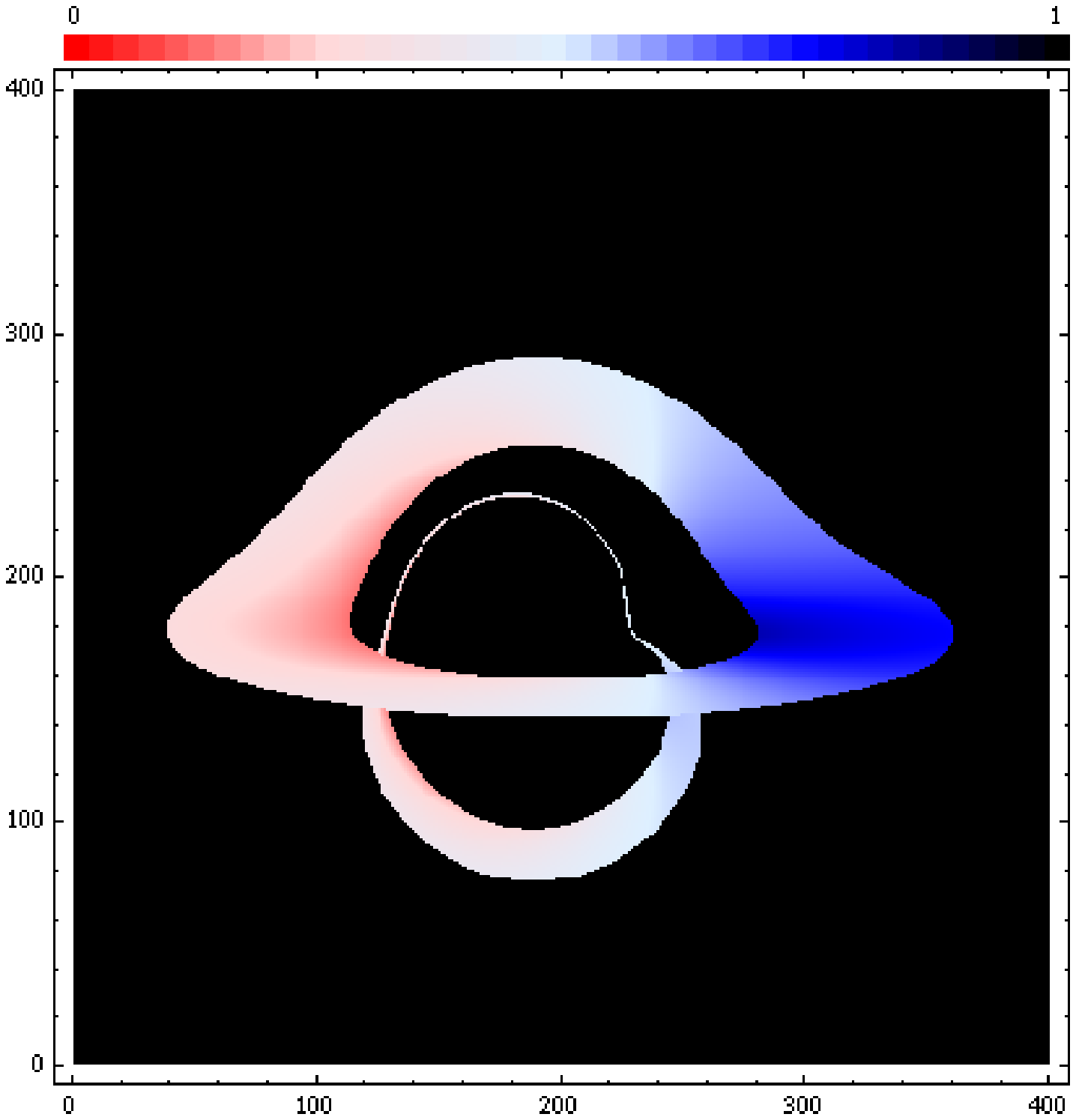}\\
     \includegraphics[width=3.6cm]{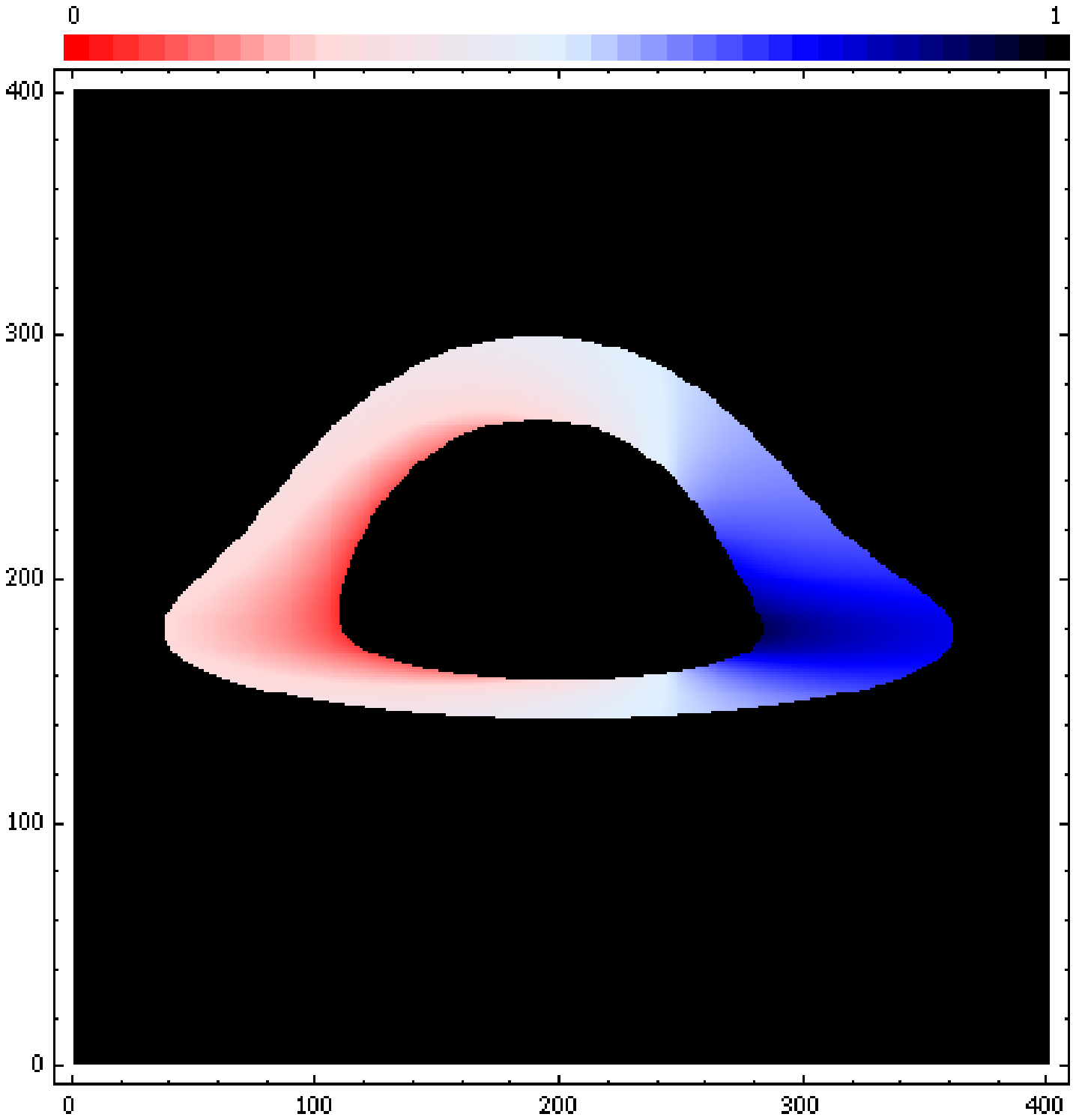}
& \includegraphics[width=3.6cm]{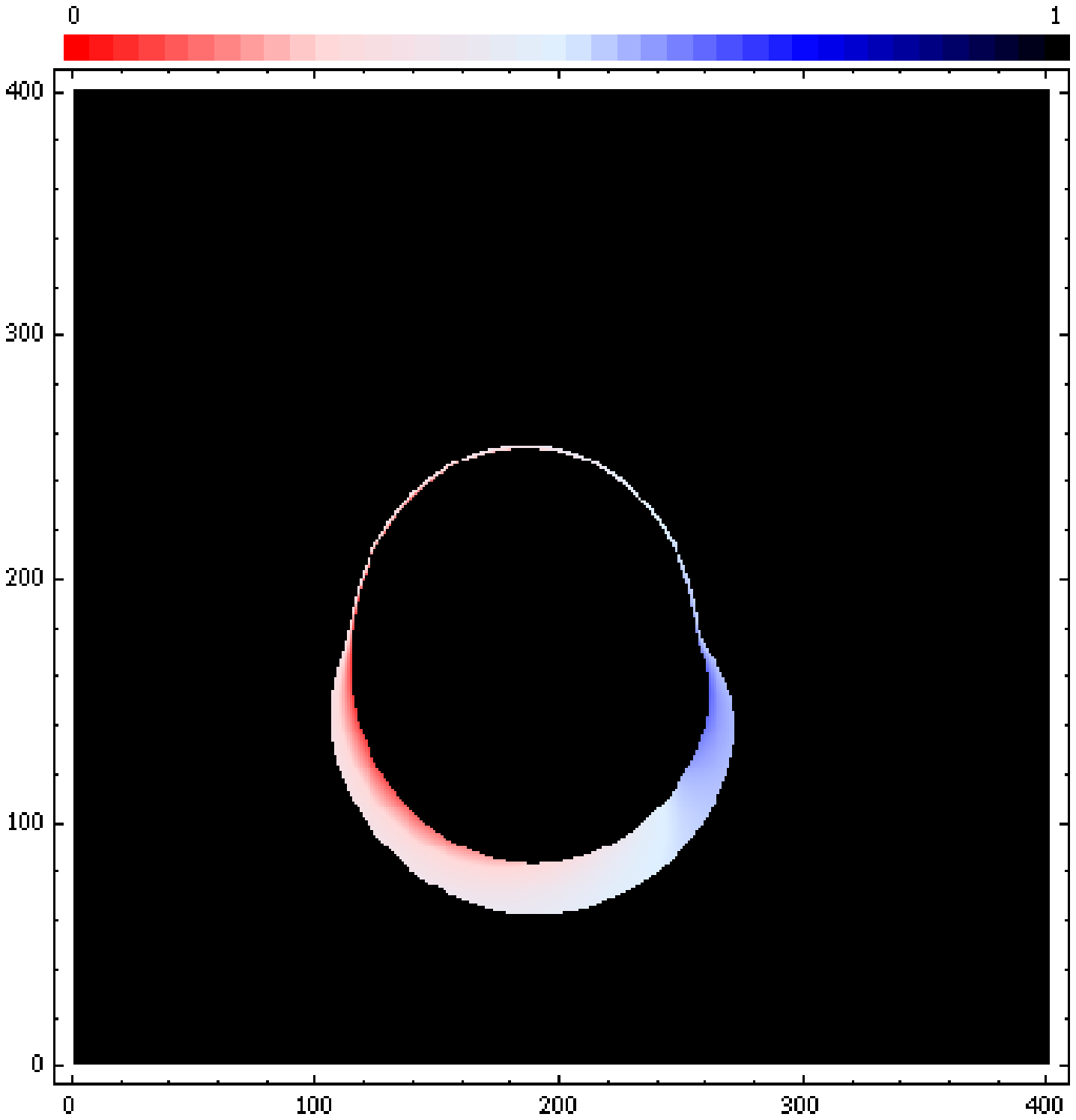}
&\includegraphics[width=3.6cm]{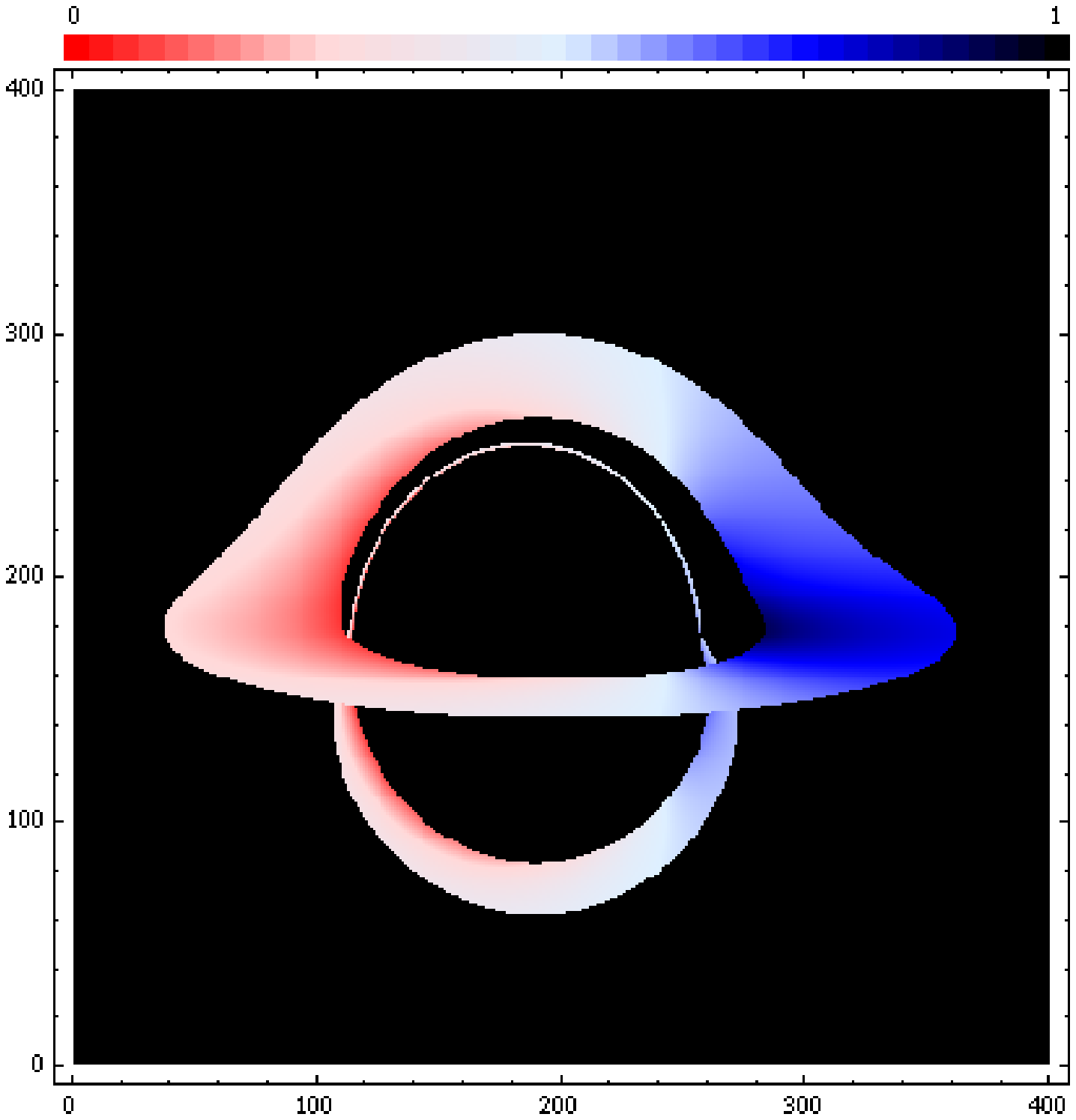}
  \end{tabular}
\caption{\label{fig24_a_i}Radiating Keplerian disc images with fixed inner and outer radii. The modified frequency shift $\bar{g}=(g-g_{min})/(g_{max}-g_{min})$, with $g_{min}=0.4$ and $g_{max}=1.5$, of the radiation emitted from the thin disk with inner radius $r_{in}=7 M$ and outer radius $r_{out}=15 M$, encoded into colors is plotted for representative values of tidal charge parameter $b=-3.0$, $0.0$ and inclination of observer $\theta_0=30^\circ$, $80^\circ$. In the left column direct images are ploted, the indirect images are ploted in the central column and the composition of direct and indirect images is plotted in the right column. The first two rows of images are plotted for the observer inclination $\theta_0=30^\circ$ and the second two rows of images are plotted for the observer inclination $\theta_0=80^\circ$. Top row images are plotted for $b=0.0$, the second row images are plotted for $b=-3.0$, the third row images are plotted for $b=0.0$ and bottom row images are plotted for $b=-3.0$.}
\end{figure}

\begin{figure}[ht]
\begin{tabular}{ccc}
     \includegraphics[width=3.6cm]{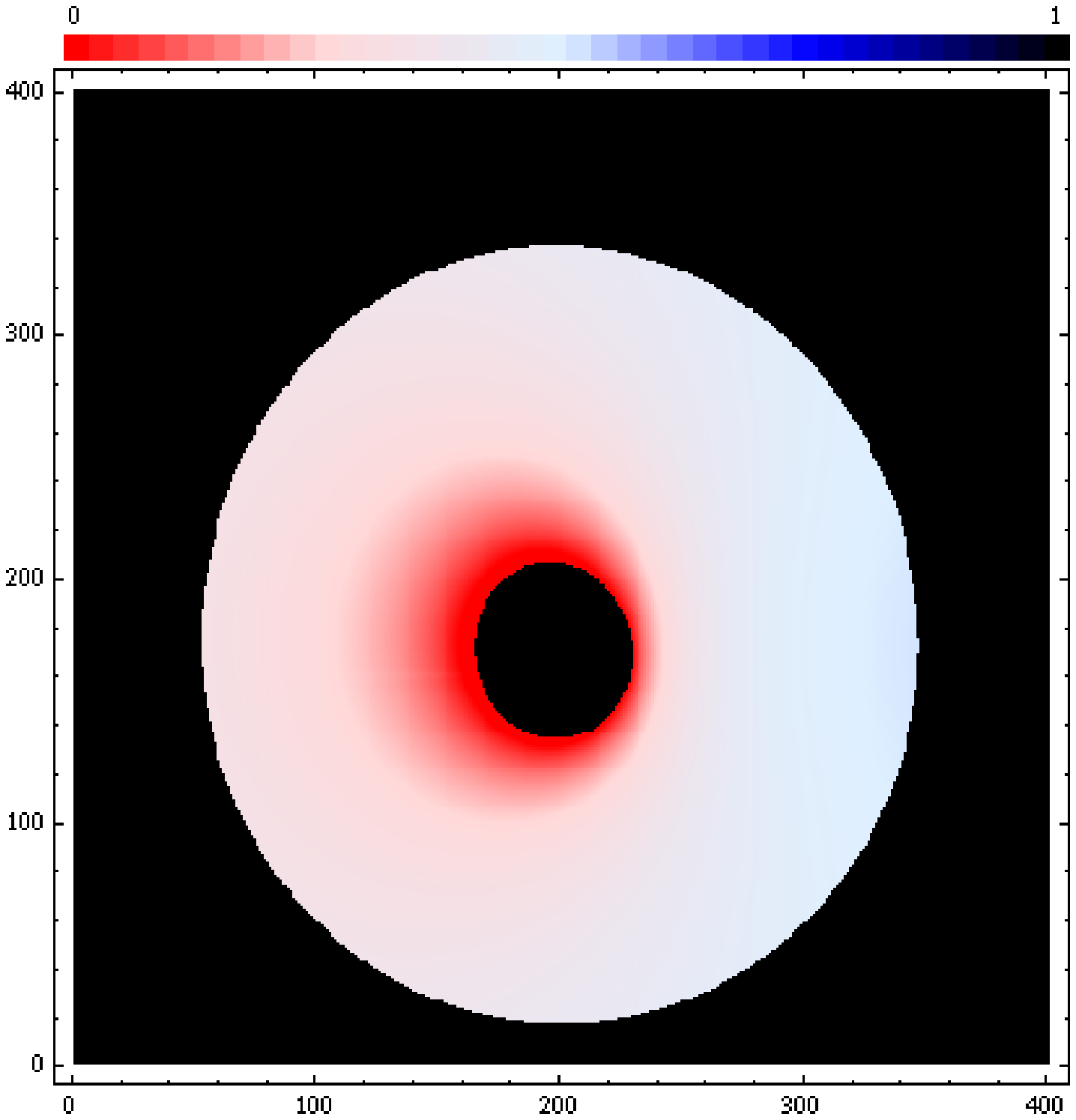}
& \includegraphics[width=3.6cm]{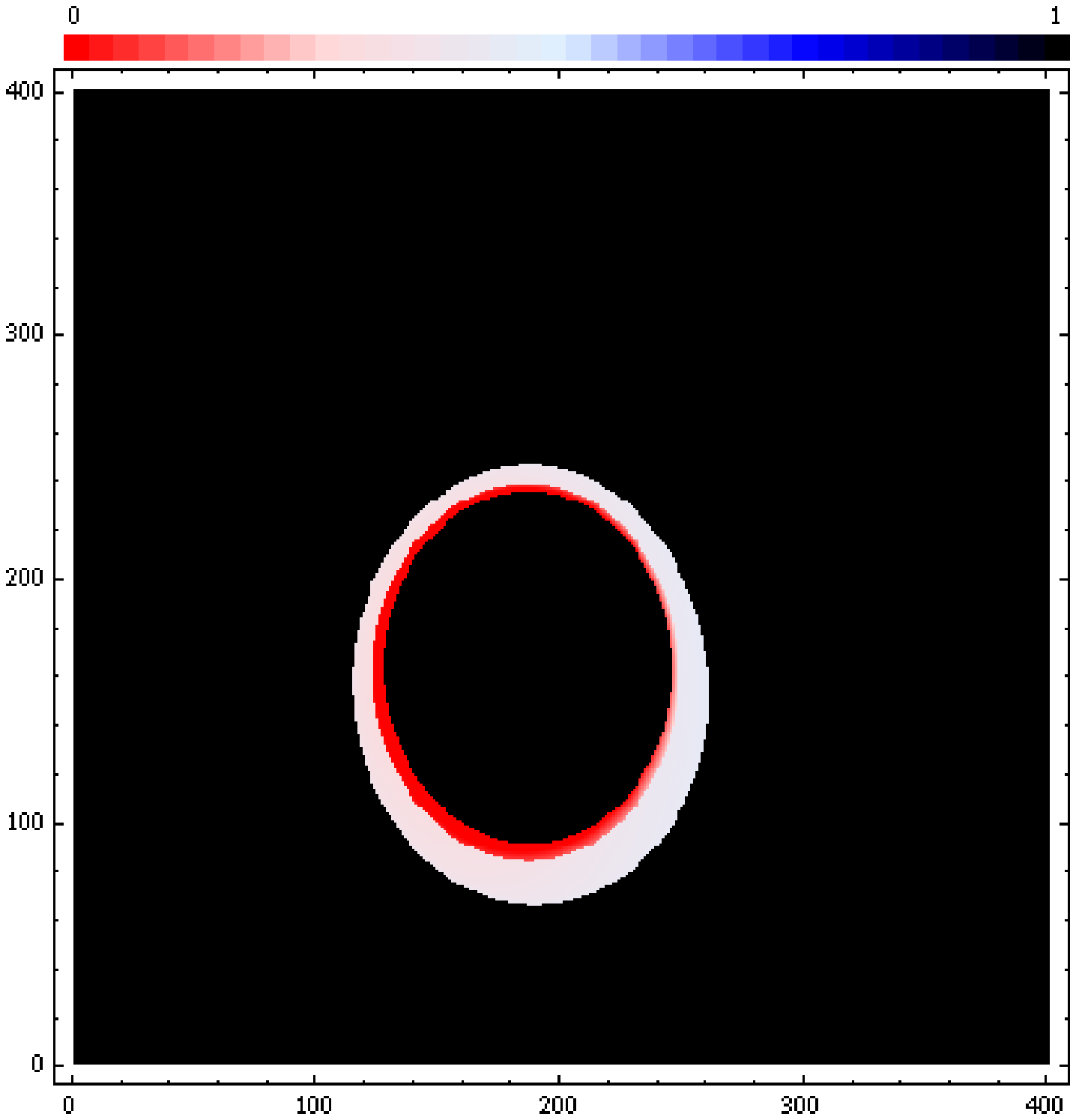}
&\includegraphics[width=3.6cm]{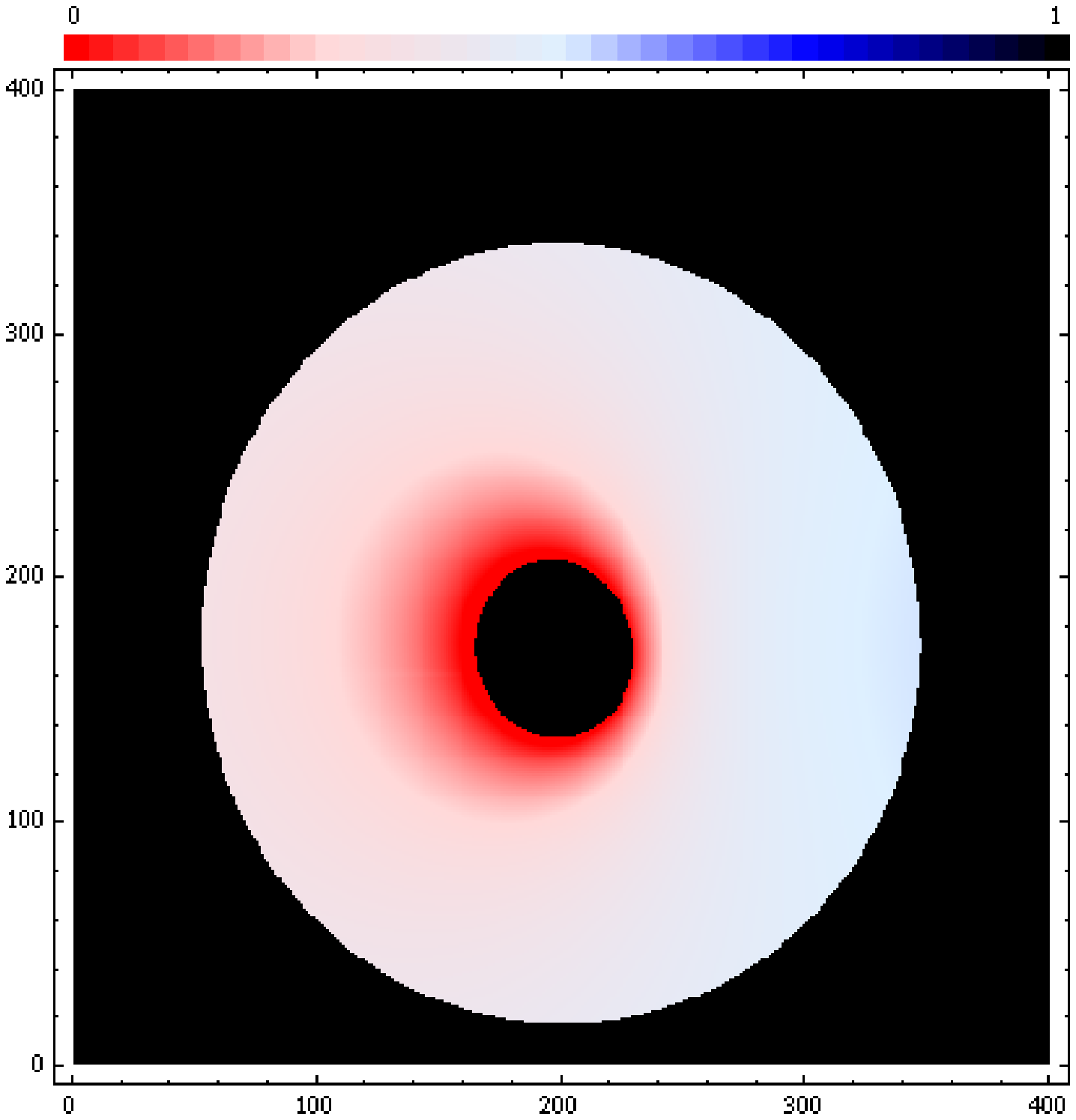}\\
     \includegraphics[width=3.6cm]{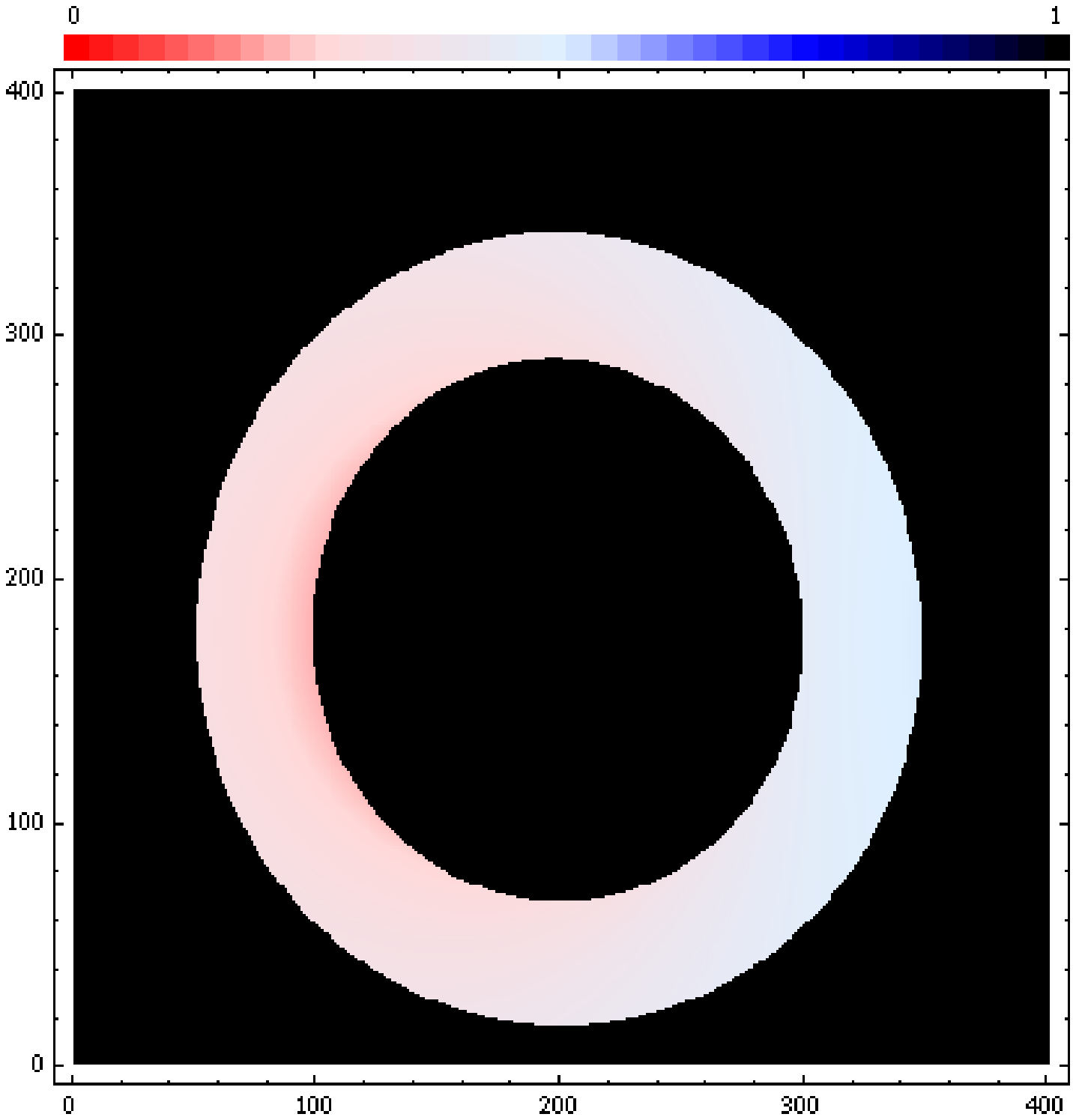}
& \includegraphics[width=3.6cm]{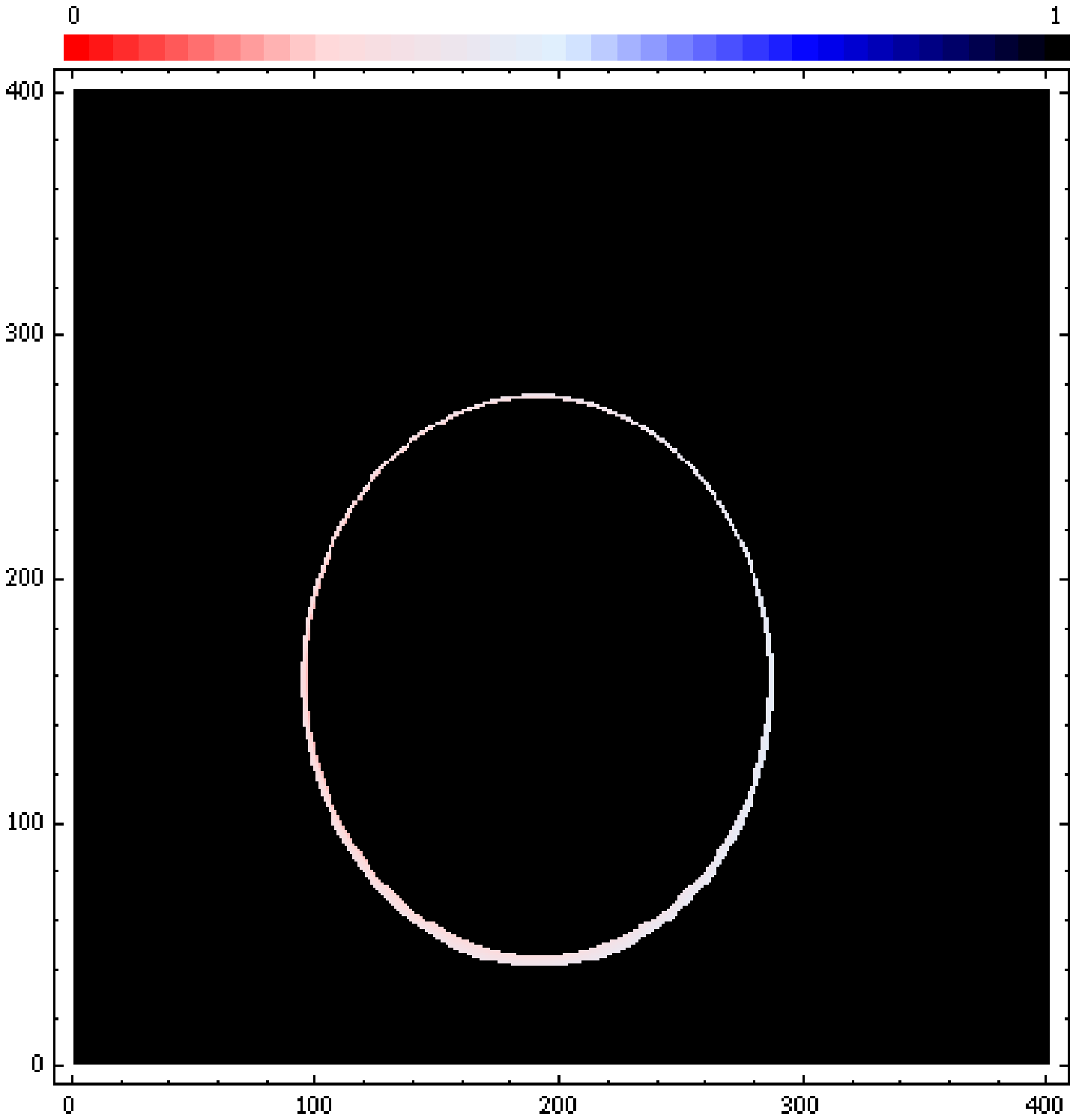}
&\includegraphics[width=3.6cm]{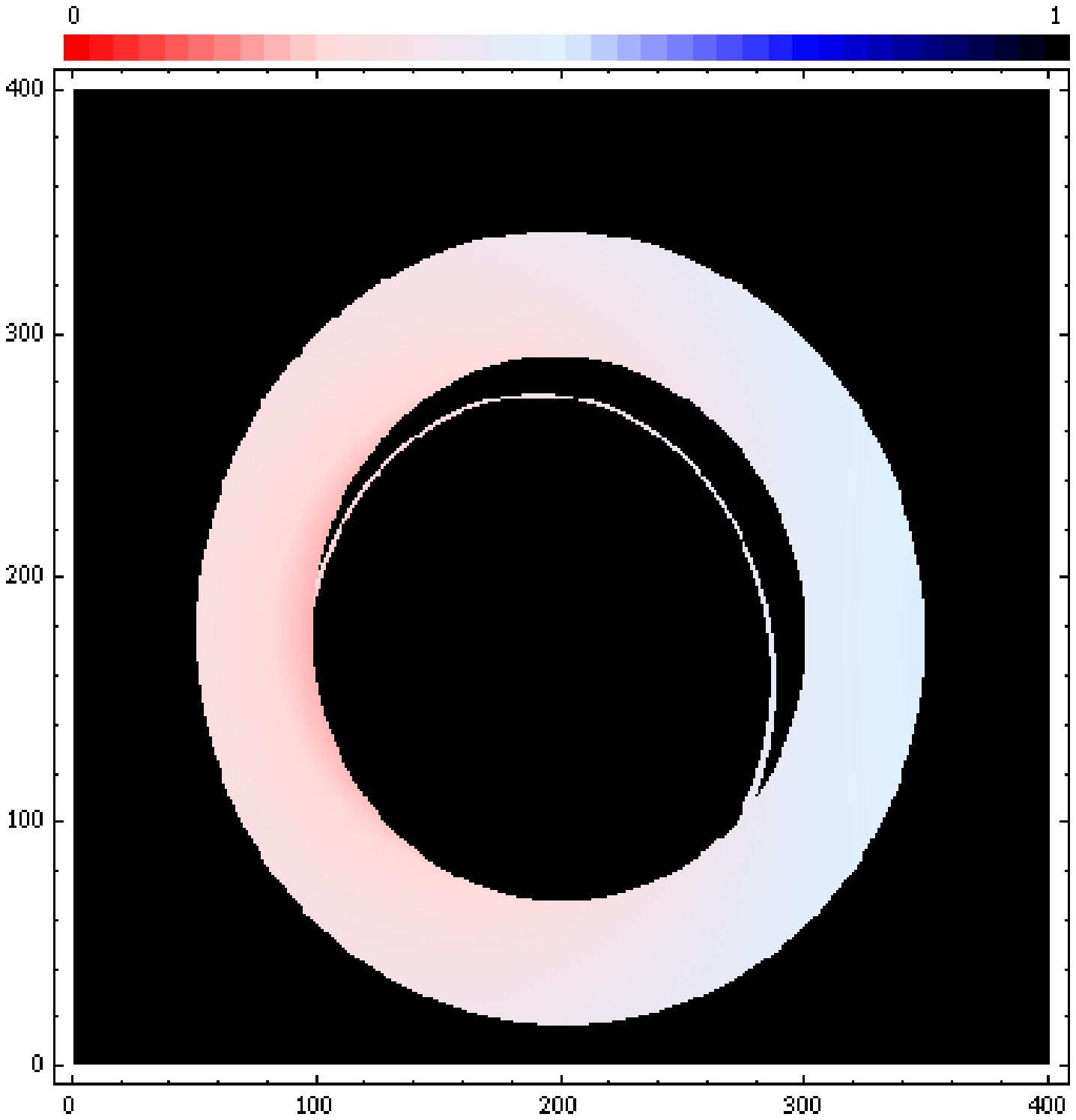}\\
     \includegraphics[width=3.6cm]{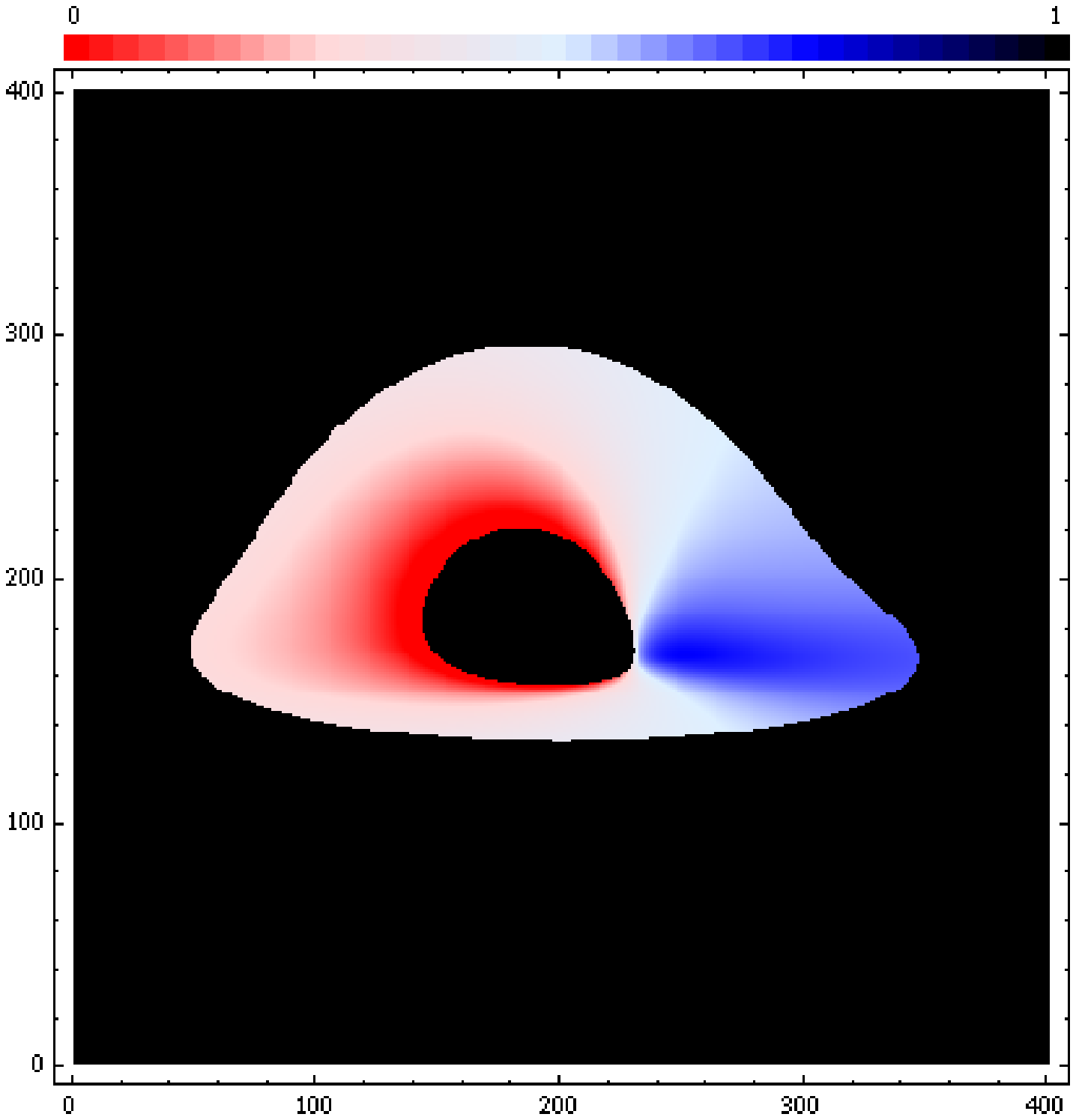}
& \includegraphics[width=3.6cm]{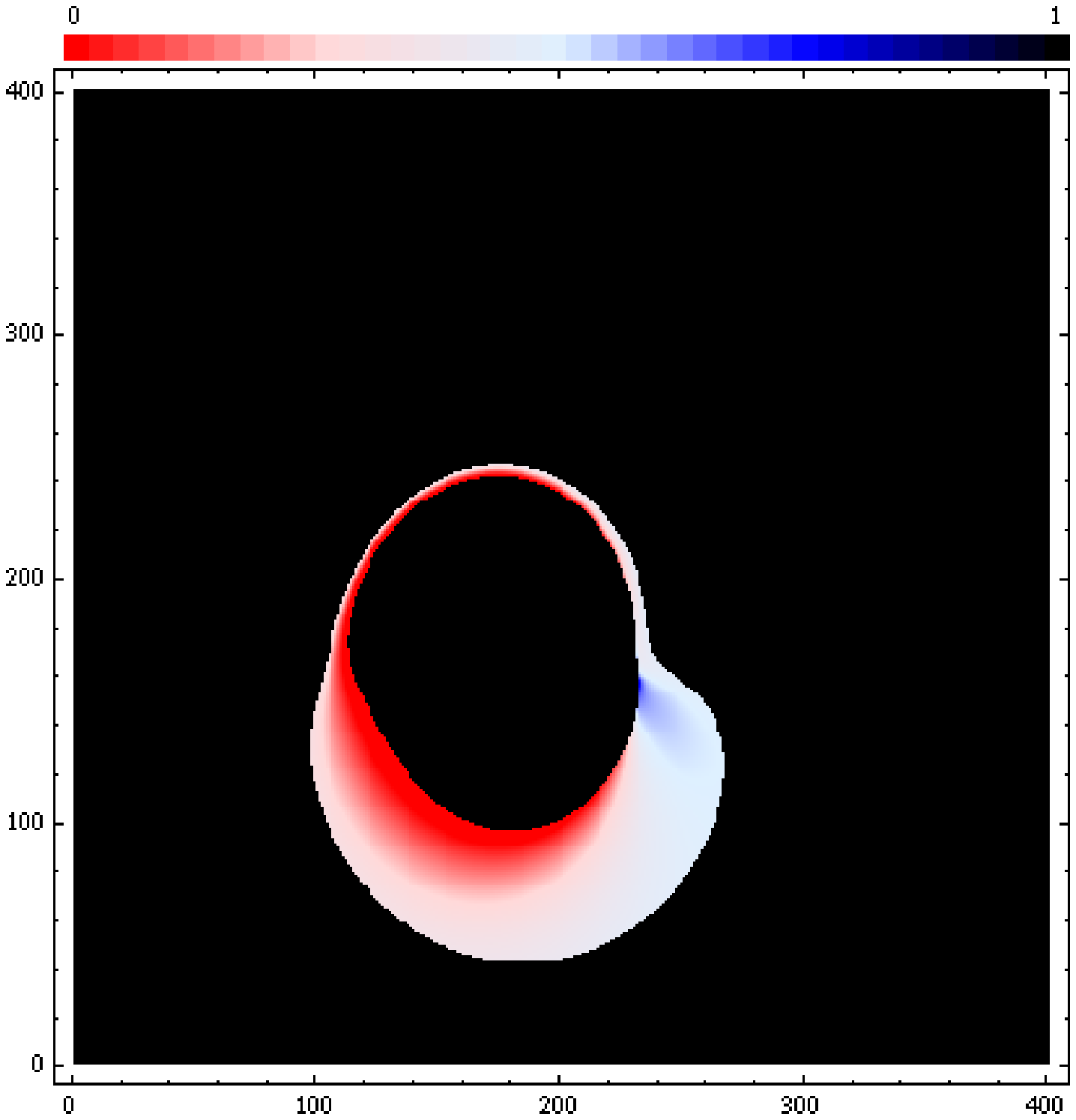}
& \includegraphics[width=3.6cm]{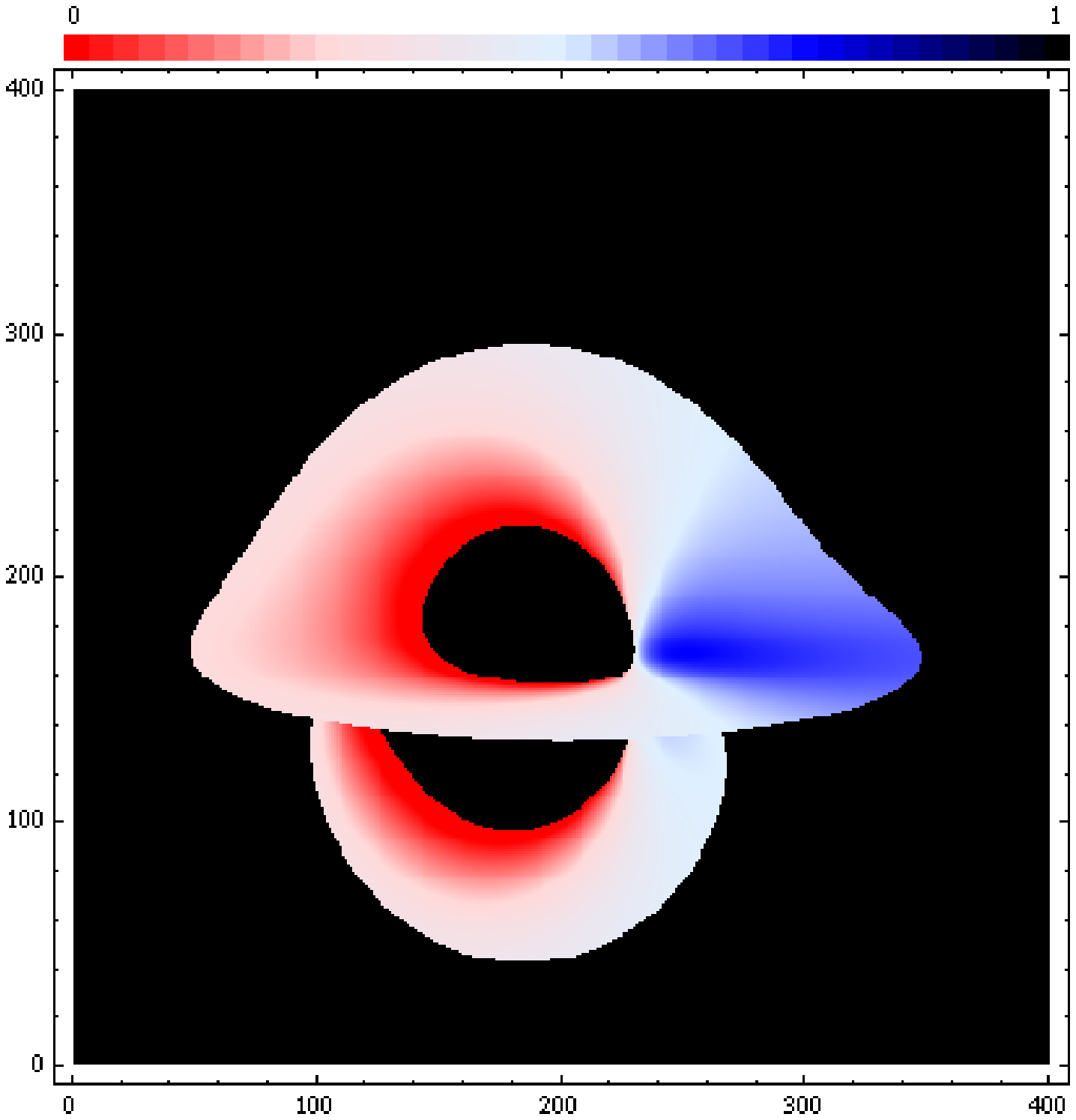}\\
     \includegraphics[width=3.6cm]{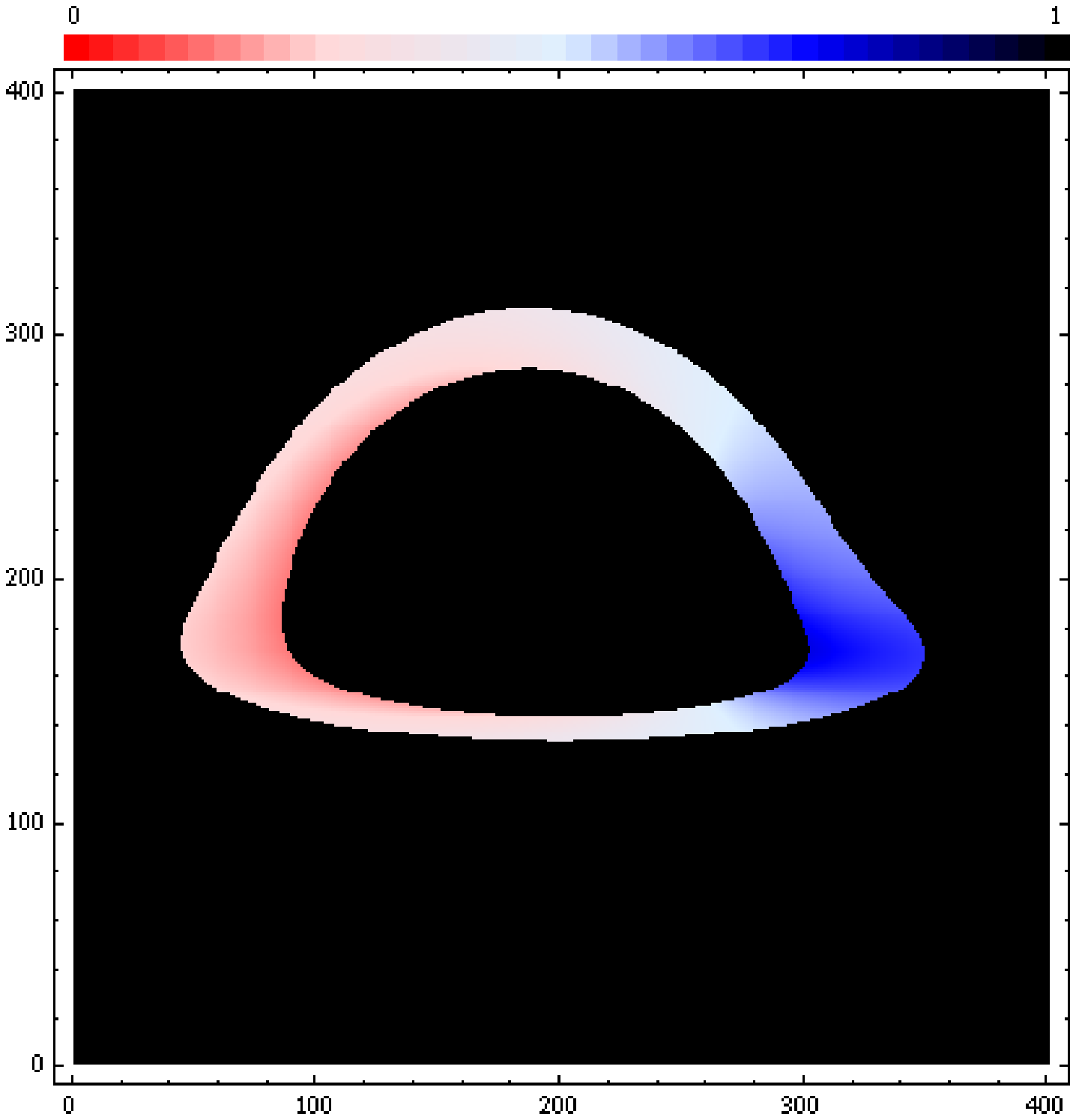}
& \includegraphics[width=3.6cm]{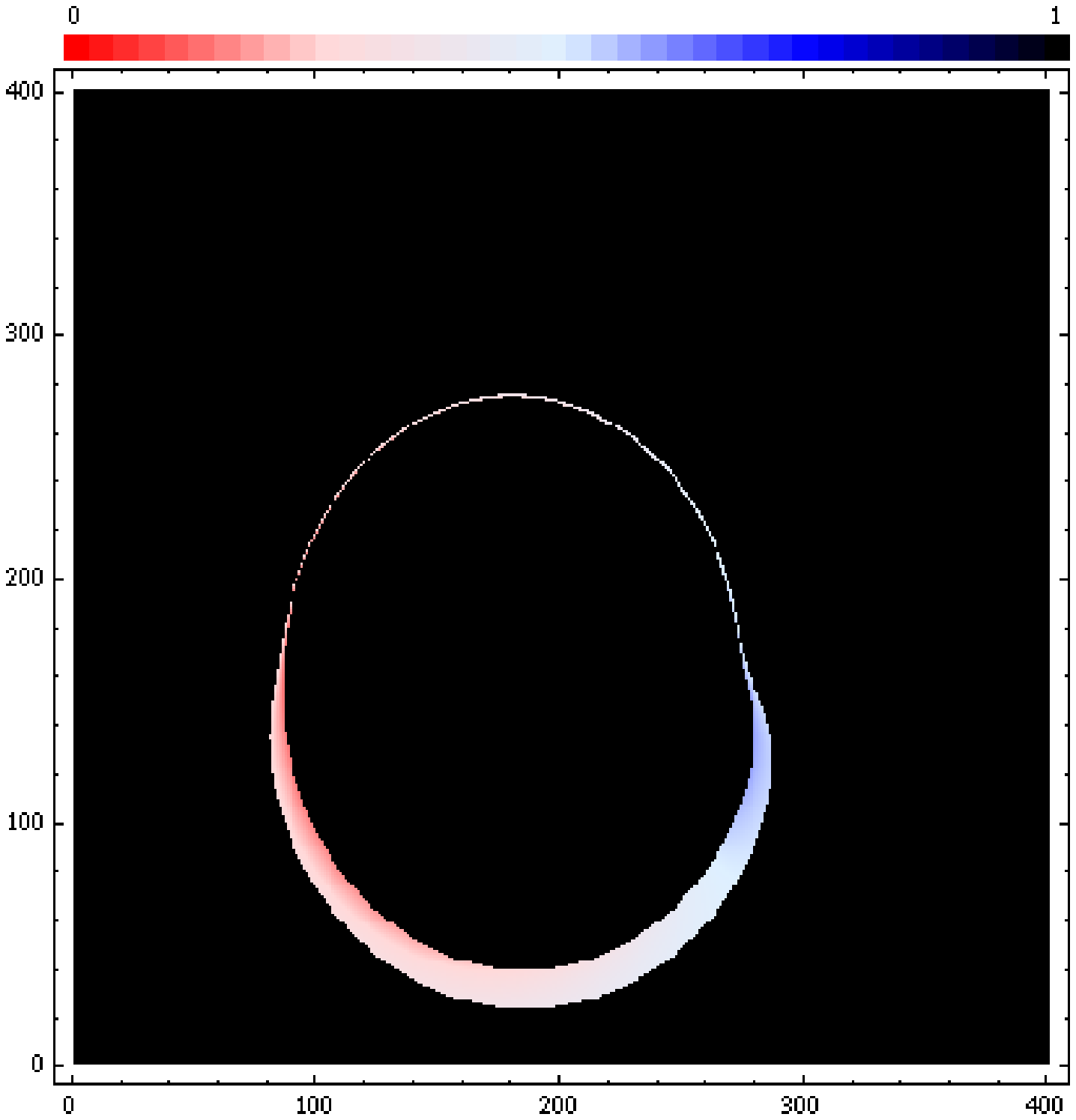}
& \includegraphics[width=3.6cm]{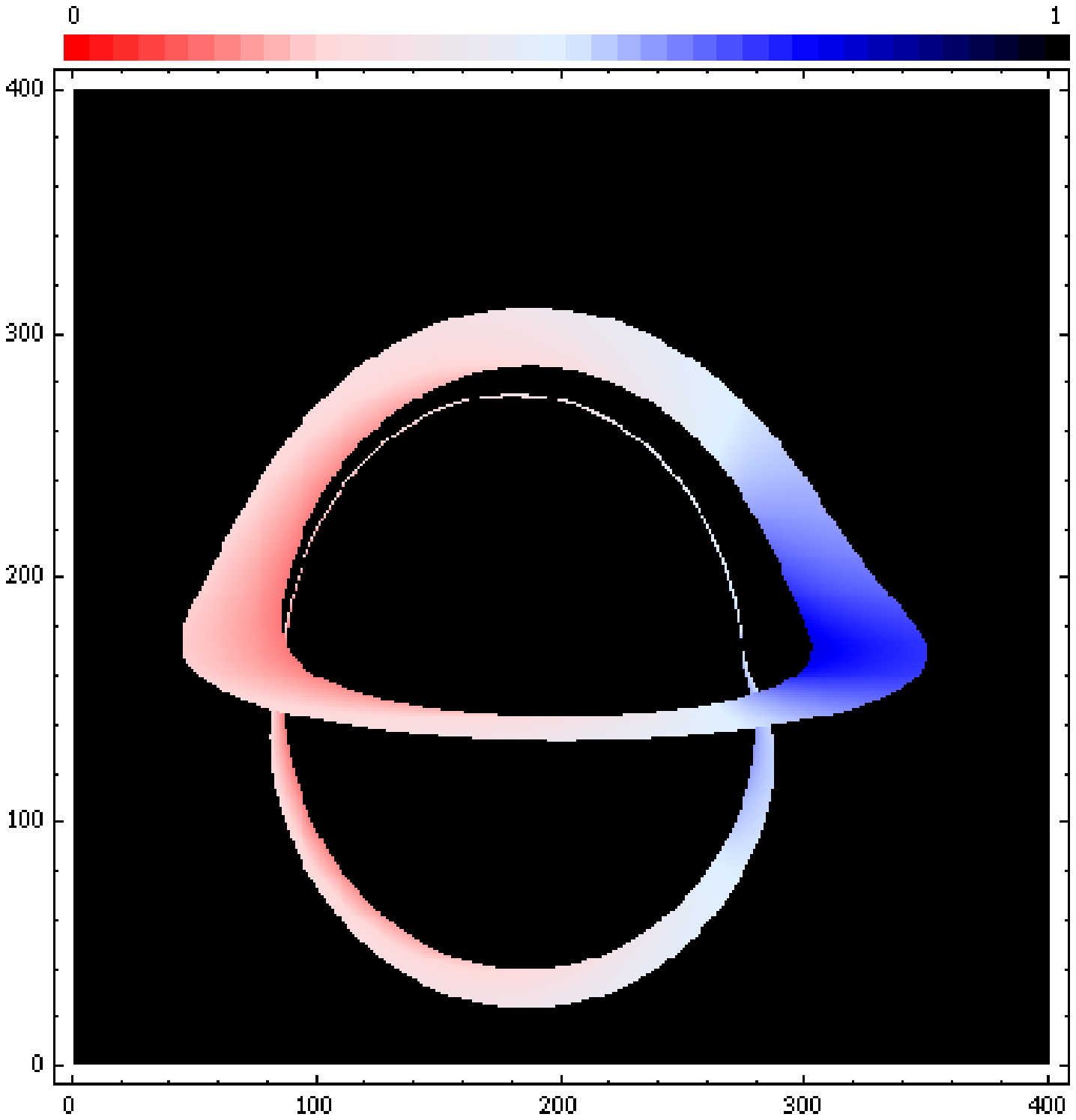}
  \end{tabular}

\caption{\label{fig26_a_i} Radiating Keplerian disc images with $r_{in}=r_{ms}$. The modified frequency shift $\bar{g}=(g-g_{min})/(g_{max}-g_{min})$, with $g_{min}=0.2$ and $g_{max}=1.8$, of the radiation emitted from the thin disk with inner radius $r_{in}=r_{ms}$ (with $r_{ms}(b=0;a=0.9981)=1.3$ and $r_{ms}(b=-3;a=0.9981)=6.3$) and outer radius $r_{out}=10$, encoded into colors is plotted for representative values of tidal charge parameter $b=-3.0$, $0.0$ and inclination of observer $\theta_0=30^\circ$, $80^\circ$. In the left column direct images are ploted, the indirect images are ploted in the central column and the composition of direct and indirect images is plotted in the right column. The first two rows of images are plotted for the observer inclination $\theta_0=30^\circ$ and the second two rows of images are plotted for the observer inclination $\theta_0=80^\circ$. Top row images are plotted for $b=0.0$, the second row images are plotted for $b=-3.0$, the third row images are plotted for $b=0.0$ and bottom row images are plotted for $b=-3.0$. 
}
\end{figure}

The influence of the frequency shift in the disc images is demonstrated in Figures \ref{fig24_a_i} and \ref{fig26_a_i}. The role of the braneworld parameter is illustrated both for small ($\theta_0=30^\circ$) and high ($\theta_0=80^\circ$) inclination angles. We consider two cases of the radiating disc extension: first one  with fixed inner and outer radii, independent of the black hole parameters, and the second one when the inner radius is identified with the marginally stable orbits, depending on the black hole parameters. 

In order to map the frequency shift $g$ into color palete we define modified frequency shift $\bar{g}=(g-g_{min})/(g_{max}-g_{min})$ where $g_{min}$ ($g_{max}$) is the minimal (maximal) value of frequency shift, which is fixed in a particular set of images.

We can see from Figs. \ref{fig24_a_i} and \ref{fig26_a_i} that the negative tidal charge has the tendency to enlarge and symmetrize the disc images.

\clearpage
\newpage
 
\section{Time delay}

For optical effects in vicinity of a black hole, the time delay in case of systems varying with time and observed along two different directions due to the light deflection in strong gravity can be important. The cordinate time that elapses from the instant of photon emission, $t_e$, to the instant of its reception, $t_o$, is integrated from the Carter equations and reads

\begin{eqnarray}
  t_o&=&t_e+\mu_{sgn}\int_{\mu_e}^{\mu_o}{a^2\mu^2\frac{\diff\mu}{\sqrt{M}}}\nonumber\\
	&&+u_{sgn}\int^{u_o}_{u_e}{\frac{2a(a-\lambda)u^3+a^2u^2+1+ab(\lambda-a)u^4}{(u/u_{+}-1)(u/u_{-}-1)\sqrt{U}}}\diff u
\end{eqnarray}
In order to succesfully integrate this formula, one must map all the turning points in $\mu$ and $u$ motion to correctly set up the signs $u_{sgn}$ and $\mu_{sgn}$.

Suppose that the two light beams, direct and indirect, are emitted at the same coordinate time $t_e$. They generally reach the observer at different coordinate times $t_o^{\mathrm{dir}}$($t_o^{\mathrm{indir}}$ resp.). By time delay we define here the difference $\Delta t\equiv t_o^{\mathrm{indir}}-t_o^{\mathrm{dir}}$.

\begin{figure}[!ht]
 \includegraphics[width=10cm]{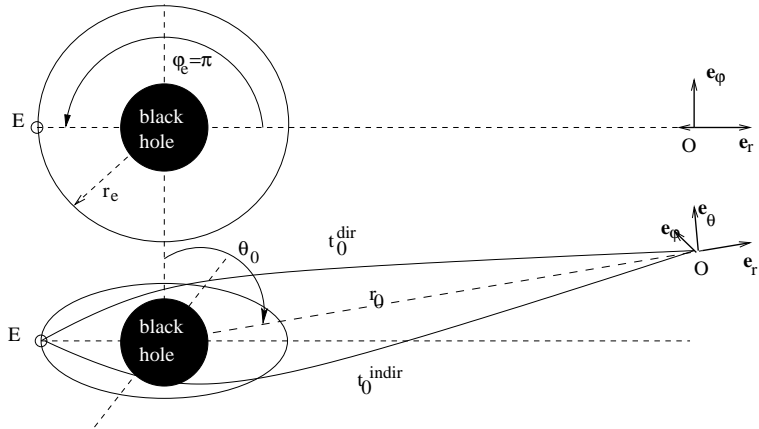}
\caption{\label{fig27}The illustration of the impact of tidal charge parameter on the time delay $\Delta t$ in case of direct and indirect photons emitted from emitter $E$ at coordinate time $t_e$ and azimuthal position $\varphi_e=\pi$. They are received at observer $O$ at coordinate times $t_o^{\mathrm{dir}}$($t_o^{\mathrm{indir}}$ resp.). The emittor is on circular geodesic in equatorial plane of braneworld Kerr black hole at radial coordinate $r=r_e$. The observer is far from the center of the black hole at $r=r_o$. Its inclination is $\theta=\theta_o$.}
\end{figure}

To demonstrate the impact of the tidal charge $b$ on the time delay we
consider the following situation (see Figure \ref{fig27}). Let the isotropicaly radiating monochromatic source orbits in the equatorial plane of the braneworld Kerr black hole at radial distance $r_e$. It can be switched on and off. When it reaches the azimuthal coordinate $\varphi=\pi$ it is switched on and we compare the coordinate times $t^{\mathrm{dir}}_o$ and $t^{\mathrm{indir}}_o$ of reception of the photons from the direct and indirect images of the source. 

\begin{figure}[!ht]
 \begin{tabular}{cc}
  \includegraphics[width=6.2cm]{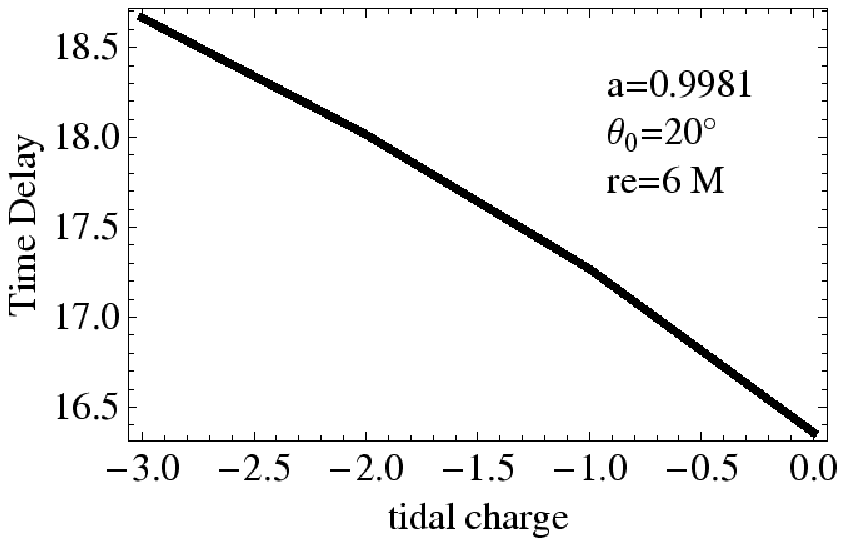}&\includegraphics[width=6.2cm]{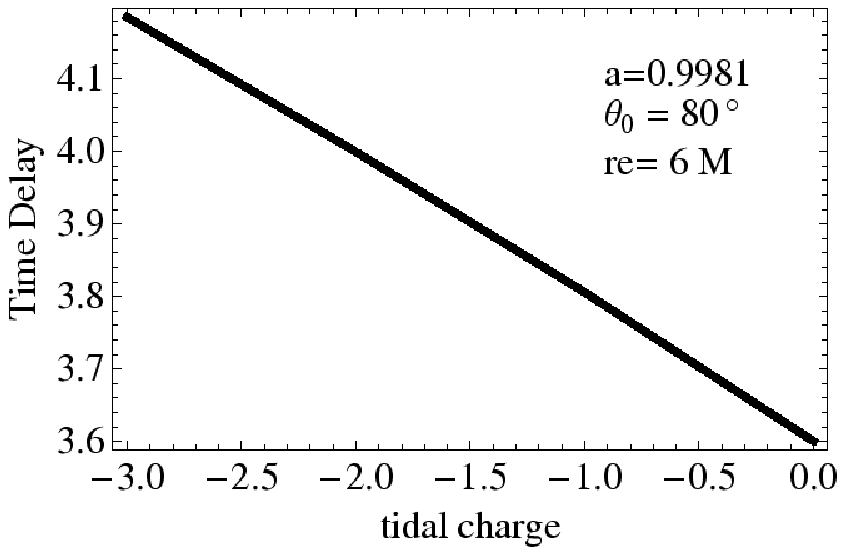}
 \end{tabular}
\caption{\label{fig28_a_b}The difference (``Time Delay``), $\Delta t = t^{\mathrm{indir}}_o -t^{\mathrm{dir}}_o$, between coordinate times of reception of direct and indirect geodesics of photons emmited at the same coordinate time $t_e$ from the azimuthal coordinate $\varphi=\pi$  is plotted as a function of tidal charge $b$. Left figure: the inclination of the observer is $\theta_0=20^\circ$. Right figure: the inclination of the observer is $\theta_0=80^\circ$. }

\end{figure}

The results are demonstrated in the Figure \ref{fig28_a_b}. We can directly see that time delay $\Delta t$ between times of reception of the direct and indirect photons emitted at the same instant from the azimuthal position $\varphi=\pi$ increases as the value of the tidal charge parameter $b$ goes to higher negative values. When $b$ is fixed, the time delay $\Delta t$ increases as the value of the inclination decreases. The same effects appear for other positions of the radiating spot ($\varphi\not= \pi$). We can see that the time delay $\Delta t$ depends strongly on the viewing angle  $\theta_0$. Therefore, it is extremely important to have a system with precisely determined viewing angle.

\section{Optical phenomena related to Sgr $A^*$}
There is an enormously growing evidence that the center of our Galaxy harbors a supermassive black hole whose position could be almost surely identified with the extremely compact radio source  Sgr $A^*$. The chain of arguments seems to be very convincing; stars orbiting an unseen mass concentration on elliptical orbits with a common focal position, the  unseen mass centered  on Sgr $A^*$ that seems to be motionless at the dynamical center of the Galaxy, extremely compact emission of the center \cite{Reid:2008:}. Recent measurements of Ghez and collaborators \cite{Ghez-etal:2008:} from the W.M. Keck 10 - meter telescopes of a fully unconstrained Keplerian orbit of the short period  star SO-2 provide the distance $R_0=8.0\pm 0.6$ kpc and black hole mass $M=(4.1\pm0.6)\times 10^6 M_\odot$. If the black hole is assumed to be at rest with respect to the Milky Way Galaxy (i.e., has no massive companion to induce its motion) as argued by Reid \cite{Reid:2008:}, the fit can be further constrained to $R_0=8.4\pm 0.4$kpc and $M=(4.5\pm 0.4)\times 10^6M_\odot$ \cite{Ghez-etal:2008:}.

Such a close and huge supermassive black hole could be clearly a very convenient object, probably the best one, for testing a wide variety of optical phenomena in strong gravity in its vicinity. The time delay of accidents happening behind the black hole and observed along two directions could be in principle easily measured. We could even expect possibility of black hole silhuette measurements. In this way the influence of the tidal charge could be properly tested and its value estimated, because for the Galaxy supermassive black hole we can determine relatively precisely the inclination angle of the observer (Solar system), although it is of course very close to $\theta_0\simeq 90^\circ$.

For non-rotating , Schwarzchild black holes, the silhuette diameter is given by the impact parameter of the photon circular orbit

\begin{equation}
	D=2\lambda_{ph}= 6\sqrt{3} M.
\end{equation}
Using the Sgr $A^*$ mass estimate $M\sim 4.5\times 10^6M_\odot$, we find $D\simeq 55\mu$arcsec while interferometer finges were reported at wavelength of $1.3$ mm and fringe spacing of $0.00005$, comparable with the expected value of $D$. Shorter wavelengths should enable detailed measurements of the black hole silhuette and relatively precise estimates of the black hole parameters due to very precise knowledge of the inclination angle. The angle can be given by the measurement of the Solar system position relative to Galaxy plane $z_\odot\sim 14pc$\cite{Yoshi:2007:}. Then $\theta_0\sim 89.9^\circ$ or more precisely, $\theta_0$ lies between the values of $89.8772^\circ$ ($z_\odot=18pc$) and $89.9318^\circ$($z_\odot=10pc$). Of course, considering the silhuette shape, it is quite enough to take $\theta_0=90^\circ$. 

\begin{figure}[h]
	\begin{center}
	\includegraphics[width=6.1cm]{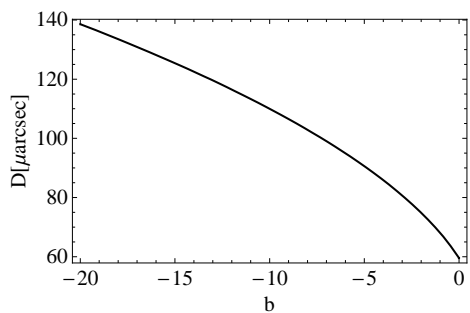}
	\end{center}
	\caption{Diameter $D$ as a function of braneworld parameter $b$ is plotted for Schwarzchild black hole of mass $M=4.5\times 10^6M_\odot$. Observer is at $r_0=8.4$kpc lying in the equatorial plane.}\label{diameter_of_silhuette_schw}
\end{figure}

In the case of spherically symmetric black holes, the influence of the tidal charge parameter $b$ on the silhuette diameter can be given by the simple formula for impact parameter of photon circular orbits that reads \cite{Stu-Hle:2002:}

\begin{equation}
	\lambda_{ph}(b)=\frac{r_{ph}^2}{\sqrt{r_{ph}-b}}M,
\end{equation}
where

\begin{equation}
	r_{ph}(b)=\frac{3}{2}\left(1+\sqrt{1-\frac{8b}{9}}\right).
\end{equation}
The resulting dependence  of the diameter $D(b)$ is illustrated in Figure \ref{diameter_of_silhuette_schw} . The diameter grows slowly with the descending of $b$; notice that its magnitude is twice the pure Schwarzchild value for $b=-12.8428$. Of course, for rotating black holes the silhuette is maximally deformed due to the influence of rotation since the viewing angle $\theta_0\sim 90^\circ$ and is given by calculations and results presented above. Testing of the combined spin and tidal charge influence would be possible with measurement precision enlarged for 1-order relative to the recently expected state mentioned above. Clearly, we can expect that the observational accuracy in near future will be high enough to measure the Sgr $A^*$ black hole silhuette implying relevant estimates of the black hole parameters.




\begin{figure}[ht]
	\begin{center}
		\includegraphics[width=6.1cm]{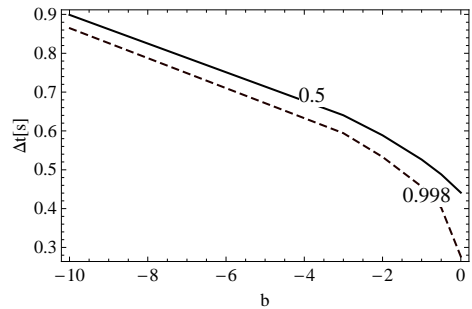}
	\end{center}
	\caption{\label{td_table4}Comparizon of time delay effect as a function of braneworld parameter $b$ between two rotating black holes with rotational parametes $a=0.5$ and $a=0.998$. For each $b$ the emitter is radiating from marginally stable orbit. The relevant values of radii $r_{ms}$ of marginally stable orbits are arranged in the Table \ref{tabulka2}. }
\end{figure}

\begin{table}[ht]
		\tbl{Table of relevant values of $r_{ms}$ used in plots oin Fig \ref{td_table4}}
		{\begin{tabular}{@{}ccccccc@{}} 
		\toprule
		$b$ & 0.0 & -0.5 & -1.0 & -2.0 & -3.0 & -10.0\\ 
		\colrule
		$r_{ms}(a=0.5)$ & 4.24M & 5.05M & 5.73M & 6.88M & 7.85M & 12.88M\\
		\colrule
		$r_{ms}(a=0.998)$ & 1.24M & 3.03M  & 3.91M & 5.22M & 6.28M & 11.44M\\
		\botrule
		\end{tabular}\label{tabulka2}}
\end{table}

Considering the time delay effects, the exact value of $\theta_0$ is crucial since it plays a fundamental role in determining the time delay effect whose scale is given by the value of $t\sim 1sec$. We illustrate the influence of the tidal charge on the time delay effects at the astrophysically important radii corresponding to marginally stable circular geodesics, i.e. in the strong gravity regime, for two representative fixed values of black hole spin (see Figure \ref{td_table4} and Table \ref{tabulka2}). We can expect importance of the regions close to $r_{ms}$ for relevant optical effects due to the idea of the low angular momentum accretion in Sgr $A^*$ advocated by B. Czerny \cite{Cze-etal:2007:}. Clearly, we can see in Figure \ref{td_table4} that the time delay effects could be well measurable and the tidal charge influence could be well tested, if the black hole spin is properly estimated.

\section{\label{sec:Conclusions}Conclusions}
One of the most promising ways of estimating influence of hypothetical hidden external dimensions, considered in the framework of the braneworld model with infinite external dimension as developed by \cite{Ran-Sun:1999:}, seems to be investigation of the optical phenomena caused by the black hole backgrounds. It is so because black holes represent the only case when the non-local influence of the bulk space on the braneworld spacetime structure can be fully described by a single, braneworld parameter called tidal charge, the sign of which can be both positive and negative, with the second possibility beeing more realistic one \cite{Ali-Gum:2005:,Dad-etal:2000:}.

Here, we focused our attention to developing a theoretical background for treating the optical phenomena in vicinity of braneworld rotating black holes and bringing general tendencies of the tidal charge effect in some basical optical phenomena. 

We have shown qualitatively how the braneworld tidal charge affects the basical optical phenomena, especially the black-hole silhuette, the accretion disc image with the frequency shift of area of the disc radiating at a specific frequency, and the time delay between the direct and indirect images of the hot spot orbiting the black hole. We have shown that these phenomena could be measured and used to put limits on the tidal charge in case of Galaxy Center Sgr $A^*$ supermassive black hole.

We generalized the approaches based on the transfer-function method as introduced and developed in Schwarzchild and Kerr backgrounds \cite{fab-rees-ste-whi:1989:,Mat-Fab-Ros:1993:,Bao-Stu:1992:,Stu-Bao:1992:,Laor:1991:,Dov-Karas-Mas-Mar:2005:,Fan-Cal-Fel-Cad:1997:,Rau-Bla:1994:} where equations of photon motion are solved in terms of the elliptic integrals (see \cite{Rau-Bla:1994:,Kra:2005:,Kra:2007:}). For purposes of the present work, the transfer-function method seems to be most efficient. Nevertheless, we prepared the ray-tracing method too, since that could be usefull in treating other optical phenomena.

 Generally, rising negative value of the tidal charge strenghtens the black hole field and suppresses the rotational phenomena, when the black-hole rotation parameter is fixed.  The magnitude of the optical phenomena grows with decreasing of the negatively-valued tidal charge, but the rotation induced asymmetry of the phenomena like the black-hole silhuette, or the accretion disc image, decreases. The black-hole silhuette is characterized by two parameters, namely the shift of the center  and ellipticity, that could be in principle measurable in the Galactic Center black-hole system Sgr $A^*$, after expected development of observational techniques that at present enable measurement of the black hole diameter, not details of the shape. 
The Galaxy center (Sgr $A^*$) seems to be also a promising candidate for testing the time delay effects both for phenomena related to the accretion disc and flares observed there , and for some expected lensing phenomena connected to the observed stars orbiting the Sgr $A^*$ central black hole. 

We have found that observable phenomena could be expected for the time-delay effects. Of special interest is comparison of time delays generated for sources in vicinity of the Sgr $A^*$ black hole (both stars and disc hot spots) and those related to weak lensing of some distant sources \cite{Zak:2003:,Sereno:2006}. 

Similarly, keeping rotational parameter fixed, the negative tidal charge has tendency to make the isoradial curve images (both direct and indirect) larger and less deformed while the positive tidal charge influence is of opposite character. On the other hand, for fixed rotational parameter of the black hole and disc radiating at the innermost part above the innermost stable orbit at $r=r_{ms}$, the negative tidal charge restricts the radiating ring image simply because the radius $r_{ms}$ grows with decreasing value of braneworld parameter $b$. Suppresion of the relativistic effects can be measurable also in the spectral line profiles generated by the inner hot part of the disc radiating at special X-ray line \cite{SS:b:RAGTime:2007:Proceedings}. 

The optical tests have to be confronted with the data obtained from quasiperiodic oscillations observed in some black-hole systems (microquasars \cite{Rem-McCli:2006:ARASTRA:}). The orbital resonance model gives good estimates of the black-hole parameters \cite{Tor-Abr-Klu-Stu:2005:,Tor:2005a:,Tor:2005b:}; this model has been recently generalized to the case of braneworld Kerr black holes \cite{Stu-Kot:2008}. It is shown that in the case of microquasar  GRS 1915+105 and Galactic Center Sgr $A^*$ black holes with the negative braneworld parameter $b$ are allowed by the observational data \cite{Stu-Kot:2008}. Detailed modelling of optical phenomena connected to the oscillating discs or orbiting (oscillating) hot spots and related resonant phenomena between the oscillation modes could be very promising in putting limits on allowed values of the tidal charge of the black hole. We plan to elaborate such modelling in future. 


\section*{Acknowledgements}
Research supported by the Czech grant MSM 4781305903 and LC 06014. One of the authors (Zden\v{e}k Stuchl\'{i}k) would like to express his gratitude to the Czech Committee for Collaboration with CERN for support.



\begin{thebibliography}{10}

\bibitem{Ali-Gum:2005:}
A.~N. Aliev and A.~E. G{\"u}mr{\"u}k{\c c}{\"u}o{\u g}lu.
\newblock Charged rotating black holes on a 3-brane.
\newblock {\em Phys. Rev. D}, 71(10):104027--+, 2005.

\bibitem{Ark-Dim-Dva:1998:}
N.~Arkani-Hamed, S.~Dimopoulos, and G.~Dvali.
\newblock The hierarchy problem and new dimensions at a millimeter.
\newblock {\em Phys. Lett. B}, 429:263--272, 1998.

\bibitem{Asch:2004:ASTRA:}
B.~Aschenbach.
\newblock Measuring mass and angular momentum of black holes with
  high-frequency quasi-periodic oscillations.
\newblock {\em Astronom. and Astrophys.}, 425:1075--1082, 2004.

\bibitem{Asch:2007:}
B.~Aschenbach.
\newblock Measurement of mass and spin of black holes with QPOs.
\newblock {arXiv:0710.3454}, 2007.

\bibitem{Bao-Stu:1992:}
G.~Bao and Z.~Stuchl{\'{i}}k.
\newblock Accretion disk self-eclipse:x-ray light curve and emmision line.
\newblock {\em Astrophys. J.}, 400:163--169, 1992.

\bibitem{Bardeen:1973:}
J.~M. Bardeen.
\newblock Timelike and null geodesics in the kerr metric.
\newblock In {\em Black holes (Les astres occlus), p. 215 - 239}, pages
  215--239, 1973.

\bibitem{Bar-Pre-Teu:1972:}
J.~M. Bardeen, W.~H. Press, and S.~A. Teukolsky.
\newblock Rotating black holes: Locally nonrotating frames, energy extraction,
  and scalar synchrotron radiation.
\newblock {\em Astrophys. J.}, 178:347--370, December 1972.

\bibitem{Bic-Stu:1976:}
J.~Bi{\v{c}}{\'{a}}k and Z.~Stuchl{\'{i}}k.
\newblock On the latitudinal and radial motion in the field of a rotating black
  hole.
\newblock {\em Bull. Astron. Inst. Czechosl.}, 27(3), 1976.

\bibitem{Cha-etal:2001:}
A.~Chamblin, H.~S. Reall, H.~Shinkai, and T.~Shiromizu.
\newblock Charged brane-world black holes.
\newblock {\em Phys. Rev. D}, 63(6):064015, 2001.

\bibitem{Cun-Bar:1973:}
C.~T. Cunningham and J.~M. Bardeen.
\newblock The optical appearance of a star orbiting an extreme kerr black hole.
\newblock {\em Astrophys. J.}, 183:237--264, 1973.

\bibitem{Cze-etal:2007:}
B. Czerny, M. Mo\'{s}cibrodska, D.Proga, T.~K. Das and A. Siemiginowska.
\newblock Low angular momentum accretion flow model of Sgr $A^*$ activity.
\newblock In {\em Proceedings of RAGtime 8/9: Workshops on black holes and
  netron stars, Opava, 15--19/19--21 September 2006/2007},p.35--44, 2007.

\bibitem{Dad-etal:2000:}
N.~Dadhich, R.~Maartens, P.~Papadopoulos, and V.~Rezania.
\newblock Black holes on the brane.
\newblock {\em Phys. Lett. B}, 487:1, 2000.

\bibitem{Fel-Cal:1972:}
F.~{de Felice} and M.~{Calvani}.
\newblock {Orbital and vortical motion in the Kerr metric.}
\newblock {\em Nuovo Cimento B Serie}, 10:447--458, 1972.

\bibitem{Dov-Karas-Mas-Mar:2005:}
M.~{Dov{\v c}iak}, V.~{Karas}, G.~{Matt}, and A.~{Martocchia}.
\newblock {Polarization of radiation from AGN accretion discs - the lamp-post
  model}.
\newblock In S.~{Hled{\'{\i}}k} and Z.~{Stuchl{\'{\i}}k}, editors, {\em RAGtime
  6/7: Workshops on black holes and neutron stars}, pages 47--54, December
  2005.

\bibitem{Emp-Mas-Rat:2002:}
R.~{Emparan}, M.~{Masip}, and R.~{Rattazzi}.
\newblock {Cosmic rays as probes of large extra dimensions and TeV gravity}.
\newblock {\em Phys. Rev. D}, 65(6):064023--+, March 2002.

\bibitem{fab-rees-ste-whi:1989:}
A.~C. {Fabian}, M.~J. {Rees}, L.~{Stella}, and N.~E. {White}.
\newblock {X-ray fluorescence from the inner disc in Cygnus X-1}.
\newblock {\em MNRAS}, 238:729--736, May 1989.

\bibitem{Fan-Cal-Fel-Cad:1997:}
C.~Fanton, M.~Calvani, F.~de~Felice, and A.~\v{C}ade\v{z}.
\newblock Detecting accretion disks in active galactic nuclei.
\newblock {\em Publ. Astron. Soc. Japan}, (49):159--169, 1997.

\bibitem{Ger-Maa:2001:}
C.~Germani and R.~Maartens.
\newblock Stars in the braneworld.
\newblock {\em Phys. Rev. D}, 64:124010, 2001.

\bibitem{Ghez:2005:}
A.~M. Ghez.
\newblock Stellar orbits and the supermassive black hole at the center of our
  galaxy.
\newblock {\em Am. Astro. Soc.}, 37:531, 2005.


\bibitem{Ghez-etal:2008:}
A.~M. Ghez, S. Salim, N.~N. Weinberg, J.~R. Lu, T. Do, J.~K. Dunn, K. Matthews, M. Morris, S. Yelda, E.~E. Becklin, T. Kremenek, M. Miloslavljevic, and J. Naiman
\newblock{Probing the properties of the Milky Way's central supermassive black hole with stellar orbits.}
\newblock{\em IAU Symposium},248, p.52--58, 2008.
\newblock{ arXiv:astro-ph 0808.2870}

\bibitem{Kar-Vok-Pol:1992:}
V.~Karas, D.~Vokrouhlicky, and A.~G. Polnarev.
\newblock In the vicinity of a rotating black hole - a fast numerical code for
  computing observational effects.
\newblock {\em MNRAS}, 259:569--575, December 1992.

\bibitem{Kerr:1963:}
R.~P. Kerr.
\newblock Gravitational field of a spinning mass as an example of algebraically
  special metrics.
\newblock {\em Phys. Rev. Lett.}, (11):26, 1963.

\bibitem{Kra:2005:}
G.~V. Kraniotis.
\newblock Frame dragging and bending of light in kerr and kerr (anti) de sitter
  spacetimes.
\newblock {\em Class. and Quant. Grav.}, 22:4391--4424, 2005.

\bibitem{Kra:2007:}
G.~V. Kraniotis.
\newblock Periapsis and gravitomagnetic precessions of stellar orbits in kerr
  and kerr-de sitter black hole spacetimes.
\newblock {\em Class. and Quant. Grav.}, 24:1775--1808, 2007.

\bibitem{Laor:1991:}
A.~Laor.
\newblock Line profiles from a disk around a rotating black hole.
\newblock {\em Astrophys. J.}, 376:90--94, 1991.

\bibitem{Maa:2004:}
R.~Maartens.
\newblock Brane-world gravity.
\newblock {\em Living Rev. Rel.}, 7:7, 2004.

\bibitem{Mat-Fab-Ros:1993:}
G.~Matt, A.~C. Fabian, and R.~R. Ross.
\newblock Iron k-alpha lines from x-ray photoionized accretion discs.
\newblock {\em MNRAS}, 262:179--186, May 1993.

\bibitem{McCli-Nar-Sha:2007:}
J.~E. McClintock, R.~Narayan, and R.~Shafee.
\newblock Estimating the spins of stellar-mass black holes.
\newblock {arXiv:0707.4492}, 2007.
\newblock To appear in Black Holes, eds. M. Livio and A. Koekemoer (Cambridge
  University Press), in press (2008).

\bibitem{MTW}
C.~W. Misner, K.~S. Thorne, and J.~A. Wheeler.
\newblock {\em Gravitation}.
\newblock W.~H. Freeman and company,~San Francisco, 1973.

\bibitem{Nar-Mcl-Sha:2007:}
R.~Narayan, J.~E. McClintock, and R.~Shafee.
\newblock Estimating the spins of stellar-mass black holes by fitting their
  continuum spectra.
\newblock {arXiv:0710.4073}, October 2007.

\bibitem{Ran-Sun:1999:}
L.~Randall and R.~Sundrum.
\newblock An alternative to compactification.
\newblock {\em Phys. Rev. Lett.}, 83(23):4690--4693, 1999.

\bibitem{Rau-Bla:1994:}
K.~P. Rauch and R.~D. Blandford.
\newblock Optical caustics in a kerr spacetime and the origin of rapid x-ray
  variability in active galactic nuclei.
\newblock {\em Astrophys. J.}, (421):46--68, 1994.

\bibitem{Reid:2008:}
M.~J. Reid.
\newblock{Is there a supermassive black hole at the center of the Milky Way?}
\newblock{arXiv:0808.2624}, 2008 

\bibitem{Rem-McCli:2006:ARASTRA:}
R.~A. Remillard and J.~E. McClintock.
\newblock X-ray properties of black-hole binaries.
\newblock {\em Ann. Rev. of Astron. and Astrophys.}, 44(1):49--92, September
  2006.

\bibitem{SS:b:RAGTime:2007:Proceedings}
J.~Schee and Z.~Stuchl\'{i}k.
\newblock Spectral line profile in brany kerr spacetime.
\newblock In {\em Proceedings of RAGtime 8/9: Workshops on black holes and
  netron stars, Opava, 15--19/19--21 September 2006/2007},p.209--220, 2007.

\bibitem{SSJ:RAGTime:2005:Proceedings}
J.~Schee, Z.~Stuchl\'{i}k, and J.~Jur\'{a}\v{n}.
\newblock Light escape cones and raytracing in kerr geometry.
\newblock In {\em Proceedings of RAGtime 6/7: Workshops on black holes and
  netron stars, Opava, 16--18/18--20 September 2004/2005}, p.143--155, 2005.

\bibitem{Sereno:2006}
M. Sereno and  F. de Luca,
\newblock{ "{Analytical Kerr black hole lensing in the weak deflection limit}"},
\newblock{\em Phys. Rev. D}, 74, 12, 2006, 
\newblock{\em arXiv:astro-ph/0609435}

\bibitem{Shi-Mae-Sas:2000:}
T.~Shiromizu, K.~Maeda, and M.~Sasaki.
\newblock The einstein equations on the 3-brane world.
\newblock {\em Phys. Rev. D}, 62:024012, 2000.

\bibitem{Stroh:2007a:}
T.~E. Strohmayer.
\newblock Raising the curtain on extreme stars.
\newblock {\em Astronomy}, 35:32--37, March 2007.

\bibitem{Stu:1981b:}
Z.~Stuchl{\'{i}}k.
\newblock Null geodesics in the kerr-newman metric.
\newblock {\em Bull. Astron. Inst. Czechosl.}, 32(6), 1981.

\bibitem{Stu:1981a:}
Z.~Stuchl{\'{i}}k.
\newblock The radial motion of photons in kerr metric.
\newblock {\em Bull. Astron. Inst. Czechosl.}, 32(1), 1981.

\bibitem{Stu-Bao:1992:}
Z.~Stuchl{\'{i}}k and G.~Bao.
\newblock Radiation from hot spots orbiting an extreme reissner-nordstr�m
  black hole.
\newblock {\em General Rel. and Grav.}, 24(9), 1992.

\bibitem{Stu-Bic-Bal:1999:}
Z.~Stuchl\'{i}k, J.~Bi\v{c}\'{a}k, and V.~Balek.
\newblock The shell of incoherent charged matter falling onto a charged
  rotation black hole.
\newblock {\em Gen. Rel. and Grav.}, 31(53), 1999.


\bibitem{Stu-Hle:2002:}
Z.~Stuchl\'{i}k and S.~Hled\'{i}k.
\newblock {Properties of the Reissner-Nordstrom spacetimes with a nonzero cosmological constant.}
\newblock {\em Acta. Phys. Slovaca}, 52, 363, 2002.

\bibitem{Stu-Kot:2008}
Z. Stuchl{\'{\i}}k  and A. Kotrlov{\'a},
\newblock{ "{Orbital resonances in discs around braneworld Kerr black holes}"},
\newblock{\em Gen. Rel. and Grav.},  doi: {10.1007/s10714-008-0709-2}, 2008,
\newblock{\em arxiv:0812.5066} 

\bibitem{Tor:2005a:}
G.~T\"{o}r\"{o}k.
\newblock A possible 3:2 orbital epicyclic resonance in qpo frequencies of sgr
  a*.
\newblock {\em Astron. and Astrophys.}, 1(440), 2005a.

\bibitem{Tor:2005b:}
G.~T\"{o}r\"{o}k.
\newblock Qpos in microquasars and sgr a* measuring the black hole spin.
\newblock {\em Astronom. Nachr.}, 856(326), 2005b.

\bibitem{Tor-Abr-Klu-Stu:2005:}
G.~T\"{o}r\"{o}k, M.~Abramowicz, W.~Klu\'{z}niak, and Z.~Stuchl\'{i}k.
\newblock {\em Astron. and Astrophys.}, 1(436), 2005.

\bibitem{Yoshi:2007:}
Y.~C. Joshi.
\newblock Displacement of the Sun from the Galactic plane.
\newblock {\em MNRAS}, 378:768--776, June 2007.
\newblock {arXiv:astro-ph/0704.0950}.

\bibitem{Zak:2003:}
A.~F.Zakharov.
\newblock The iron $K_{\alpha}$-line as a tool for analysis of black hole characteristics.
\newblock {\em Publications of the Astronomical Observatory of Belgrade}, 76, 147--162, 2003.
\newblock {arXiv: astro-ph/0411611}.

\end{thebibliography}

\end{document}